\documentclass{aa}  
\usepackage{graphicx}
\usepackage{txfonts}
\usepackage[colorlinks=true,citecolor=blue,linkcolor=blue,urlcolor=cyan]{hyperref}
\usepackage{pdflscape}
\newcommand{\nodata}{}
\renewcommand\labelitemii{$\m@th\bullet$}
\begin{document} 

\title{$\chi^{1}$ Fornacis cluster DANCe.}
\subtitle{Census of stars, structure, and kinematics of the cluster with \textit{Gaia}-EDR3
\thanks{Tables \ref{tab_members}, \ref{tab_prob}, \ref{tab_isochrone}, and \ref{tab_phyiscal_properties} are only available in electronic form at the CDS via anonymous ftp to cdsarc.u-strasbg.fr (130.79.128.5) or via http://cdsweb.u-strasbg.fr/cgi-bin/qcat?J/A+A/}\fnmsep 
\thanks{Based on observations collected at the European Southern Observatory under ESO programmes 0106.D-0429(A) and 0106.D-0429(B).}}

\author{
P.A.B.~Galli \inst{1}
\and
H.~Bouy \inst{1}
\and
J.~Olivares\inst{1}
\and 
N.~Miret-Roig\inst{2}
\and
L.M.~Sarro\inst{3}
\and
D.~Barrado\inst{4}
\and
A.~Berihuete\inst{5}
}

\institute{
Laboratoire d’Astrophysique de Bordeaux, Univ. Bordeaux, CNRS, B18N, allée Geoffroy Saint-Hillaire, F-33615 Pessac, France\\
\email{phillip.galli@u-bordeaux.fr}
\and
University of Vienna, Department of Astrophysics, T{\"u}rkenschanzstrasse 17, 1180 Wien, Austria
\and
Depto. de Inteligencia Artificial, UNED, Juan del Rosal, 16, 28040 Madrid, Spain
\and
Centro de Astrobiolog\'ia, Depto. de Astrof\'isica, INTA-CSIC, ESAC Campus, Camino Bajo del Castillo s/n, 28692 Villanueva de la Ca\~nada, Madrid, Spain
\and
Dpto. Estadística e Investigación Operativa, Universidad de Cádiz, Campus Río San Pedro s/n, 11510 Puerto Real, Cádiz, Spain
}

\date{Received September 15, 1996; accepted March 16, 1997}

 \abstract
{The $\chi^{1}$ Fornacis cluster (Alessi~13) is one of a few open clusters of its age and distance in the Solar neighbourhood that ought to benefit from more attention as it can serve as a cornerstone for numerous future studies related to star and planet formation. }
{We take advantage of the early installment of the third data release of the \textit{Gaia} space mission in combination with archival data and our own observations, to expand the census of cluster members and revisit some properties of the cluster. }
{We applied a probabilistic method to infer membership probabilities over a field of more than 1\,000~deg${^2}$ to select the most likely cluster members and derive the distances, spatial velocities, and physical properties of the stars in this sample. }
{We identify 164 high-probability cluster members (including 61 new members) covering the magnitude range from 5.1 to 19.6~mag in the G-band. Our sample of cluster members is complete down to 0.04~M$_{\odot}$. We derive the distance of $108.4\pm0.3$~pc from Bayesian inference and confirm that the cluster is comoving with the Tucana-Horologium, Columba, and Carina young stellar associations. We investigate the kinematics of the cluster from a subsample of stars with measured radial velocities and we do not detect any significant expansion or rotation effects in the cluster. Our results suggest that the cluster is somewhat younger (about 30~Myr) than previously thought. Based on spectroscopic observations, we argue that the cluster is mass-segregated and that the distribution of spectral types shows little variation compared to other young stellar groups.}
{In this study, we deliver the most complete census of cluster members that can be done with \textit{Gaia} data alone and we use this new sample to provide an updated picture on the 6D structure of the cluster. }

\maketitle

\section{Introduction}\label{section1}

The young stellar clusters of the Solar neighbourhood constitute a primary laboratory for the investigation of the early stages of star and planet formation, as well as to connect models to observations and visualise the local structure of the Galaxy. Significant progress has been made in recent years to discover and characterise nearby young stellar clusters thanks to the precise astrometry delivered by the \textit{Gaia} satellite \citep[see e.g.][]{CantatGaudin2018,Sim2019,Meingast2019,CastroGinard2020}. One important contribution to the field has been the recognition of the $\chi^{1}$~Fornacis cluster (also known as Alessi~13) as one of a few open clusters of its distance ($\sim100$~pc) and age ($\sim40$~Myr), with more than a hundred members, making it a major constituent of the Solar neighbourhood \citep{Zuckerman2019}. In the following, we refer to the $\chi^{1}$~Fornacis cluster (XFOR) using the same terminology that is also used to name the young stellar groups in the Solar neighbourhood after either a constellation or the most prominent cluster member; we emphasise that both names used in the literature refer to the same cluster. 

The XFOR cluster is listed in the catalogue of open clusters compiled by \citet{Dias2002}, with only a few cluster members, as it was poorly studied in the pre-\textit{Gaia} era. The cluster was reanalysed by \citet{Yen2018} based on data from the first data release of the \textit{Gaia} catalogue \citep[\textit{Gaia}-DR1,][]{GaiaDR1} where the authors confirmed nine members. The subsequent study conducted by \citet{CantatGaudin2018}, using the second data release of the \textit{Gaia} catalogue \citep[\textit{Gaia}-DR2, ][]{GaiaDR2}, identified 48 cluster members. 
 
More recently, \citet[][hereafter ZKK2019]{Zuckerman2019} revisited the cluster using different selection criteria to search for cluster members based on \textit{Gaia}-DR2 data and increased the sample of cluster members to 108 stars. The authors argue that the cluster is coeval and co-moving with the Tucana-Horologium (THA) and Columba (COL) young stellar associations by comparing their positions in the colour-magnitude diagram and assigned the age of  about 40~Myr to the cluster. Thus, the cluster appears to be significantly younger than 525~Myr as reported previously by \citet{Kharchenko2013}, but still in good agreement with the age of about 30~Myr estimated by \citet{Mamajek2016} from the saturated X-ray emission of the stars. Another interesting finding reported by ZKK2019 is the abundance of M-dwarfs in the cluster with mid-infrared excess emission indicative of debris discs with warm dust that could be associated with rocky planet formation, making the XFOR cluster a promising target in the search for exoplanets. 

The early installment of the third data release of the \textit{Gaia} space mission \citep[\textit{Gaia}-EDR3,][]{GaiaEDR3} represents a significant advance with respect to its predecessor delivering more precise astrometry and photometry, along with reduced systematic errors that allow for more precise studies. Given the close proximity and young age of the XFOR cluster, we believe that it will become an important cornerstone to many studies related to star formation from models to the physical properties of the stars (e.g. as done  for the Pleiades). In this paper, we revisit the census of cluster members using \textit{Gaia}-EDR3 data and combine our results with spectroscopic observations to revisit the cluster properties. This study is conducted as part of the Dynamical Analysis of Nearby Clusters \citep[DANCe, ][]{Bouy2013} project and follows a similar methodology applied to other stellar clusters in this series \citep{Olivares2019,Miret-Roig2019,Galli2020a,Galli2020b,Galli2021}.

The paper is organised as follows. In Section~\ref{section2}, we perform a new membership analysis of the cluster based on the methodology developed by our team for the DANCe project, which allows us to identify 164 cluster members and confirming most of the members previously identified by ZKK2019 as well as adding new ones to the list. Section~\ref{section3} describes the results obtained from our observations to determine the spectral type of the stars from low-resolution spectroscopy. Then, in Sect.~\ref{section4}, we use our new sample of cluster members to revisit several properties of the cluster (distance, kinematics, age, spatial distribution, and initial mass function). Finally, we present our conclusions in Sect.~\ref{section5}.

\section{Membership analysis}\label{section2}

We selected the most likely members of the XFOR cluster based on Gaia-EDR3 data using the same methodology applied by our team to other young stellar groups \citep[see e.g.][]{Miret-Roig2019,Galli2020a,Galli2020b,Galli2021}. We summarise the main steps of our membership analysis in this section and refer the reader to the original papers \citep{Sarro2014,Olivares2019} for further details on the methodology.  

To begin with, we downloaded the \textit{Gaia}-EDR3 catalogue in the region defined by $215^{\circ}\leq l \leq 260^{\circ}$ and $-68^{\circ}\leq b \leq -45^{\circ}$ in Galactic coordinates. This input catalogue contains 2\,057\,639 sources and includes all cluster members previously identified by ZKK2019. The representation space (i.e. the set of observables) we use in the membership analysis includes both the astrometric and photometric features of the \textit{Gaia}-EDR3 catalogue. It is defined by $\mu_{\alpha}\cos\delta$, $\mu_{\delta}$, $\varpi$, $G_{RP}$, and $G-G_{RP}$, and it also includes 1\,633\,156 sources of the input catalogue with complete data. We do not use the $G_{BP}$ magnitude in our analysis because the flux in this band is likely to be overestimated for red sources \citep[see e.g.][]{Fabricius2021}. In addition, we corrected the published G-band photometry of the sources with a six-parameter astrometric solution in the \textit{Gaia}-EDR3 catalogue fainter than $G=13$~mag, using the formula presented in \citet{Riello2021}, as explained in Appendix~A of \citet{GaiaEDR3}.

We modelled the field population using Gaussian mixture models (GMM)\footnote{A Gaussian Mixture Model is a model that describes the probability distribution of data points as a linear combination of a finite number of Gaussian distributions.} with a random sample of 10$^{6}$ sources from the input catalogue. We tested the field model with different numbers of GMM components (i.e. 120, 140, 160, 180, 200, 220, and 240) and took the model with 160 components that returns the minimum Bayesian information criterion \citep[BIC,][]{Schwarz1978} value. The cluster was modelled using GMM in the space of proper motions and a principal curve in the photometric space, with a spread at any point given by a multivariate Gaussian to define the cluster sequence. Both the cluster locus in the space of proper motions and the principal curve in the photometric space are initialised based on an initial list of members. We use the sample of cluster members identified by ZKK2019 as initial list for the first iteration of our membership analysis. The method computes membership probabilities for all sources in the catalogue and classifies them into members and non-members based on a probability threshold, $p_{in}$, which is predefined by the user. The cluster members that result from this process serve as input for the next iteration of the algorithm and this procedure is repeated until the list of cluster members remains fixed after successive iterations. Then we use the field and cluster models inferred by our method to generate synthetic data and define the optimum probability threshold $p_{opt}$ for our membership analysis based on the performance of our classifier \citep[see][]{Olivares2019}. Finally, we classify the sources in our catalogue as members ($p\geq p_{opt}$) and non-members. 

In Table~\ref{tab_membership}, we compare the results of our membership analysis using different values for the user-defined probability threshold $p_{in}$. The true positive rate (TPR, i.e. the fraction of cluster members recovered by the method) and contamination rate (CR, i.e. the fraction of field stars classified as cluster member by the method) given for each solution are derived from the synthetic data (based on the cluster and field models inferred in the previous steps) and, therefore, these represent only estimates of the indicators that remain useful in the absence of the true distributions. We see little variation in the number of cluster members among the solutions obtained from different $p_{in}$ thresholds. However, we note that high $p_{in}$ values ($p_{in}=0.8$ and $p_{in}=0.9$) lead to low values of the optimum probability threshold, $p_{opt}$, and, consequently, to the inclusion of stars with lower membership probabilities in our solution, while the least conservative threshold of $p_{in}=0.5$ yields the highest $p_{opt}$ value. The solution obtained with $p_{in}=0.6$ offers the best compromise between the two probability thresholds. We therefore use this solution to define the list of XFOR cluster members derived in this study (see more comments below).

\begin{table}[!h]
\centering
\caption{Comparison of results for the membership analysis using different probability thresholds.
\label{tab_membership}}
\begin{tabular}{ccccc}

\hline\hline
$p_{in}$&$p_{opt}$&$N_{\star}$&TPR&CR\\
\hline
0.5& 0.74 & 160 & 0.932 $\pm$ 0.025 & 0.093 $\pm$ 0.035 \\
0.6& 0.62 & 161 & 0.947 $\pm$ 0.033 & 0.087 $\pm$ 0.017 \\
0.7& 0.74 & 156 & 0.934 $\pm$ 0.026 & 0.087 $\pm$ 0.046 \\
0.8& 0.43 & 162 & 0.940 $\pm$ 0.029 & 0.068 $\pm$ 0.017 \\
0.9& 0.20 & 155 & 0.960 $\pm$ 0.033 & 0.061 $\pm$ 0.019 \\
\hline\hline
\end{tabular}
\tablefoot{We provide the optimum probability $p_{opt}$, number of cluster members, true positive rate (TPR), and contamination rate (CR) for each solution using different probability thresholds $p_{in}$.}
\end{table}

\begin{figure*}
\begin{center}
\includegraphics[width=0.33\textwidth]{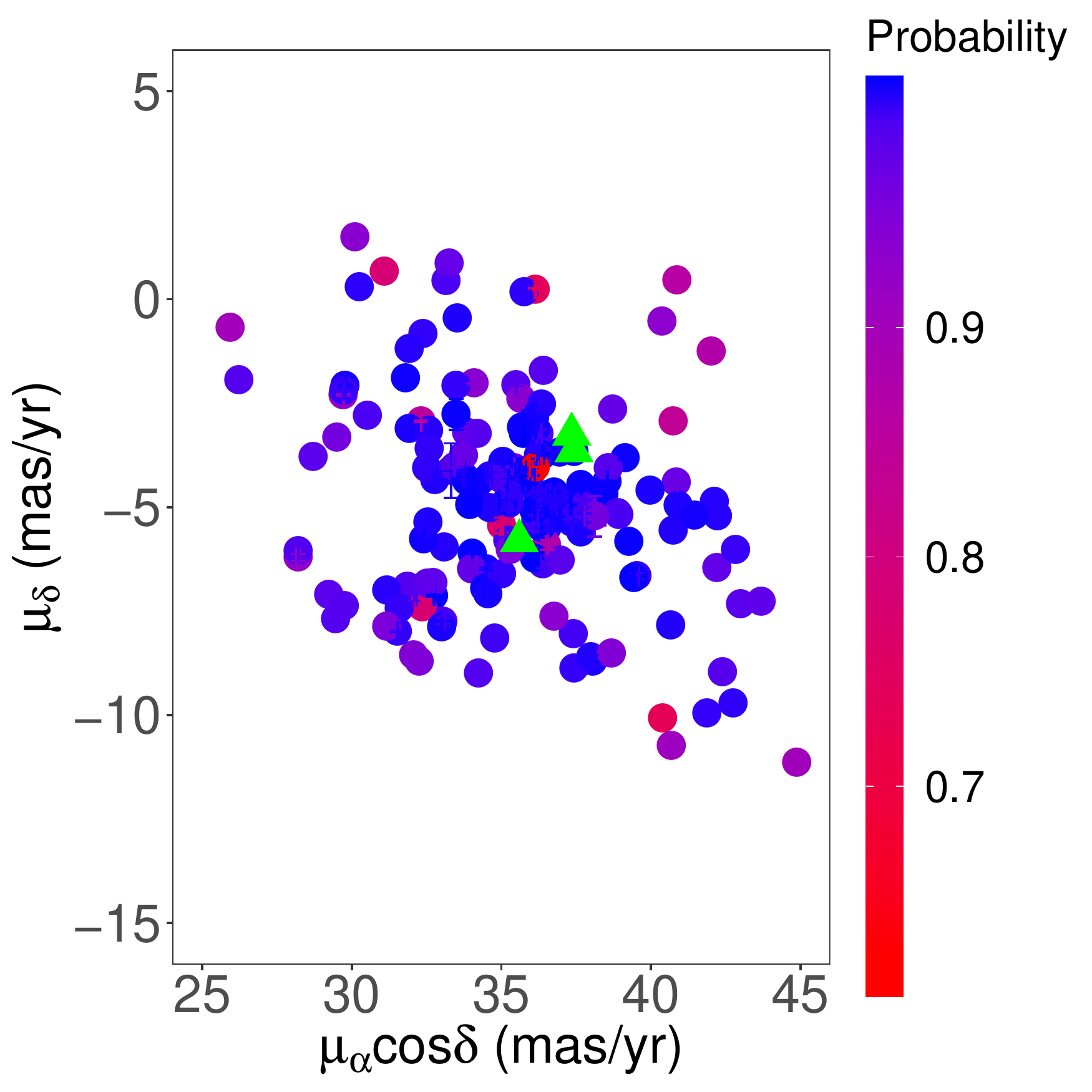}
\includegraphics[width=0.33\textwidth]{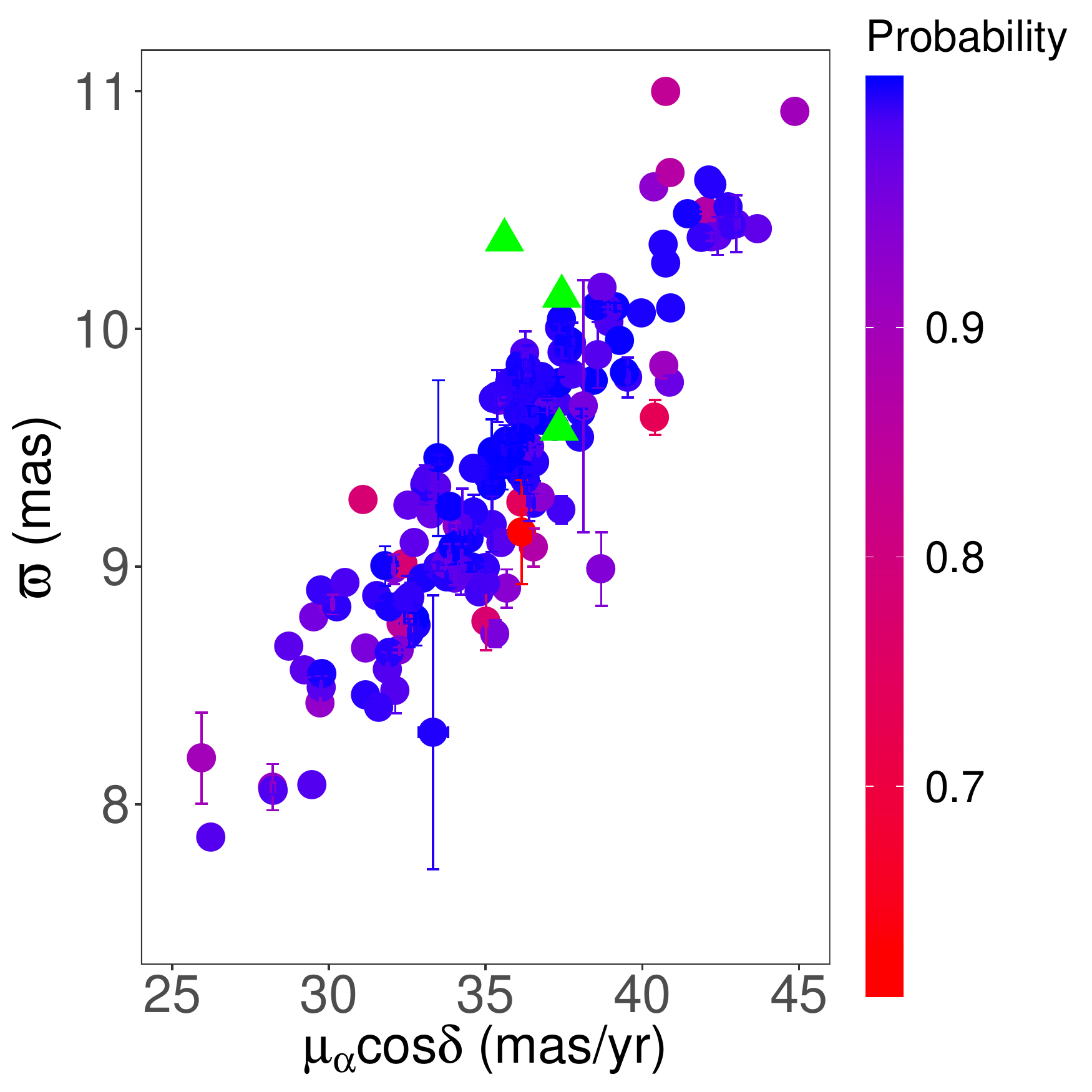}
\includegraphics[width=0.33\textwidth]{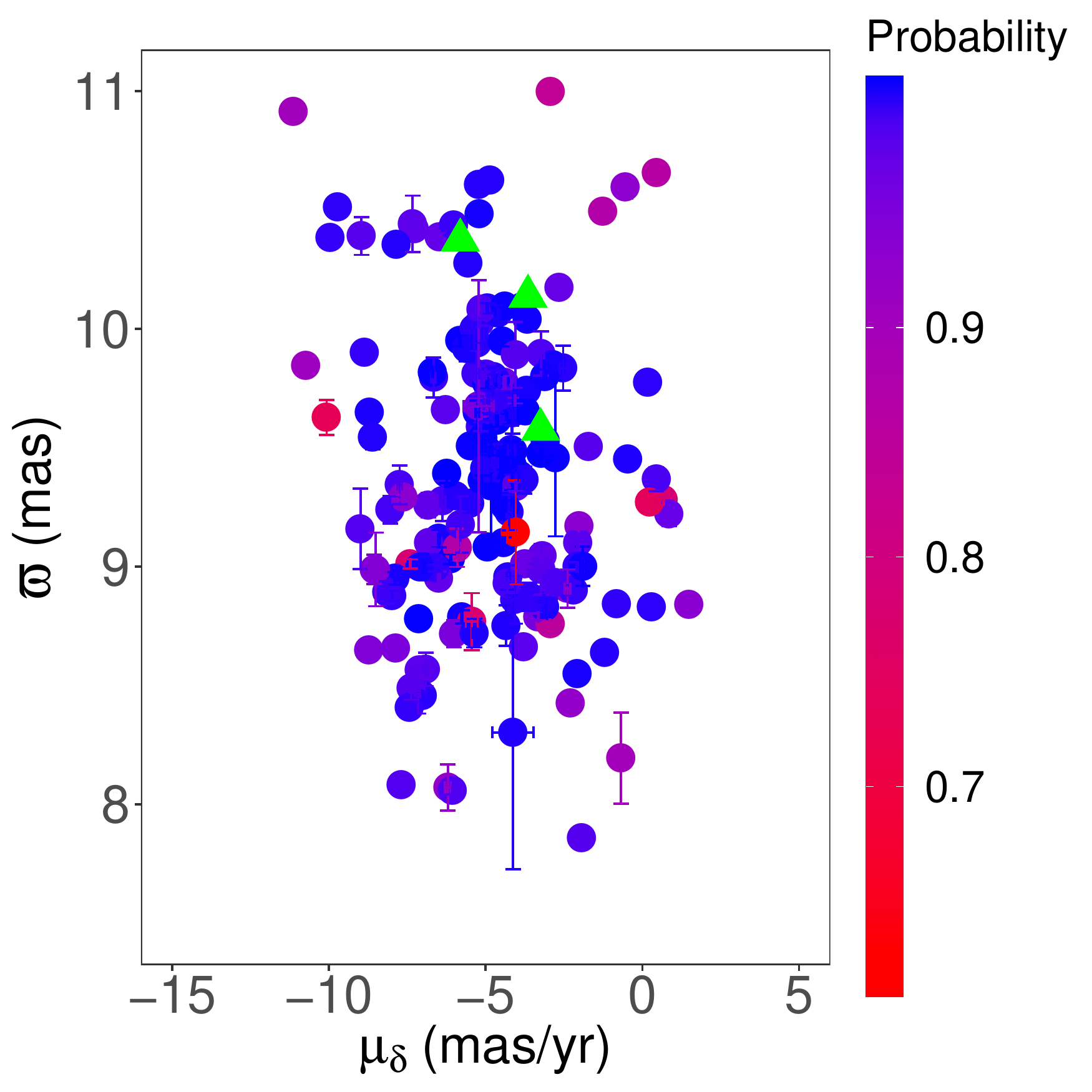}
\caption{
Proper motions and parallaxes of the 164 cluster members identified in our membership analysis. The stars are colour-coded based on their membership probability that is scaled from zero to one. The three sources marked with green triangles denote the components of binary systems that we added manually to the sample, as explained in the text of Sect.~\ref{section2}.
}\label{fig_pm_plx}
\end{center}
\end{figure*}

\begin{figure}
\begin{center}
\includegraphics[width=0.50\textwidth]{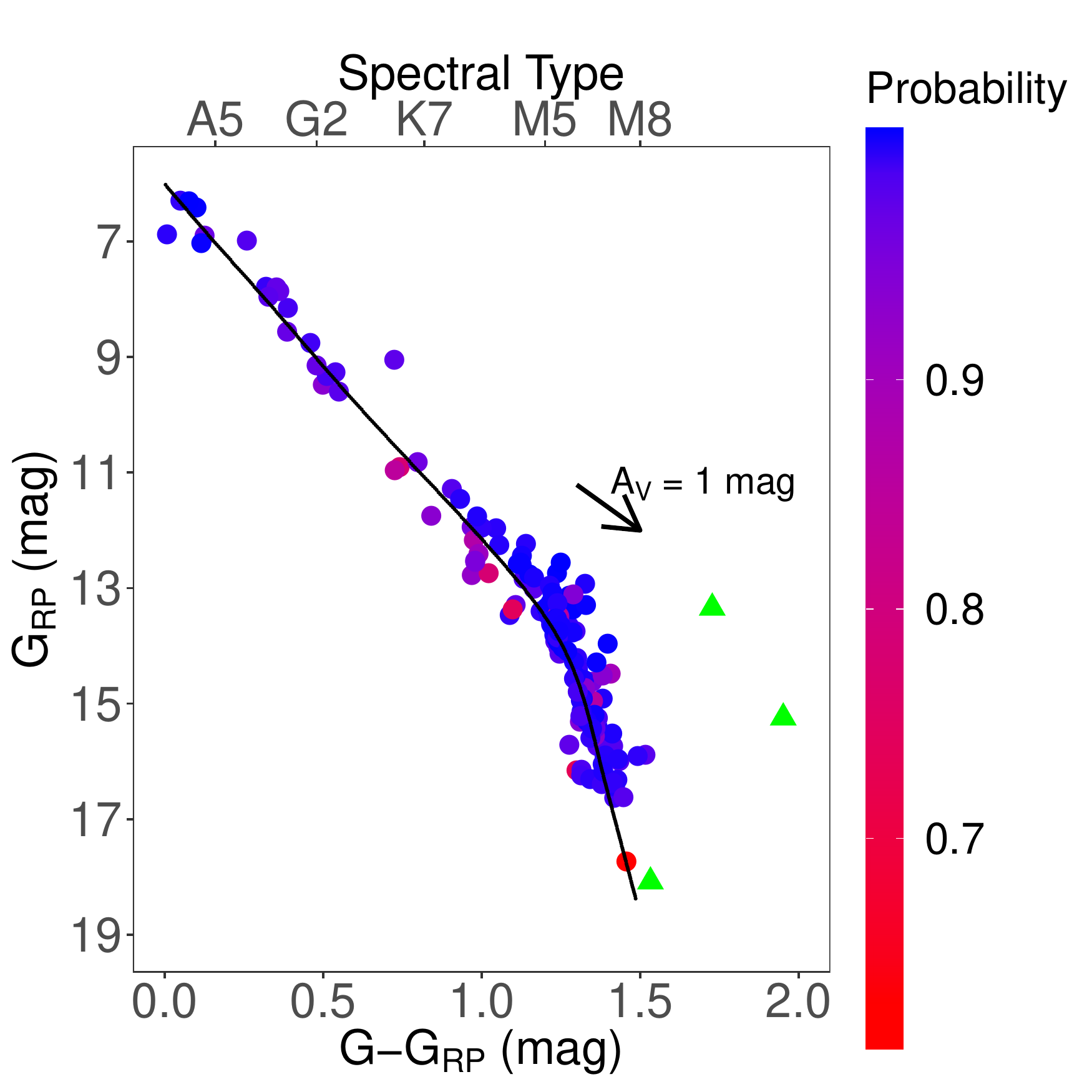}
\caption{
\label{fig_venn}
Colour-magnitude diagram of the cluster members selected in our membership analysis. The stars are colour-coded based on their membership probability that is scaled from zero to one. The black solid line shows the empirical isochrone of the cluster (see Table~\ref{tab_isochrone}). The arrow indicates the extinction vector of $A_{V}=1$~mag converted to the \textit{Gaia} bands using the relative extinction values given by \citet{Wang2019}. The three sources marked with green triangles denote the components of binary systems that we added manually to the sample, as explained in the text of Sect.~\ref{section2}.
}\label{fig_CMD}
\end{center}
\end{figure}

We note that only one component of three binary systems in that sample was directly identified by our methodology, so we add the missing components of these systems, namely:  \textit{Gaia}~EDR3~5053156704532702720 (032317-343547), \textit{Gaia}~EDR3~4854879573671981056 (032215-354719), and \textit{Gaia}~EDR3~4854540344270707200 (032513-370909) to the list. These sources have proper motions and parallax that are consistent with membership in the cluster (see Figure~\ref{fig_pm_plx}) and they were rejected by our method due to their photometry. \textit{Gaia} EDR3 5053156704532702720 (032317-343547) is the faintest cluster member and it is located in a region of the colour-magnitude diagram where our model is less efficient because it is trained on only a few sources in this colour regime (i.e. candidate members from the literature). On the other hand, we see that \textit{Gaia} EDR3 4854879573671981056 (032215-354719) and \textit{Gaia} EDR3 4854540344270707200 (032513-370909) are significantly above the empirical isochrone of the cluster (see Figure~\ref{fig_CMD}), which results from a less reliable photometry. Indeed, we verified that all transits in the $G_{RP}$ band for these two sources are affected by blending effects (as indicated by the fields \texttt{phot\_rp\_n\_obs} and \texttt{phot\_rp\_n\_blended\_transits} in the \textit{Gaia}-EDR3 catalogue).  

We also searched for other astronomical objects (among the rejected sources) that exhibit astrometric features consistent with membership in the cluster but are automatically discarded by our method due to their relative position with respect to the cluster isochrone in the colour-magnitude diagram. We find one white dwarf candidate, namely, \textit{Gaia} EDR3 4856170984438879232 \citep[see][]{Fusillo2019}, that has proper motion ($\mu_{\alpha}\cos\delta=30.218\pm1.099$~mas/yr, $\mu_{\delta}=-5.781\pm1.339$~mas/yr) and parallax ($\varpi=8.270\pm0.979$~mas) consistent with the other cluster members in the sample. However, based on the \citet{Bergeron1995} evolutionary sequences for white dwarfs\footnote{The evolutionary cooling sequences for white dwarfs used in this study were downloaded from \href{http://www.astro.umontreal.ca/~bergeron/CoolingModels}{http://www.astro.umontreal.ca/~bergeron/CoolingModels}.}, we conclude that this source is much older (about 9~Gyr) than the XFOR cluster (see discussion in Sect.~\ref{section4.4}) and it is unlikely to be a cluster member. 
 
Our final list of cluster members consists of 164 stars (see Table~\ref{tab_members}). We also provide in Table~\ref{tab_prob} the membership probability for all sources in the input catalogue used for the membership analysis derived from the different $p_{in}$ values investigated in our study. This offers subsequent studies the opportunity to select cluster members with criteria that are more specific to their scientific objectives than the ones used in this paper. Our new sample of cluster members includes stars in the magnitude range from $G=5.1$ to $G=19.6$~mag. The absolute magnitude of the stars range from about $G_{abs}=-0.3$ to $G_{abs}=14.7$~mag. Figures~\ref{fig_pm_plx} and \ref{fig_CMD} show the proper motions, parallaxes, and colour-magnitude diagram of the selected sources. The empirical isochrone of the cluster derived in our membership analysis is given in Table~\ref{tab_isochrone}. The observed correlation between $\mu_{\alpha}\cos\delta$ and $\varpi$ illustrated in Figure~\ref{fig_pm_plx} can be explained by projection effects. Cluster members located at different sky positions (but with similar space velocities) display proper motions and parallaxes that vary in proportion to each other. 

The XFOR cluster is located in a sky region that overlaps with the nearby THA and COL young stellar associations. We compiled the lists of known members given in the literature for THA and COL \citep{Kraus2014,Gagne2018a,Gagne2018b,Gagne2018c}, cross-matched them with the list of stars identified in our membership analysis, and confirmed that our sample of XFOR cluster members is not contaminated by known members of these two stellar groups. The published lists of stars in THA and COL suggest that these stellar groups are closer to Earth than the XFOR cluster by about 50~pc. In a recent study, \citet{Gagne2021} proposed that many of the nearby moving groups in the Solar neighbourhood are more extended than previously thought and may be fragments of dissolving open clusters. In particular, the authors associate THA and COL with the IC~2602 and Platais~8 open clusters, respectively. Thus, if this scenario is correct, then future studies may reveal the existence of additional THA and COL members in the vicinity of the XFOR cluster.

Our membership analysis confirms 103 stars (out of 107 sources, see Figure~\ref{fig_venn}) that were previously identified as cluster members by ZKK2019.\footnote{ZKK2019 identified 107 candidate members with \textit{Gaia}-DR2 data, but one of them, \textit{Gaia} EDR3 4860808041586101632 (032559-355730), is not resolved by the \textit{Gaia} satellite, so they refer to 108 cluster members throughout that study. Here, we count this system as one single entry in our tables.} Only four stars (\textit{Gaia} EDR3 4860808316464009856, \textit{Gaia} EDR3 5046516311431900160, \textit{Gaia} EDR3 4852758100345682816, and \textit{Gaia} EDR3 5048052294816437120) from that study cannot be recovered by our methodology. One of them, namely, \textit{Gaia} EDR3 4860808316464009856 (CD-36 1289), was classified as a ``core-member'' of the cluster by ZKK2019 based on its proper motion ($\mu_{\alpha}\cos\delta=34.6\pm1.0$~mas/yr and $\mu_{\delta}=-1.8\pm1.0$~mas/yr), given in the UCAC4 catalogue \citep{Zacharias2013}. Indeed, the \textit{Gaia}-EDR3 proper motion ($\mu_{\alpha}\cos\delta=25.255\pm0.054$~mas/yr, $\mu_{\delta}=-1.366\pm0.080$~mas/yr) is not consistent with membership in the cluster, which explains the reason it was rejected by our method. The same conclusion holds if we use the \textit{Gaia}-DR2 measurements in the membership analysis. Thus, we discard this star from our study for consistency with the rest of our membership analysis where we use for all stars the same source of astrometric data (i.e. \textit{Gaia}-EDR3) for both proper motion and parallax. The other three sources rejected in this study are classified as ``potential members'' by ZKK2019. Our membership analysis shows that they exhibit low membership probabilities ($p<0.13$) that are below the adopted probability threshold (see Table~\ref{tab_membership}). 

\begin{figure}[!h]
\begin{center}
\includegraphics[width=0.49\textwidth]{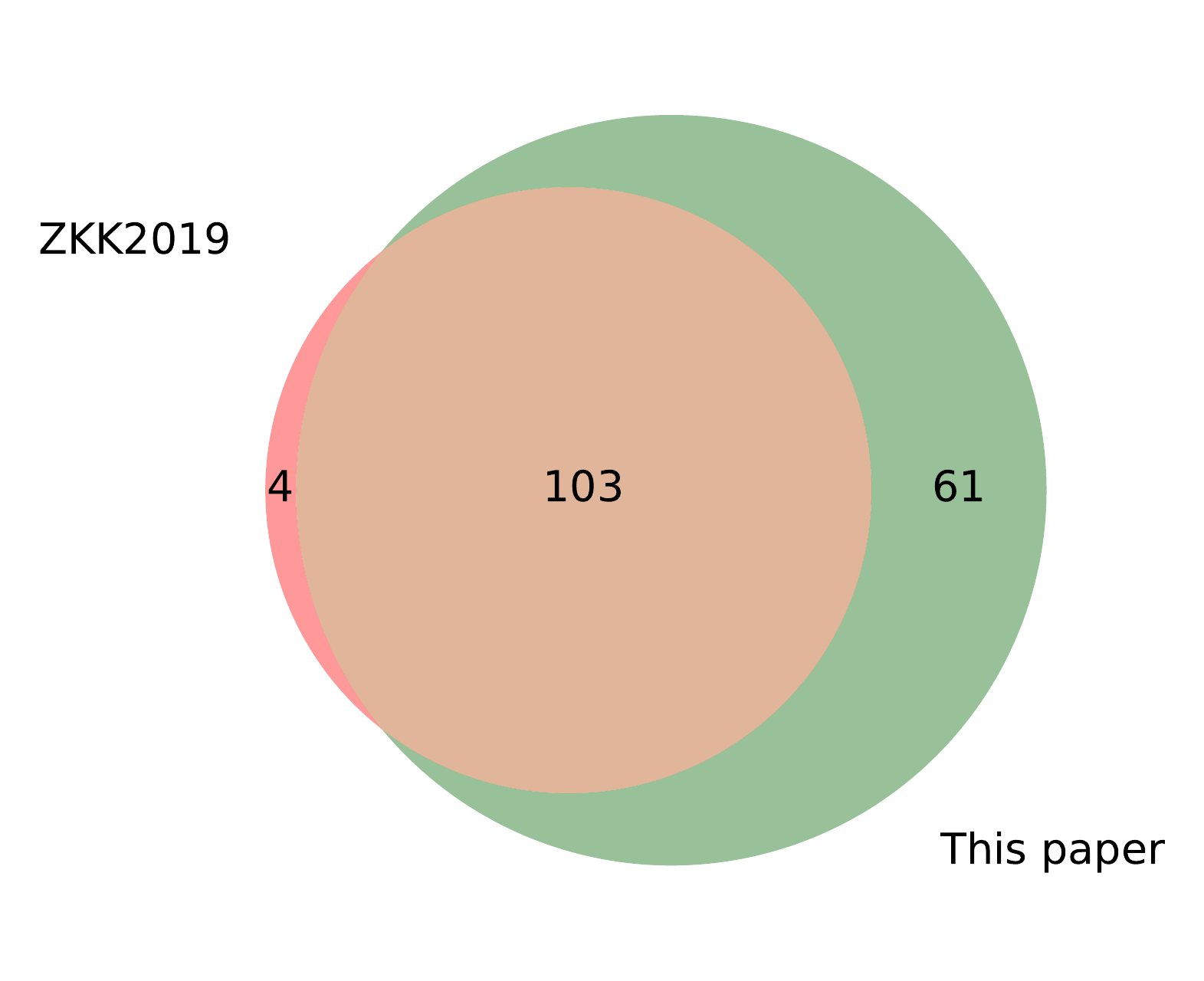}
\caption{
\label{fig_venn}
Venn diagram comparing the sample of cluster members identified in our membership analysis with the results obtained by ZKK2019. The area of each circle is proportional to the number of stars in the samples. 
}\label{fig_venn}
\end{center}
\end{figure}

In this study, we identified 61 new cluster members, which represents an increase of about 60\% in the census of XFOR stars with respect to the number of confirmed cluster members from the literature. In the following, we use this new sample of stars to re-visit some properties of the cluster.

\section{Spectroscopic observations}\label{section3}

We conducted a low-resolution spectroscopy study of our new sample of cluster members with the EFOSC2 spectrograph \citep{Buzzoni1984} mounted at the ESO/NTT telescope, operating at La Silla (ESO, Chile). The observations were taken with the 1.0\arcsec slit, grism Gr\#16, and filter OG530, yielding a spectral coverage from 6015{\AA} to 10320{\AA} and resolution of about 16{\AA}. The exposure times ranged from a few seconds up to 55~min depending on the magnitude of the target, so that the signal-to-noise ratio (S/N) is often better than 50 for most spectra collected in our campaign. We performed our observations in designated visitor mode (i.e. remote observations) during the nights of November 23 to November 30, 2020 (programme: 106.D-0429, PI: Galli). We observed 91 stars from our sample of cluster members identified in this study. Two of the new cluster members identified in this paper, namely \textit{Gaia} EDR3 4755198025592857088 (025251-435842) and \textit{Gaia} EDR3 4755198025592857216 (025251-435845), form a close binary system and we observed both components in our campaign.  

The spectra were reduced with the EFOSC2 pipeline using the EsoRex command line tool, which performs a bias subtraction, flat-fielding, along with wavelength and flux calibration. We observed five spectrophotometric standards (Feige110, CD-34\,241, LTT1020, LTT2415, and LTT3218) during the nights allocated to our programme and took the closest one (in time and airmass) to each science frame to flux calibrate the spectrum. The reduced spectra were then used to derive the spectral type of the stars as explained below. 

We take advantage of the PyHammer v2.0 facility \citep{Roulston2020} to perform the spectral classification of the stars in our sample. This code is based on the IDL spectral typing suite called `The Hammer'   developed by \citet{Covey2007} and recently re-written in Python programming language. PyHammer automatically assigns spectral types to the input spectra by measuring spectral indices of atomic and molecular lines and comparing them to templates of an empirical stellar library \citep[see][]{Kesseli2017}. The input spectrum is compared to each template by computing the chi-squared value between the two spectra. The best matching spectral type is defined as the one that minimises the chi-squared value between the spectral indices from the template and input spectrum. 

To investigate the robustness of our spectral classification, we implemented a chi-squared minimisation routine that uses  the entire spectrum (instead of spectral indices). We ran this comparison between input spectrum and templates using the same spectral library of the PyHammer code. The spectral types derived from the two methods are consistent among themselves within one subclass for M stars and two subclasses for the other stars in our sample. We adopted these numbers as uncertainties of our spectral classification and report the spectral types derived directly with PyHammer as our final results. 

We complemented the spectral classification of cluster members with additional measurements that are available in the literature, mostly for the stars of spectral types A and F of our sample. Table~\ref{tab_members} lists the spectral classification of the stars and the references from the literature. By combining the spectral types derived in this study from EFOSC2 observations and literature results, we were able to provide the spectral classification for 100 stars that we use in the forthcoming analysis (see Sect.~\ref{section4}). The spectral type of the stars ranges from B6.5 to M8.

\section{Properties of the cluster}\label{section4}

\subsection{Radial velocities}\label{section4.1}

The scarcity of radial velocities for the stars in our sample is the main limitation to access the 3D space of motion of the XFOR cluster. We proceed as follows to complement the \textit{Gaia}-EDR3 astrometry with radial velocity data. Firstly, we searched the CDS databases for published radial velocities in the literature for the stars in our sample. We found radial velocity information for 24 stars that have been collected from \citet{Gontcharov2006}, \citet{Kordopatis2013}, \citet{Shkolnik2017}, and the \textit{Gaia}-EDR3 catalogue itself \footnote{The \textit{Gaia}-EDR3 catalogue does not contain new radial velocity measurements. The radial velocities of Gaia-DR2 have been added to the Gaia-EDR3 catalogue for user convenience.}. Secondly, we searched for high-resolution spectra in public archives and derived the radial velocities ourselves (see below). We found spectroscopic data for four stars in our sample (one of them with multiple spectra). The results of our search in public archives are summarised in Table~\ref{tab_spectra}. 

We use the reduced spectra downloaded from the archives to derive the radial velocity of the stars with available data. We use the iSpec facility \citep{BlancoCuaresma2014} to cross-correlate the spectrum of each star with masks of different spectral types (A0, F0, G2, K0, K5, and M5). The iSpec routines implement the cross-correlation algorithm to generate the velocity profile of the star \citep[see e.g.][]{Baranne1996}, and fit a Gaussian/Voigt to the cross-correlation function (CCF) to derive its radial velocity with the associated uncertainty. We inspected the CCFs individually and discarded the radial velocity measurements that result from a poor fit to the CCF that can arise, for instance, due to the low S/N of the collected spectra or a mismatch between the spectral type of the stars and the adopted mask. We compute the radial velocity for each star from the closest mask available in iSpec to its spectral type. The scatter in the radial velocities that results from using masks of different spectral types ranges from 0.1 to 0.3~km/s for the stars with available spectra in our sample. We add the radial velocity scatter obtained from different masks in quadrature to the formal uncertainty on the radial velocity given by iSpec for each star. This procedure is likely to overestimate the uncertainties derived in this study, but it accounts for the observed fluctuation on radial velocity results obtained with masks of different spectral types (which often do not precisely match the spectral type of the star). 

We note that two stars for which we derived the radial velocities ourselves, namely \textit{Gaia} EDR3 4860643905115917312 (032807-355444) and \textit{Gaia} EDR3 5056560315790459776 (033121-303059), have known radial velocities (see comparison in Table~\ref{tab_comp_RV}). The published radial velocities for these two sources are less precise than the ones obtained in this study, so we use the results derived by us in the upcoming analysis. The median precision of the radial velocities derived in this paper from high-resolution spectra in public archives is 0.4~km/s, while the radial velocities published in the literature (including those from the \textit{Gaia}-EDR3 catalogue) have a median precision of 1.9~km/s. 

\begin{figure}[!h]
\begin{center}
\includegraphics[width=0.50\textwidth]{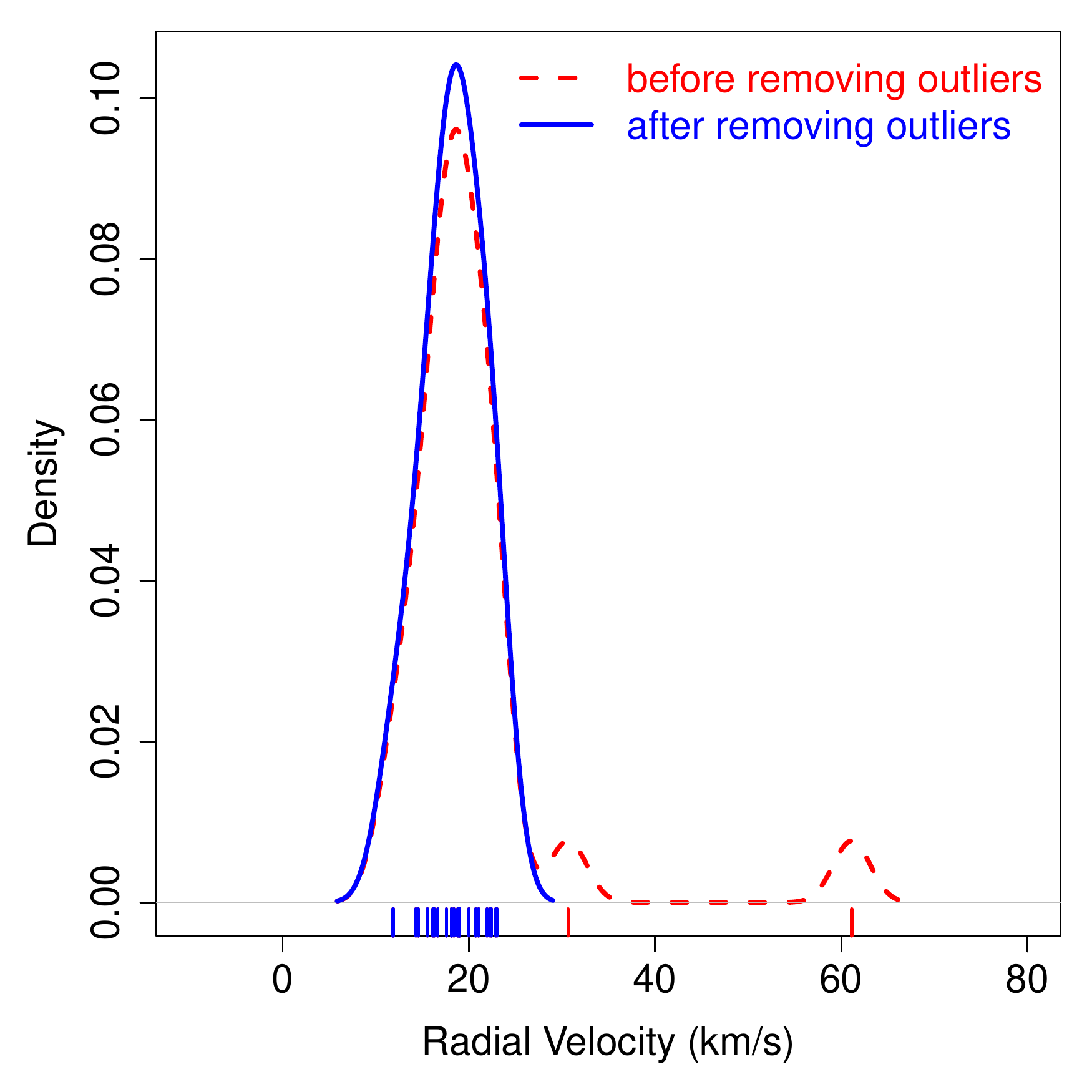}
\caption{
Kernel density estimate of the distribution of radial velocities using a kernel bandwidth of 2~km/s that roughly corresponds to the median precision of the radial velocities in our sample. The tick marks in the horizontal axis indicate the individual radial velocity measurements of the stars. 
}\label{fig_RV}
\end{center}
\end{figure}

By combining the radial velocities from the literature with the ones derived in this paper we end up with a sample of 28~stars with available radial velocity information. Two sources in this sample, namely \textit{Gaia} EDR3 5056578663891978880 (033036-302006) and \textit{Gaia} EDR3 5058417459651237504 (031426-300700), have poor radial velocities ($V_{r}=30.5\pm 6.2$~km/s and $V_{r}=-4.6\pm17.5$~km/s) with uncertainties that greatly exceed the observed velocity dispersion of a few km/s in young stellar groups. We have therefore discarded these measurements from our analysis. We also note the existence of discrepant radial velocity measurements in this sample as illustrated in Figure~\ref{fig_RV}. We applied the interquartile range (IQR) criterion to identify the outliers in the distribution of radial velocities by rejecting the measurements that lie 1.5$\times$IQR below the first quartile or above the third quartile. We discard the radial velocity of two stars based on this method. However, we do not remove these sources from our list of cluster members because their proper motions, parallaxes, and photometry are consistent with membership in the cluster ($p\geq0.9)$. The discrepant radial velocity is more likely to be affected by other reasons (e.g. undetected binarity) and this requires further investigation with a greater store of data. In addition, we verified that three stars (with radial velocity data) are binary systems and we decided to also discard these measurements from the forthcoming analysis (see Sect.~\ref{section4.2}). Altogether, this reduces the sample of cluster members with radial velocity information to 21 stars. The radial velocities used in this paper are given in Table~\ref{tab_members}.

\begin{table*}[!h]
\centering
\caption{Properties of the data downloaded from public archives. 
\label{tab_spectra}}
\begin{tabular}{ccccl}

\hline\hline
Star identifier& Other name&$N$&Instrument&Programme identifier\\
\hline\hline
\textit{Gaia} EDR3 4755745925980823296 & 025903-423245 & 1 & UVES \citep{UVES} & 097.C-0444\\
\textit{Gaia} EDR3 4854308003720230528 & 031603-372522 & 1 & FEROS \citep{FEROS} & 090.C-0815\\
\textit{Gaia} EDR3 4860643905115917312 & 032807-355444 & 1 & FEROS \citep{FEROS} &090.C-0815\\
\textit{Gaia} EDR3 5056560315790459776 & 033121-303059 & 5 & FEROS \citep{FEROS} & 082.A-9011, 084.A-9004, 086.A-9006\\
\hline
\end{tabular}
\tablefoot{We provide for each star the number of spectra downloaded from the archives, spectrograph, and the ESO programme identification of the corresponding observation. }
\end{table*}

\begin{table*}[!h]
\centering
\caption{Comparison of the radial velocities derived in this paper with the literature results. 
\label{tab_comp_RV}}
\begin{tabular}{cccccc}

\hline\hline
&&This paper&Literature&\\
Star identifier&Other name&$V_{r}$&$V_{r}$&Reference\\
&&(km/s)&(kms/)&\\
\hline\hline
\textit{Gaia} EDR3 4860643905115917312&032807-355444&$22.42\pm0.74$&$20.2\pm2.0$&ZKK2019\\
\textit{Gaia} EDR3 5056560315790459776&033121-303059&$18.99\pm0.37$&$16.4\pm2.0$&\citet{Kordopatis2013}\\
\hline

\end{tabular}
\tablefoot{See note added in proof in ZKK2019 for the radial velocity of \textit{Gaia} EDR3 4860643905115917312 (032807-355444).}
\end{table*}

\subsection{Distance and spatial velocity}\label{section4.2}

We converted the proper motion and parallax of the stars in our sample into distance and 2D tangential velocities using Bayesian inference following the online tutorials available in the \textit{Gaia} archive \citep{Luri2018}. We used the \textit{Kalkayotl} package \citep{Olivares2020} to investigate the dependency of our results on the choice of the prior for the distance. The \textit{Kalkayotl} code implements two families of priors for the distance that are either based on statistical probability density distributions (hereafter, statistical priors) or astrophysical assumptions (hereafter, astrophysical priors). We performed our calculations with two statistical priors that implement the Uniform and Gaussian probability density distributions, and another two astrophysical priors that are based on the surface brightness profile of young star clusters in the Large Magellanic Clouds \citep{Elson1987} and the profile distribution of globular clusters observed by \citet{King1962}. Our analysis confirms that our distance results are independent of our choice of priors for the distance thanks to the good precision of the \textit{Gaia}-EDR3 parallaxes. Thus, the individual distances listed in Table~\ref{tab_members} refer to the solution obtained with the Uniform prior, which is the least informative (i.e. most simple) prior investigated in our analysis. 

Then we computed the 3D position of the stars from their distances in a $XYZ$ reference system that has its origin at the Sun where $X$ points to the Galactic centre, $Y$ points to the direction of Galactic rotation, and $Z$ points to the Galactic North pole. We converted the 2D tangential velocities and radial velocities of the stars into the $UVW$ components of the Galactic velocity in the same reference system as before using the transformation outlined by \citet{Johnson1987}. The individual distances and spatial velocity of the stars are given in Table~\ref{tab_members}.  

The distance to the XFOR cluster determined from Bayesian inference in this study is $108.4\pm0.3$~pc and the distances of individual stars in the cluster range from $91.1\pm0.2$~pc to $125.7\pm0.6$~pc. To gain confidence in our results, we recomputed the distances (as previously explained) after correcting the \textit{Gaia}-EDR3 parallaxes for the colour-dependent bias (i.e. zero-point correction) following the recipe outlined by \citet{Lindegren2021a}, and we confirmed that our results are not significantly affected by this correction. The proposed functions to correct for the parallax bias are only indicative (as explained by the authors themselves in their study) and they do not bring about significant changes in our case due to the close proximity of the cluster. We therefore report our distance results based on the original parallaxes given in the \textit{Gaia}-EDR3 catalogue (i.e. without applying any correction). 

We derived the cluster space motion from the procedure described in Section~7.2 of \citet{Perryman1998} to account for the existence of correlated errors in the velocity components that result from the transformation of proper motions, parallaxes, and radial velocities into 3D velocities. The mean space motion is obtained by averaging the measured velocities and the uncertainty is computed from the mean of the covariance matrices of individual stars. This yields a mean space motion of $U=-12.8\pm0.6$~km/s, $V=-21.6\pm0.8$~km/s, $W=-4.9\pm1.4$~km/s, and a standard deviation of $\sigma_{U}=1.6$~km/s, $\sigma_{V}=1.7$~km/s, $\sigma_{W}=2.6$~km/s. The median uncertainties in the $UVW$ velocity components are 0.3, 0.5, and 1.1~km/s, implying that the internal velocity dispersion of the cluster is resolved.  

The space motion of the XFOR cluster can now be compared with the space motion of the COL and THA associations. The relative motion of XFOR with respect to the space motions of COL and THA given by \citet{Gagne2018a} in the sense XFOR `minus' COL or THA are $(\Delta U,\Delta V, \Delta W)=(-0.9,-0.3,0.8)\pm(1.2,1.5,1.6)$~km/s and $(\Delta U,\Delta V, \Delta W)=(-3.0,-0.7,-3.9)\pm(1.1,1.1,1.6)$~km/s, respectively. This comparison shows that the space motion of XFOR is consistent with COL and THA within $3\sigma$ of the reported uncertainties in the spatial velocities when the correlations are taken into account. We therefore confirm the previous finding of ZKK2019 based on the space motion of only four core members that the XFOR cluster is comoving with the COL and THA young stellar associations. 

\begin{figure*}
\begin{center}
\includegraphics[width=0.33\textwidth]{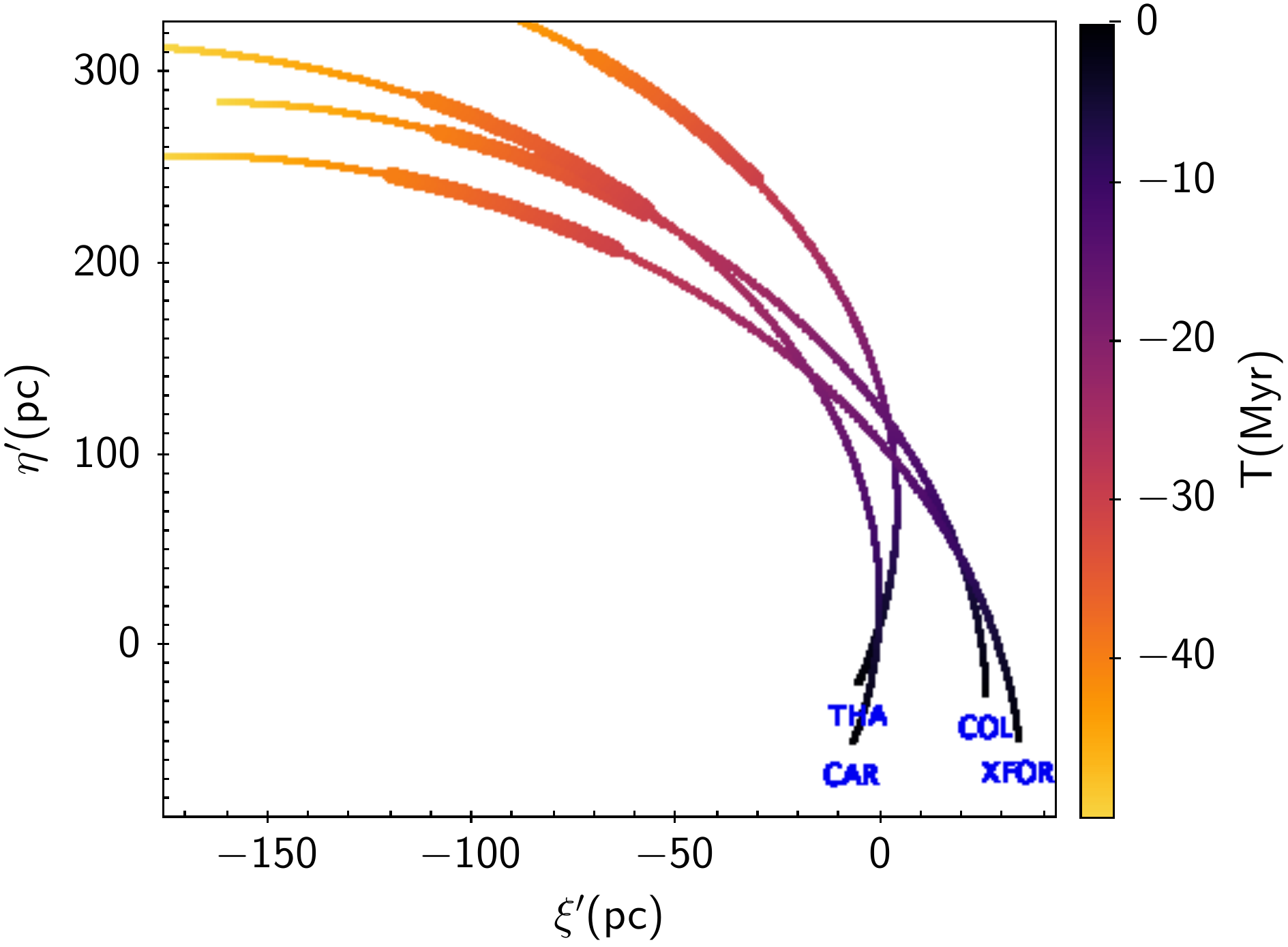}
\includegraphics[width=0.33\textwidth]{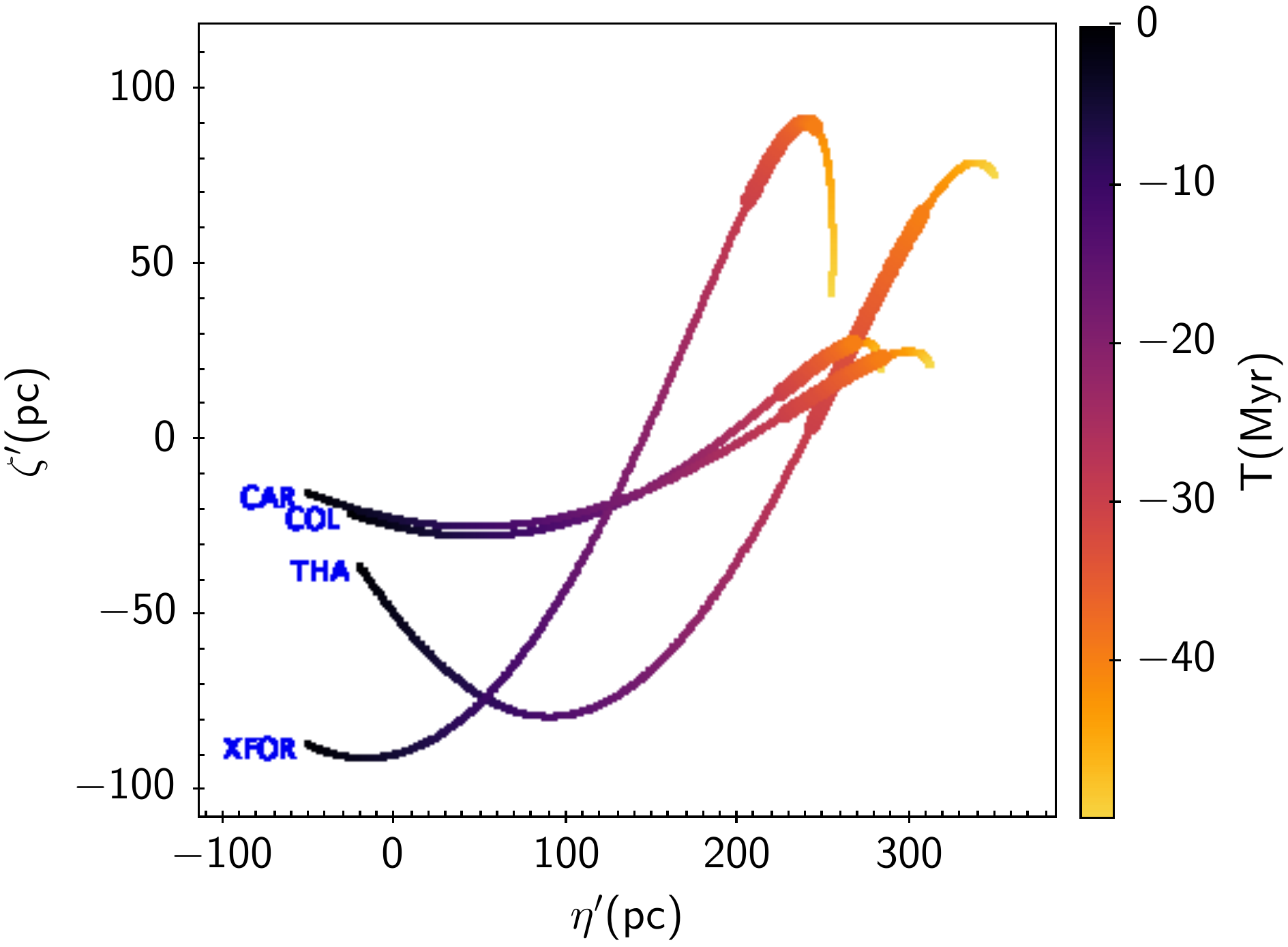}
\includegraphics[width=0.33\textwidth]{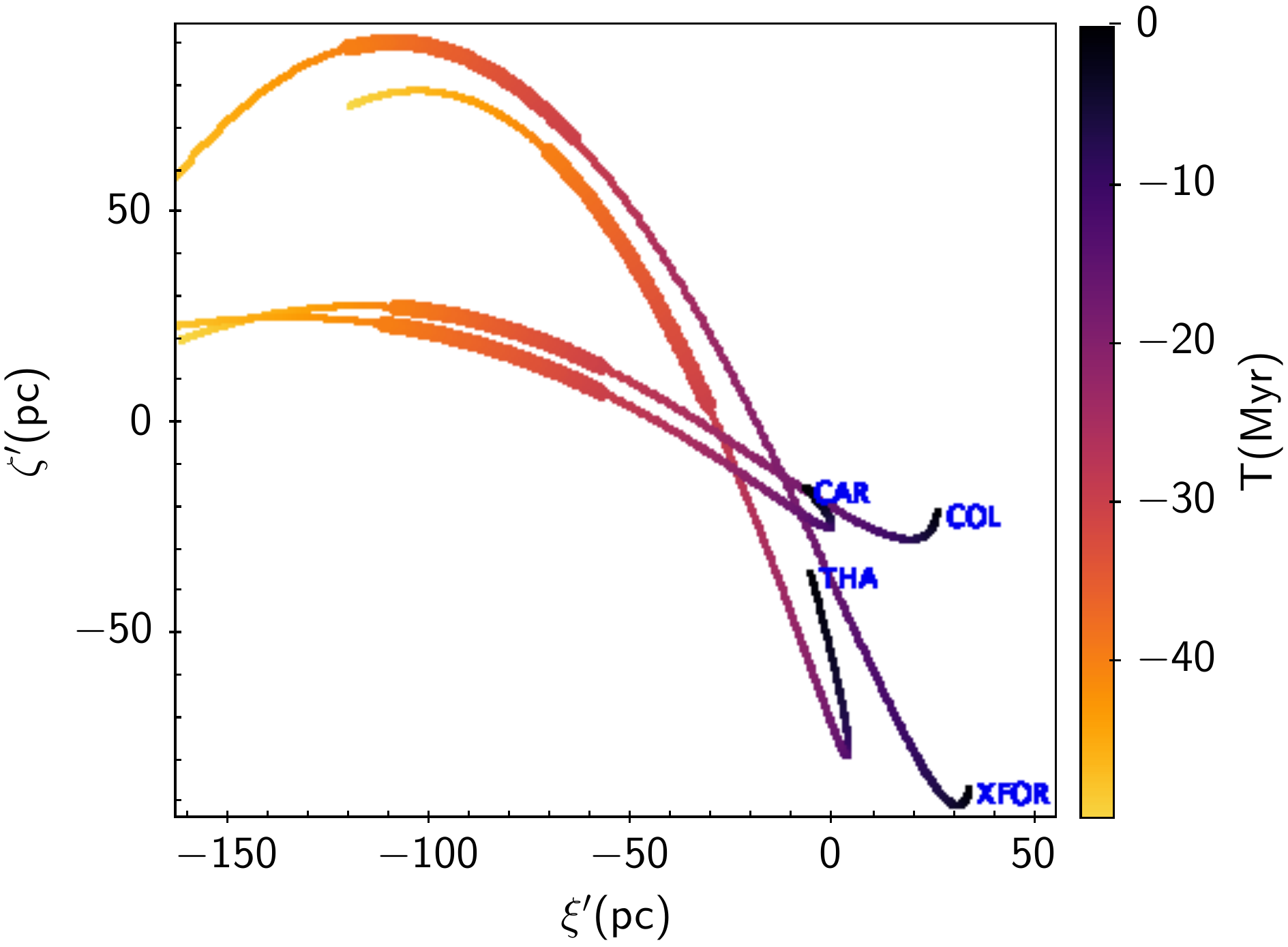}
\caption{
Orbits of the young stellar groups integrated back in time in a curvilinear coordinate system using a 3D axisymmetric galactic potential. The orbits are colour-coded based on the integration time used in our calculations. The thicker lines roughly indicate the assumed age ($\sim$30-40~Myr) of these young stellar groups given in the literature (see also Sect.~\ref{section4.4}). \vspace{0.6cm}
}\label{fig_traceback}
\end{center}
\end{figure*}

Finally, we performed a simplified traceback analysis by integrating the orbit of the XFOR cluster back in time and comparing its location in the past with the other young stellar groups of the Solar neighbourhood. Here, we follow a similar procedure as described in \citet{MiretRoig2018} which consists of integrating the centroid of each stellar group. Thus, we use the UVW velocity derived in this section and the mean XYZ position of the stars to define the space motion and present-day location of the XFOR cluster. For the other young stellar groups, we use the values given in Table~9 of \citet{Gagne2018a}. Our calculations use the heliocentric curvilinear coordinate system $(\xi^\prime,\eta^\prime,\zeta^\prime)$ defined by \citet{Asiain1999} and the same 3D axisymmetric Milky Way potential as in \citet{MiretRoig2020} to integrate the equations of motion. This model is based on the \citet{Allen1991} potential which consists of a spherical central bulge, a disc, and a massive spherical halo, but with updated parameters taken from Table~1 of \citet{Irrgang2013}. As illustrated in Figure~\ref{fig_traceback}, the minimum distance of the XFOR cluster to the THA and COL stellar groups occurred at $t=-10.6$~Myr ($\sim$24~pc) and $t=-19.1$~Myr ($\sim$18~pc), respectively. Interestingly, the closest approach of the XFOR cluster occurred with the Carina (CAR) young stellar group at $t=-18.6$~Myr with a separation of only $\sim$7~pc. The relative space motion of the XFOR cluster and CAR \citep[see][]{Gagne2018a} is $(\Delta U,\Delta V, \Delta W)=(-2.1,0.3,0.6)\pm(0.9,1.3,1.7)$~km/s indicating that these two young stellar groups are also comoving. Altogether, this suggests that XFOR, THA, COL, and CAR may have a common origin and that they could be the remnant of a much larger association. 

\subsection{Internal motions}\label{section4.3}

The 3D velocity vectors of the cluster members with useful radial velocity measurements (see Sect.~\ref{section4.1}) are mostly parallel to each other. The non-trivial question that arises at this stage is whether the observed velocities result from random or organised motions within the cluster and their quantitative importance. In this section, we investigate the existence of potential expansion and rotation effects in the XFOR cluster following the procedure applied by \citet{Rivera2015} and \citet{Galli2019} in the Taurus star-forming region. 

First, we computed the unit position vector $\mathbf{\hat{r}_{\star}}=\mathbf{r}_{\star}/|\mathbf{r}_{\star}|$ of each star that represents its distance to the centre of the cluster. Then we computed the relative velocity $\delta \mathbf{v_{\star}}$ of each star with respect to the mean velocity of the cluster (see Sect.~\ref{section4.2}). The dot product ($\mathbf{\hat{r}_{\star}} \cdot \delta \mathbf{v_{\star}} $) is large and positive (negative) if the cluster is undergoing expansion (contraction). Similarly, the cross product ($\mathbf{\hat{r}_{\star}} \times \delta \mathbf{v_{\star}}$), which is related to angular momentum, is large (small) when the cluster has significant (negligible) rotation. It is important to mention that the dot and cross products are not exactly the expansion and rotation velocities of the cluster, but they can be used as a proxy for these quantities to access the existence (or absence) of these motions within the cluster. 

We computed $\mathbf{\hat{r}_{\star}} \cdot \delta \mathbf{v_{\star}} $ and $\mathbf{\hat{r}_{\star}} \times \delta \mathbf{v_{\star}} $ for each star in our sample individually and then we took the average and standard error of the mean of the resulting values to characterise the cluster as a whole. In doing so, we find that $\mathbf{\hat{r}_{\star}} \cdot \delta \mathbf{v_{\star}} =0.5\pm 0.4$~km/s and $\mathbf{\hat{r}_{\star}} \times \delta \mathbf{v_{\star}}=(0.1,-0.1,0.4)\pm(0.3,0.4,0.3)$~km/s, showing that both quantities are consistent with zero within $1\sigma$ of the reported uncertainties. Thus, we conclude that the effects of a potential expansion and rotation of the cluster are negligible and that the observed velocity dispersion is most likely to be dominated by random motions within the cluster. However, we should note that we reached the present conclusion based on a small fraction of cluster members with complete data for this analysis (3D positions and 3D velocities); thus, it is clear that more study is clearly warranted when the radial velocity of a more significant number of cluster members becomes available in the future. 

\subsection{Isochronal ages}\label{section4.4}
The first step of our analysis to revisit the age of the cluster consists of constructing the Hertzsprung-Russell (HR) diagram of the cluster. Thus, we fit the spectral energy distribution (SED) of the stars with the Virtual Observatory SED Analyzer \citep[VOSA,][]{VOSA} to derive the effective temperatures and bolometric luminosities. We used the BT-Settl grid of models \citep{Allard2012} to fit the SEDs and leave the extinction $A_{V}$ as a free parameter in the fit varying in the range of 0 to 10~mag. We provide the input photometry to the VOSA facility to avoid erroneous cross-matches with the system interface. We cross-matched our list of cluster members with the 2MASS \citep{2MASS}, AllWISE \citep{AllWISE}, Pan-STARRS \citep{PanSTARRS}, APASS \citep{APASS}, and TYCHO2 \citep{TYCHO2} catalogues to retrieve the photometric data for our stars using the precomputed cross-matched tables available in the \textit{Gaia} archive. In doing so, we derived effective temperatures and bolometric luminosities for 154~stars in our sample with available photometry for the SED analysis with VOSA. 

We compared the effective temperatures obtained from the SED fit with the ones inferred from the spectral type of the sample of 100~stars with available spectral classification (see Section~\ref{section3}). We used the tabulated values from \citet{Pecaut2013} to convert the spectral type of the stars to effective temperatures. The mean difference between the effective temperatures in the two data sets is 136$\pm$19~K, which is consistent with the uncertainty obtained in our spectral classification. Similarly, we compared the bolometric luminosities derived from VOSA with the results we computed, using the J-band photometry from the 2MASS catalogue and the bolometric corrections given by \citet{Pecaut2013} for a sample of 96~stars with available spectral classification and 2MASS photometry. The results of this comparison confirm the robustness of our results obtained with different methods. In the following, we restrict our discussion to the results obtained directly from the VOSA  service because this method allows us to construct the HR diagram with a much more significant number of cluster members. 

\begin{figure*}[!h]
\begin{center}
\includegraphics[width=0.49\textwidth]{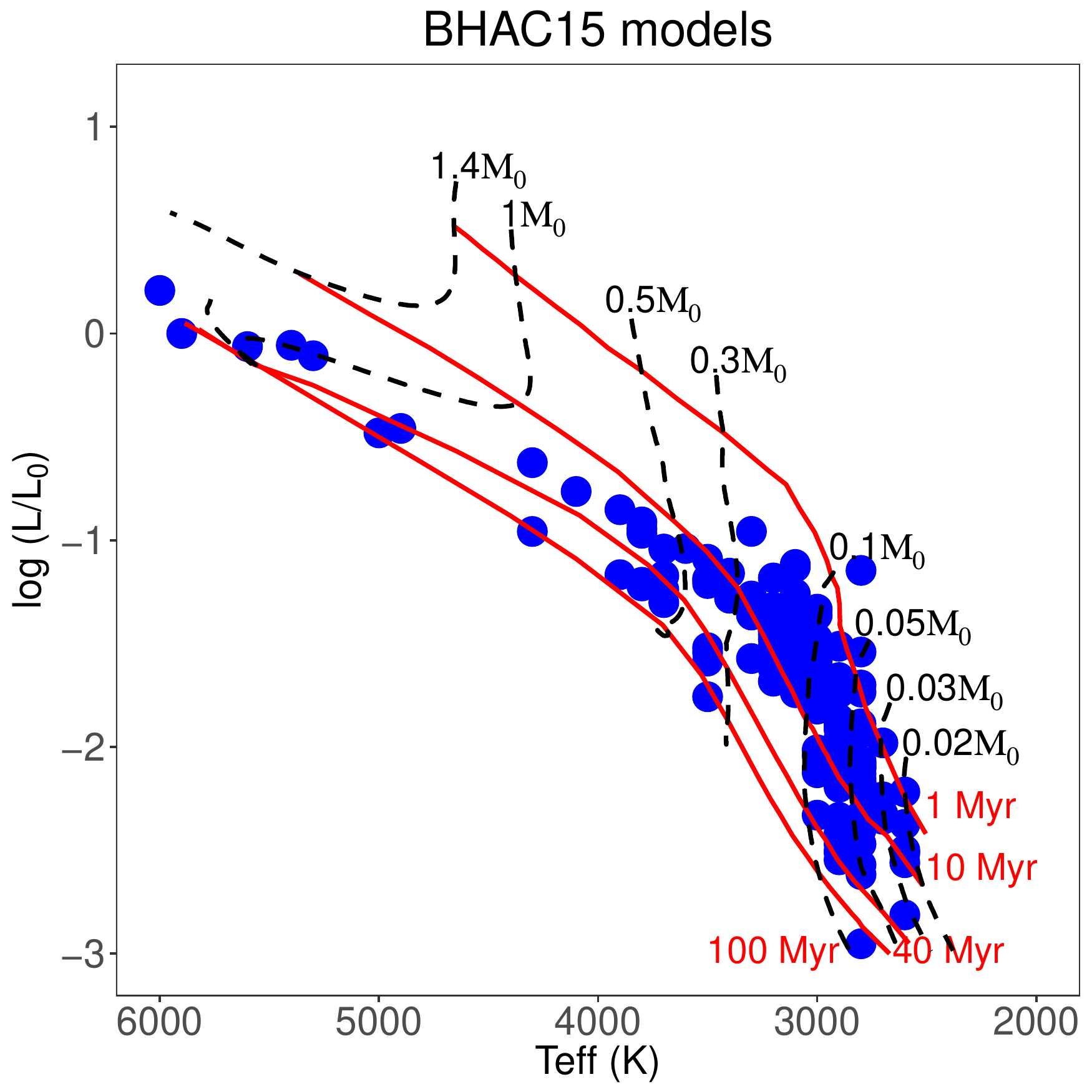}
\includegraphics[width=0.49\textwidth]{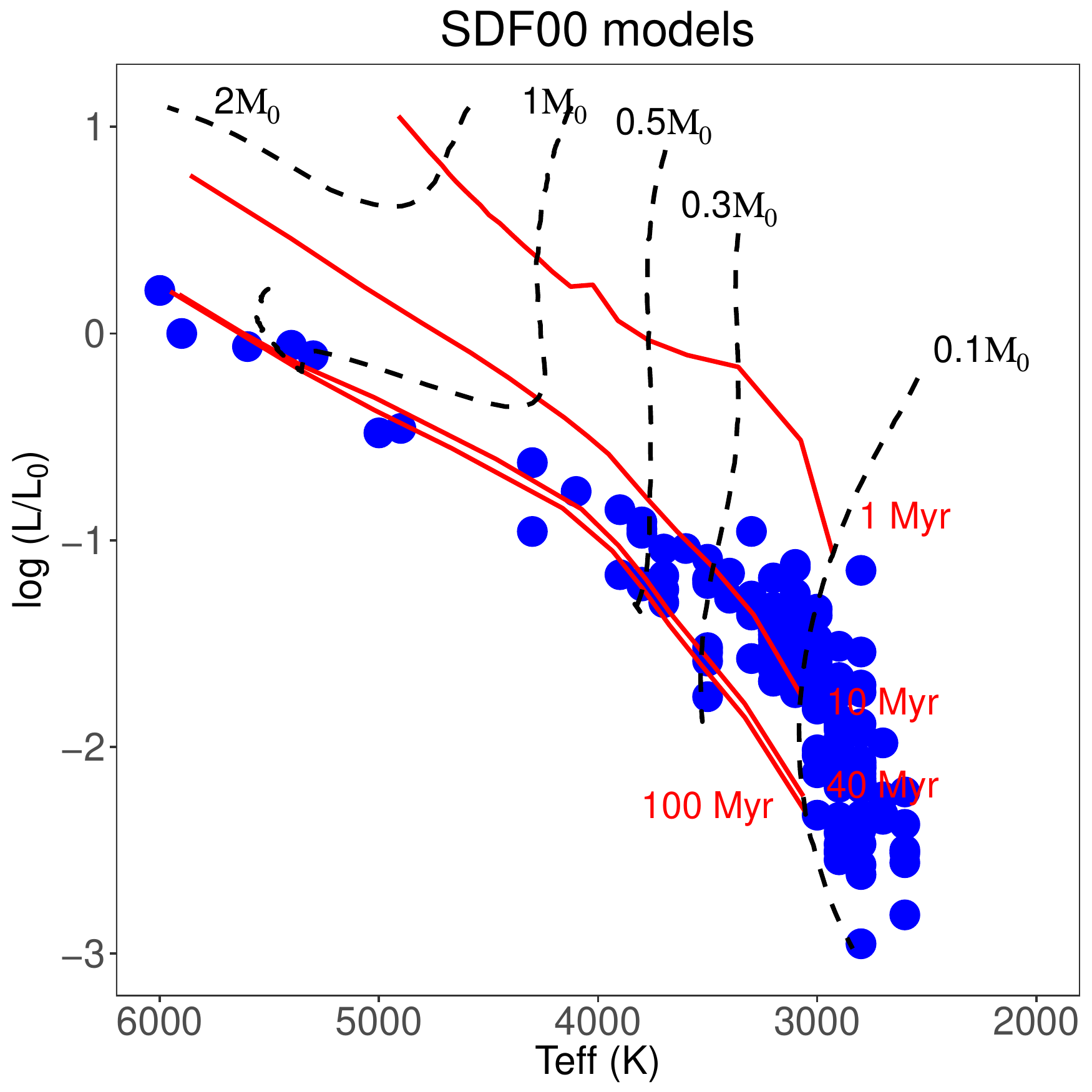}
\includegraphics[width=0.49\textwidth]{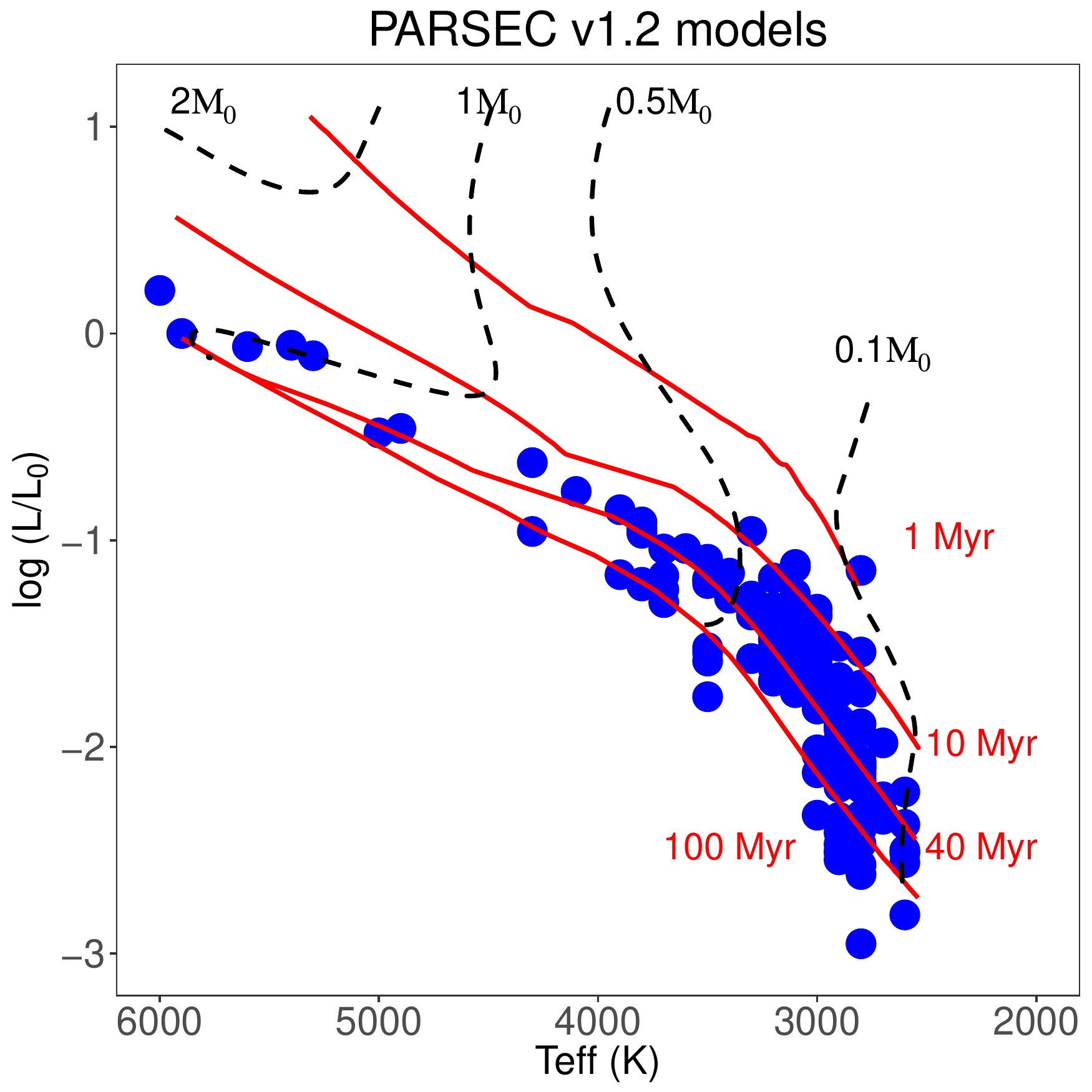}
\includegraphics[width=0.49\textwidth]{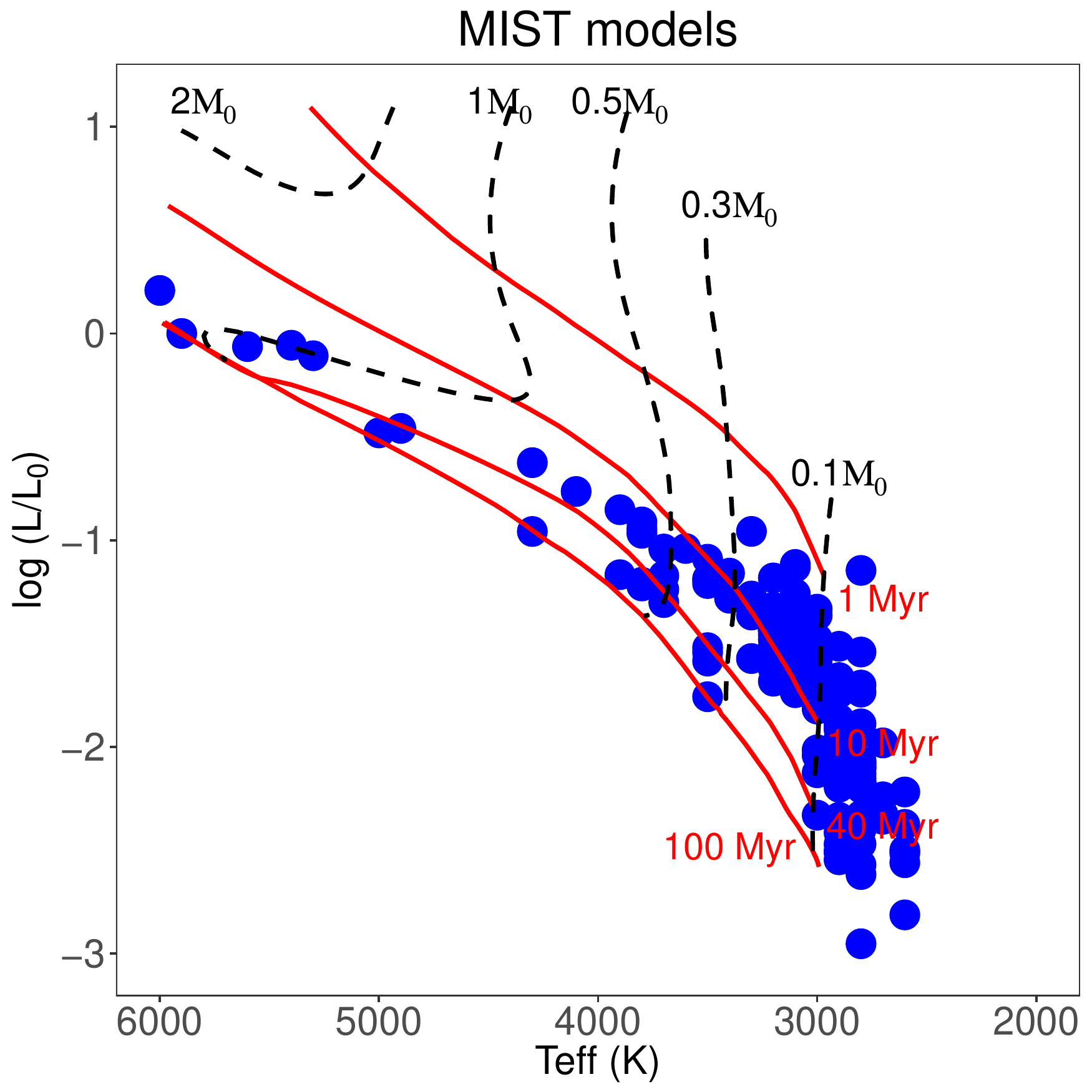}
\caption{
HR diagram of the XFOR cluster compared to the grid of isochrones and tracks computed from different models. The solid red lines and black dashed lines indicate the isochrones and tracks for each model, with the age (in Myr) and mass (in M$_{\odot}$) values indicated in the panels.   \vspace{1.0cm}
}\label{fig_HRD}
\end{center}
\end{figure*}

Figure~\ref{fig_HRD} presents the resulting HR diagram. We compare the position of the stars in the HR diagram with four grids of pre-main sequence star models: \citet[][BHAC15]{BHAC15}, \citet[][SDF00]{SDF00}, PARSEC~v1.2S \citep{PARSEC}, and MIST \citep{MIST}, with stellar rotation $(\nu=0.4\nu_{crit})$. Our sample covers the mass range from about 0.02 to 4~M$_{\odot}$, but we show in Figure~\ref{fig_HRD} only the mass domain that includes most cluster members and where the models overlap for a better comparison. The lower mass limit is consistent with a brown dwarf of spectral type $\sim$ M8 and magnitude of $G\sim18$~mag. We note from the HR diagram that many cluster members appear above the 10~Myr isochrone given by the BHAC15, SDF00 and MIST models. This suggests that the cluster can be somewhat younger than previously thought. However, the fraction of young disc-bearing stars in the cluster is rather low to support the hypothesis of the cluster being only a few Myr old (see Section~\ref{section4.5}) and some of these sources are probably binary or high-order multiple systems that will require further investigation in future studies. The observed spread in the HR diagram for the least massive stars in the sample ($<0.3M_{\odot}$) indeed translates into a broad range of ages, but the stars with higher masses give a much tighter constrain and clearly discard an age younger than 10~Myr for the cluster. 

\begin{figure}
\begin{center}
\includegraphics[width=0.48\textwidth]{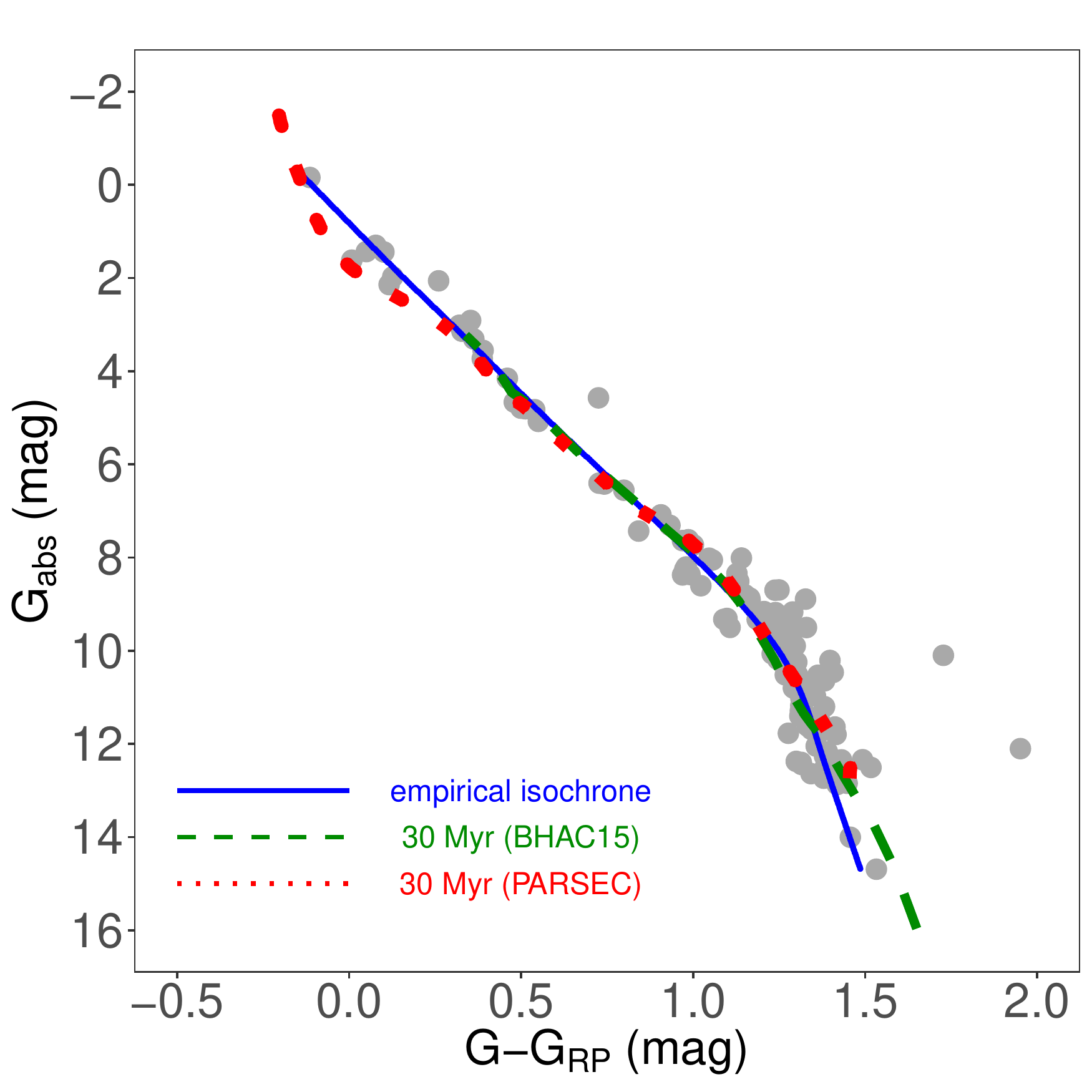}
\caption{
Absolute colour-magnitude diagram of the cluster members identified in our membership analysis (see Sect.~\ref{section2}). The lines indicate the empirical isochrone of the cluster and the 30~Myr isochrones computed from the BHAC15 and PARSEC models. The three stars located above the cluster sequence are \textit{Gaia} EDR3 5056560315790459776 (033121-303059), \textit{Gaia} EDR3 4854540344270707200 (032513-370909), and \textit{Gaia} EDR3 4854879573671981056 (032215-354719).
}\label{fig_CMD_isochrones}
\end{center}
\end{figure}

\begin{figure}
\begin{center}
\includegraphics[width=0.48\textwidth]{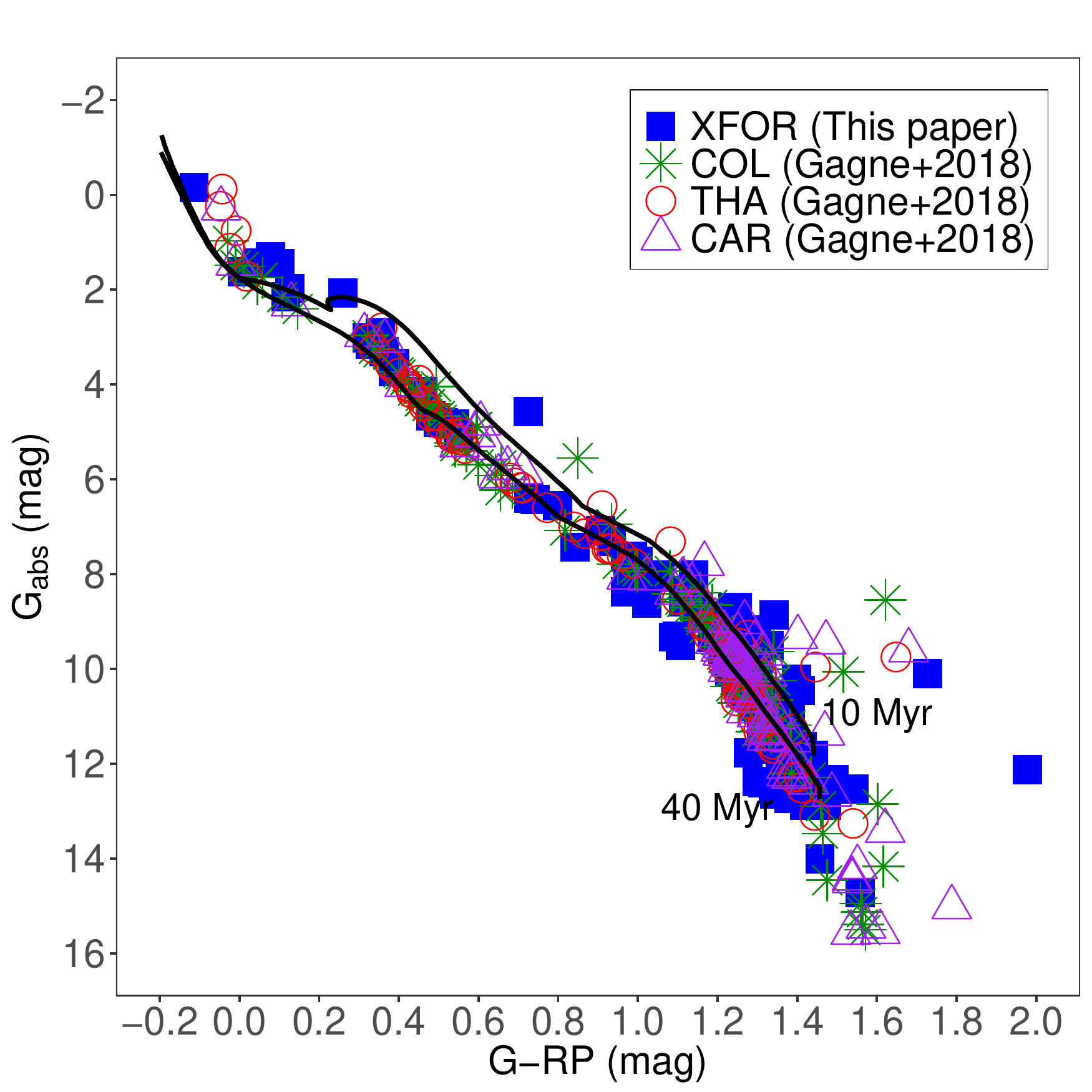}
\caption{
Absolute colour-magnitude diagram derived from \textit{Gaia}-EDR3 photometry. The different colours and symbols indicate the XFOR cluster members identified in this study and known members of the COL, THA, and CAR stellar groups published in the literature \citep{Gagne2018a,Gagne2018b,Gagne2018c}. The solid lines mark the 10~Myr and 40~Myr isochrones from the PARSEC models.  
}\label{fig_CMD_COL_THA}
\end{center}
\end{figure}

From a visual inspection in the colour-magnitude diagram (see Figure~\ref{fig_CMD_isochrones}), it is apparent that the empirical isochrone of the cluster -- derived in our analysis and converted to absolute magnitudes at the distance of the cluster ($108.4\pm0.3$~pc, see Sect.~\ref{section4.2}) -- is more consistent with an age of about 30~Myr when compared to the isochrones of the BHAC15 and PARSEC models. This age is consistent with the median age of 35~Myr inferred from the PARSEC models based on interpolating the stars in the HR diagram between the isochrones. It is also in good agreement with the age estimate given by \citet{Mamajek2016} based on the X-ray emission of a few cluster members. As illustrated in Figure~\ref{fig_CMD_COL_THA} the age of 30~Myr is also still consistent with the assumed age for the COL \citep[42$^{+6}_{-4}$~Myr,][]{Bell2015}, THA \citep[45$\pm$4~Myr,][]{Bell2015}, and CAR \citep[45$^{+11}_{-7}$~Myr,][]{Bell2015} stellar groups, supporting the idea that they are all coeval (see also ZKK2019). 

At this stage it seems premature to assign a definitive and unambiguous age for the cluster given the important age variation that we observe from the HR diagram depending on our choice of evolutionary model (see Figure~\ref{fig_HRD}). In the remainder of this paper, we assume an age of 30~Myr for the cluster, but we believe that more study using alternative methods, such as the traceback \citep[see][]{MiretRoig2020} and lithium depletion boundary \citep[see][]{Binks2014} methods, is required to provide a more robust result for the age.

\subsection{Spatial distribution}\label{section4.5}

\begin{figure*}
\begin{center}
\includegraphics[width=1.0\textwidth]{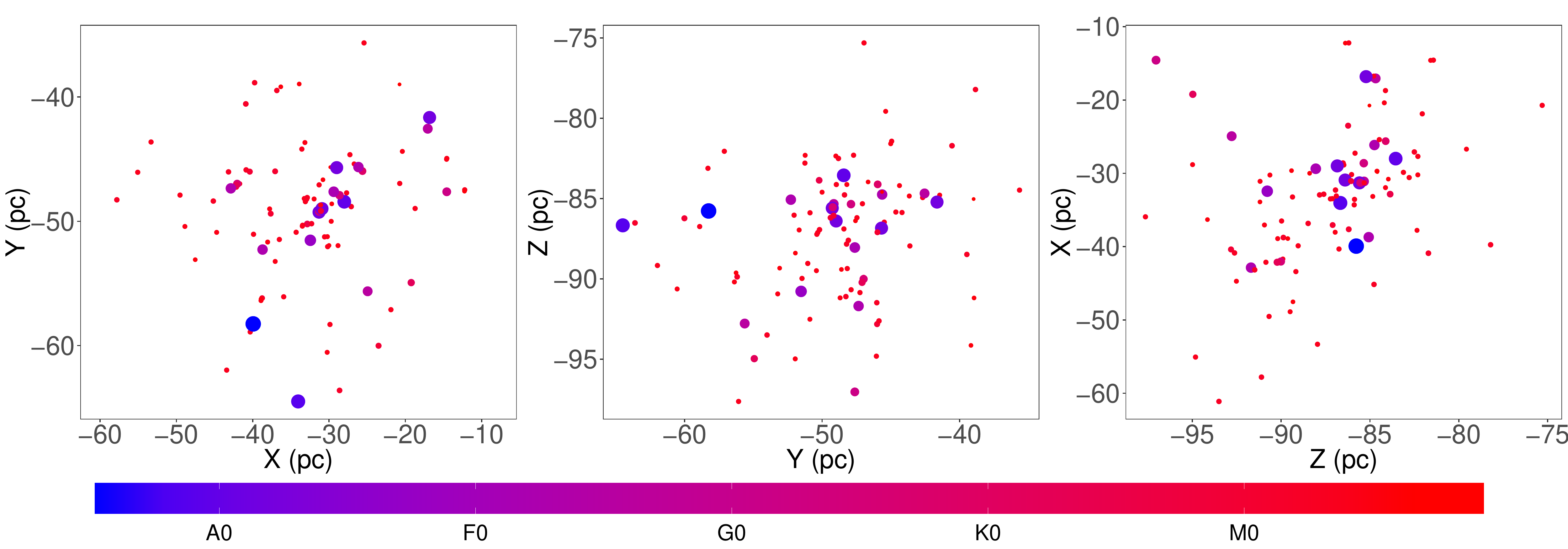}
\caption{
3D spatial distribution of the stars in the cluster and the effects of mass segregation. The size and colour of the symbols are scaled with the spectral type of the stars derived in this study (see Sect.~\ref{section3}) in the sense that small red symbols denote the M dwarfs in our sample. 
}\label{fig_mass_segregation}
\end{center}
\end{figure*}

Figure~\ref{fig_mass_segregation} shows the spatial distribution of the stars from our sample of cluster members with available spectral classification (see Sect.~\ref{section3}). Using the spectral type of the stars as a proxy for the stellar mass, we can confirm from our observations that the cluster is mass segregated with the more massive stars located preferentially in the central portions of the cluster and the M dwarfs spread across multiple locations (see also ZKK2019). For example, we note that nine stars among the 15 cluster members with spectral types B, A, and F in our sample are located within 10~pc of the 3D median position of the cluster $(X,Y,Z)=(-32,-49,-87)$~pc. The three most dispersed early-type cluster members are \textit{Gaia} EDR3 4843613397714473088 (035629-385744), \textit{Gaia} EDR3 4949081572411575552 (025352-401139), and \textit{Gaia} EDR3 4949158198924394496 (025048-395556) which are located farther than 15~pc from the cluster centre. The scatter of XYZ positions is $(\sigma_{X},\sigma_{Y},\sigma_{Z})=(5.6,3.8,3.0)$~pc for the B, A, and F stars in our sample (excluding the three most dispersed stars beyond 15~pc of the centre), and $(\sigma_{X},\sigma_{Y},\sigma_{Z})=(10.0,5.3,4.1)$~pc for the remaining cluster members of spectral types G, K, and M. 

\begin{figure}
\begin{center}
\includegraphics[width=0.47\textwidth]{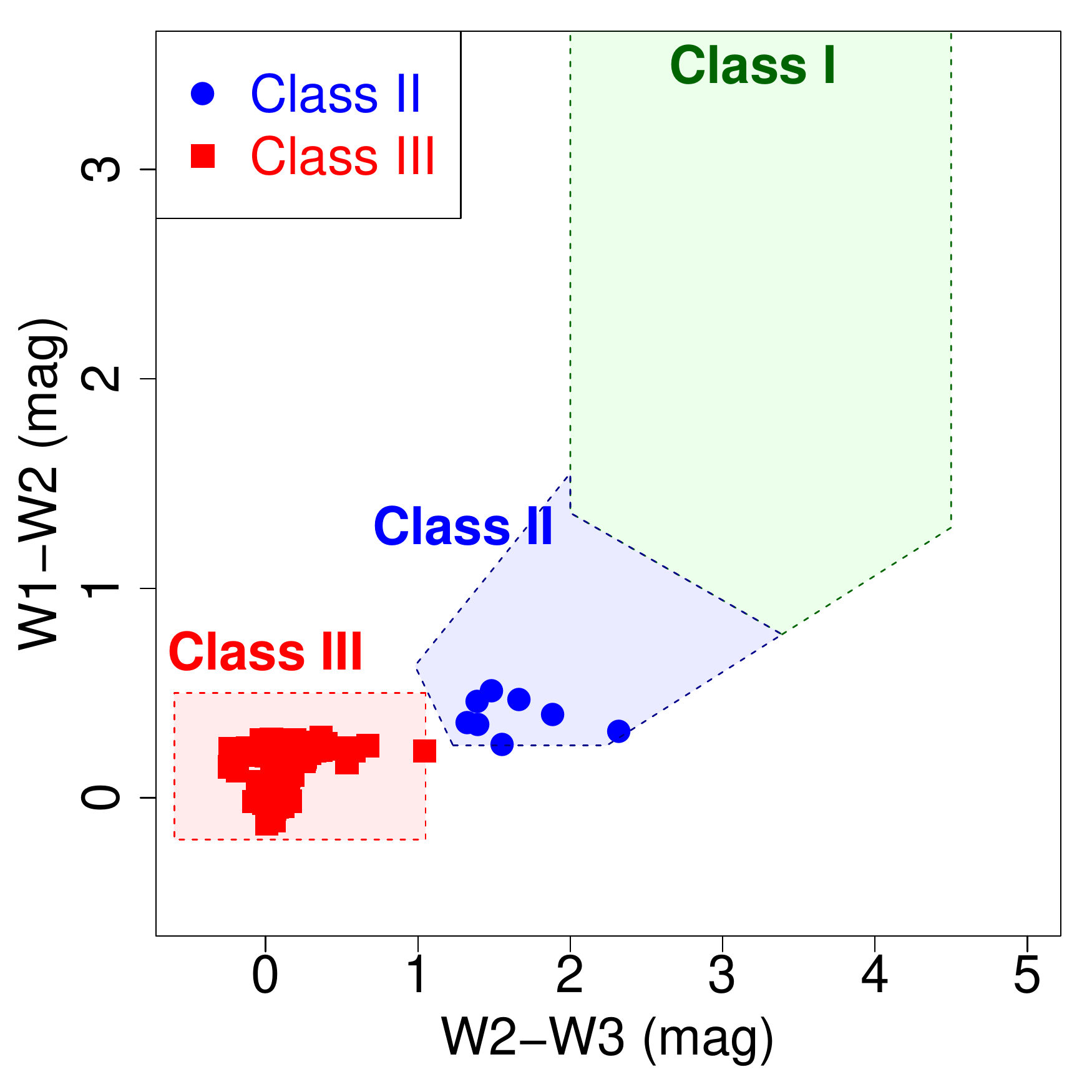}
\caption{
Colour-colour diagram of cluster members using infrared photometry from the AllWISE catalogue. This is one of the diagrams used in the classification scheme developed by \citet{Koenig2014} to distinguish the SED subclasses based on infrared excess emission. 
}\label{fig_koenig2014}
\end{center}
\end{figure}

\begin{figure*}
\begin{center}
\includegraphics[width=0.33\textwidth]{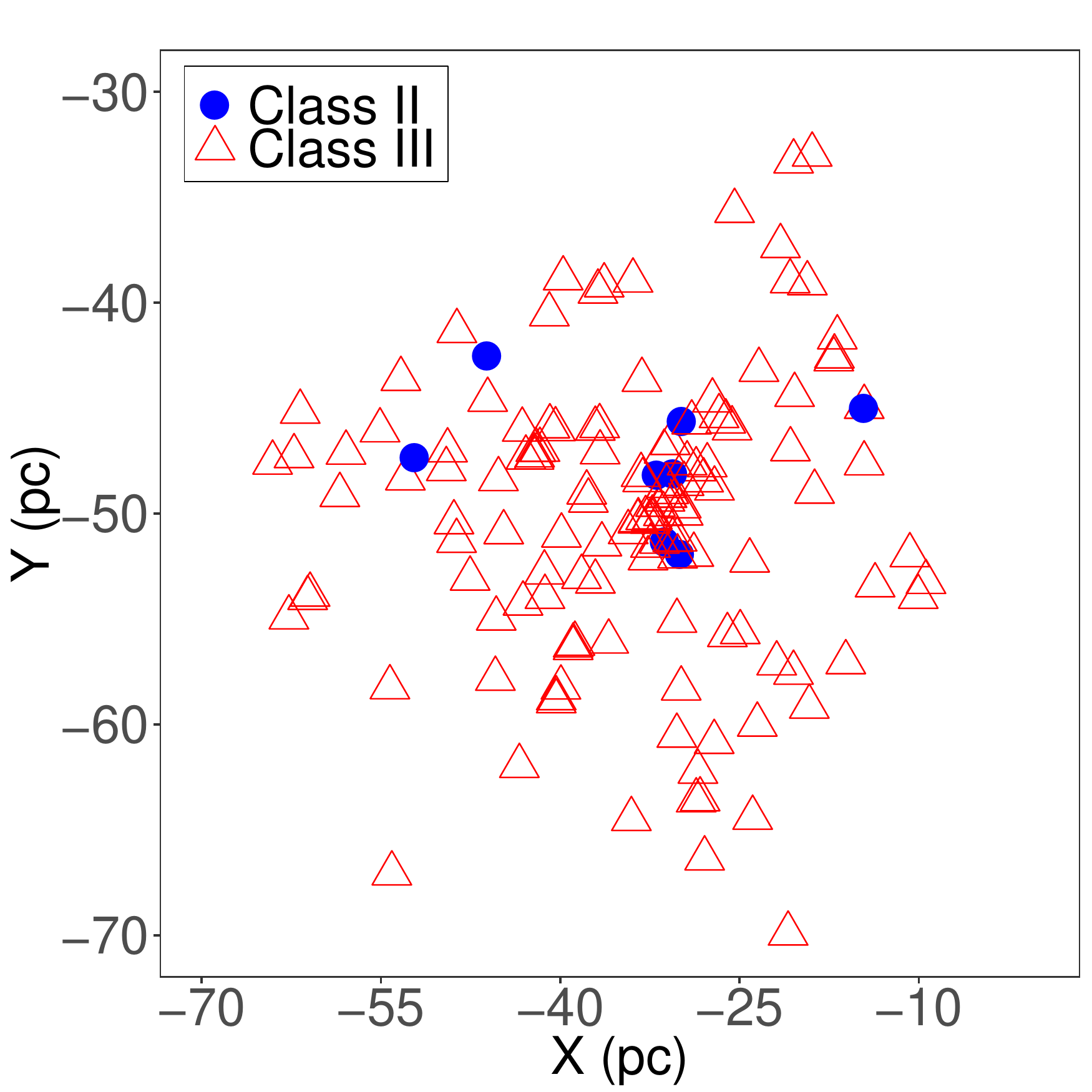}
\includegraphics[width=0.33\textwidth]{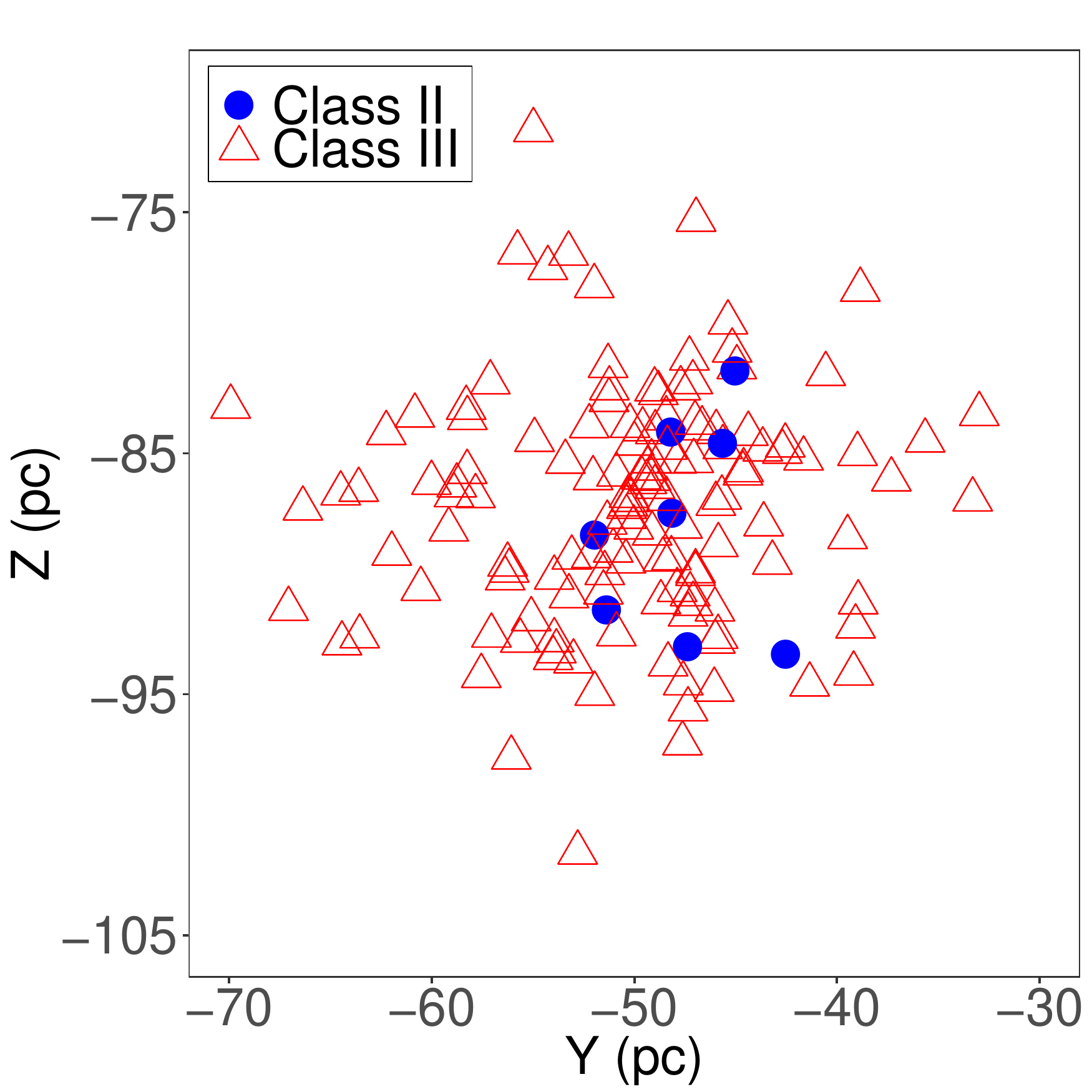}
\includegraphics[width=0.33\textwidth]{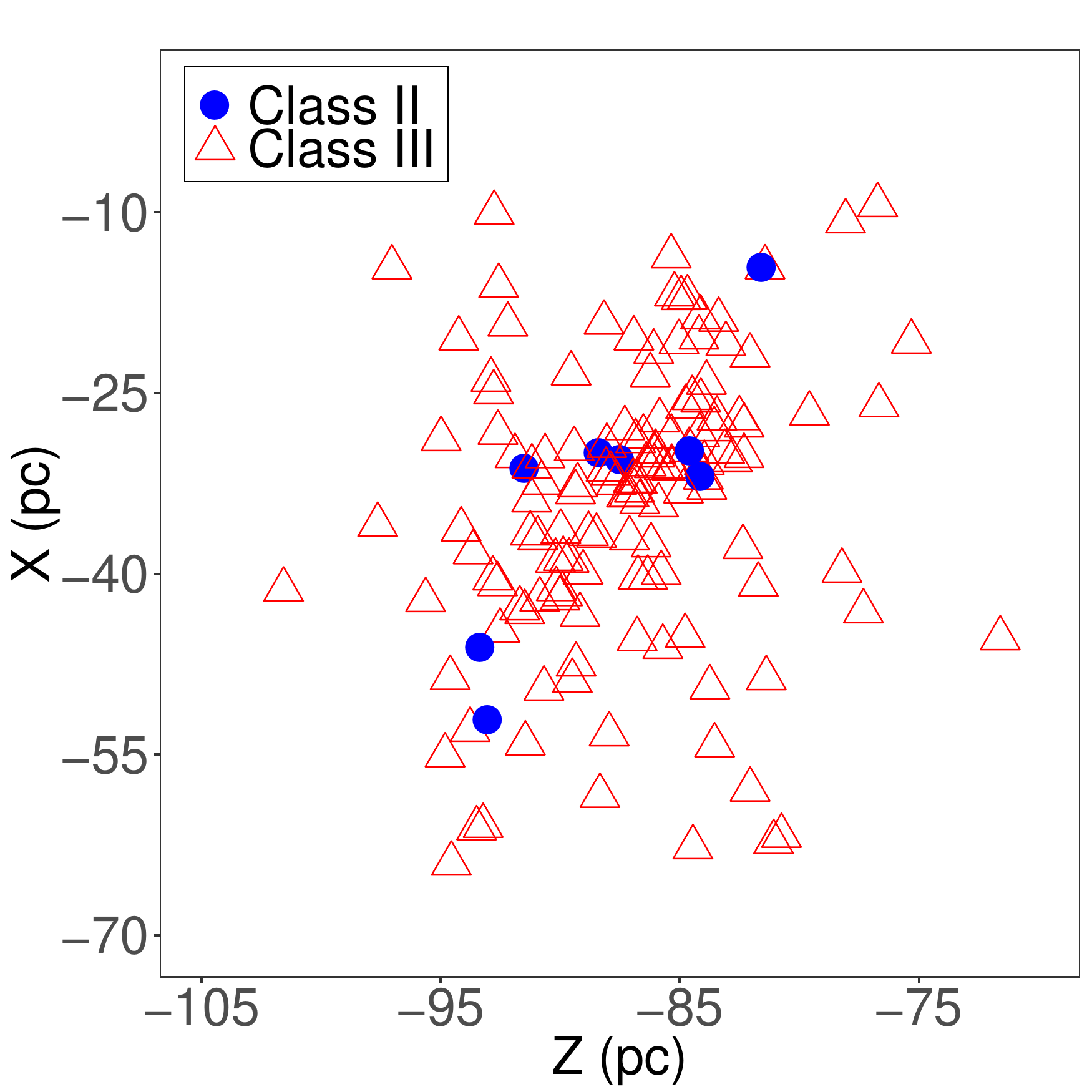}
\caption{
3D spatial distribution of the cluster members identified in our membership analysis with SED classification derived in this study (see Sect.~\ref{section4.5}). The different colours and symbols denote the SED subclasses. 
}\label{fig_XYZ}
\end{center}
\end{figure*}

We now compare the spatial distribution of the stars with different SED subclasses in our sample. We use the classification scheme proposed by \citet{Koenig2014} to identify Class~I and Class~II sources based on infrared colours from the AllWISE catalogue. We classified 142~stars in our sample of cluster members with available photometric data for this analysis (see Figure~\ref{fig_koenig2014}) and we found only eight Class~II stars (i.e. 5\% of the sample) which are listed in Table~\ref{tab_members}. These sources are late-type stars as confirmed from our observations and results derived from the SED fit. Five of them are listed in Tables~2 and 3 of ZKK2019. The remaining three sources, namely \textit{Gaia} EDR3 4755745925980823680 (025902-423220), \textit{Gaia} EDR3 5060663418310359424 (032443-273323), and \textit{Gaia} EDR3 5081640790199917696 (033852-264615), are new members identified in this study. The remaining cluster members are likely to be Class~III stars, given the young ages that we observe from the HR diagram and they are the dominant population in our sample (as expected for a 30~Myr old cluster). As illustrated in Figure~\ref{fig_XYZ}, the Class~II stars are mostly concentrated in the central portions of the cluster. 

\subsection{Initial mass function}\label{section4.6}
In this section, we use our new sample of XFOR stars to discuss the initial mass function of the cluster. The stellar masses inferred from evolutionary models can vary significantly depending on the grid of models that is used in the analysis and some models do not cover the entire mass domain of our sample (see Figure~\ref{fig_HRD}). So, to overcome this difficulty, we used the distribution of spectral types in the sample derived from our observations (see Sect.~\ref{section3}) as a proxy for the initial mass function.  

We estimate the completeness limits of our sample of XFOR stars from the completeness limits of the \textit{Gaia}-EDR3 catalogue, which is the only source of data used in the membership analysis. The \textit{Gaia}-EDR3 catalogue has a high completeness at $G=19$~mag \citep[see Sect.~2.2 of][]{Fabricius2021} and provides proper motion and parallax (i.e. five and six-parameter solutions) for about 98.3\% of the detected sources up to this magnitude \citep[see Tables 4 and 5 of][]{Lindegren2021b}. We therefore consider our sample to be complete up to $G=19$~mag. We used the 3D dust maps produced by \citet{Lallement2019} to estimate the extinction at the position of the stars in our sample. We verified that the maximum observed interstellar extinction for our sample (within the $1\sigma$ uncertainty of the reddening values delivered by the 3D maps) is $A_{G}=0.04$~mag. Thus, the completeness limit corrected for the maximum observed extinction at the distance of the cluster ($d=108.4\pm0.3$~pc, see Sect.~\ref{section4.2}) yields the absolute magnitude of $G_{abs}=13.79$~mag. Now, if we assume that the cluster has an age of 30~Myr and we use the BHAC15 grid models to convert the absolute magnitude into mass, we may conclude that our sample is complete down to 0.04~M$_{\odot}$.

\begin{figure}
\begin{center}
\includegraphics[width=0.49\textwidth]{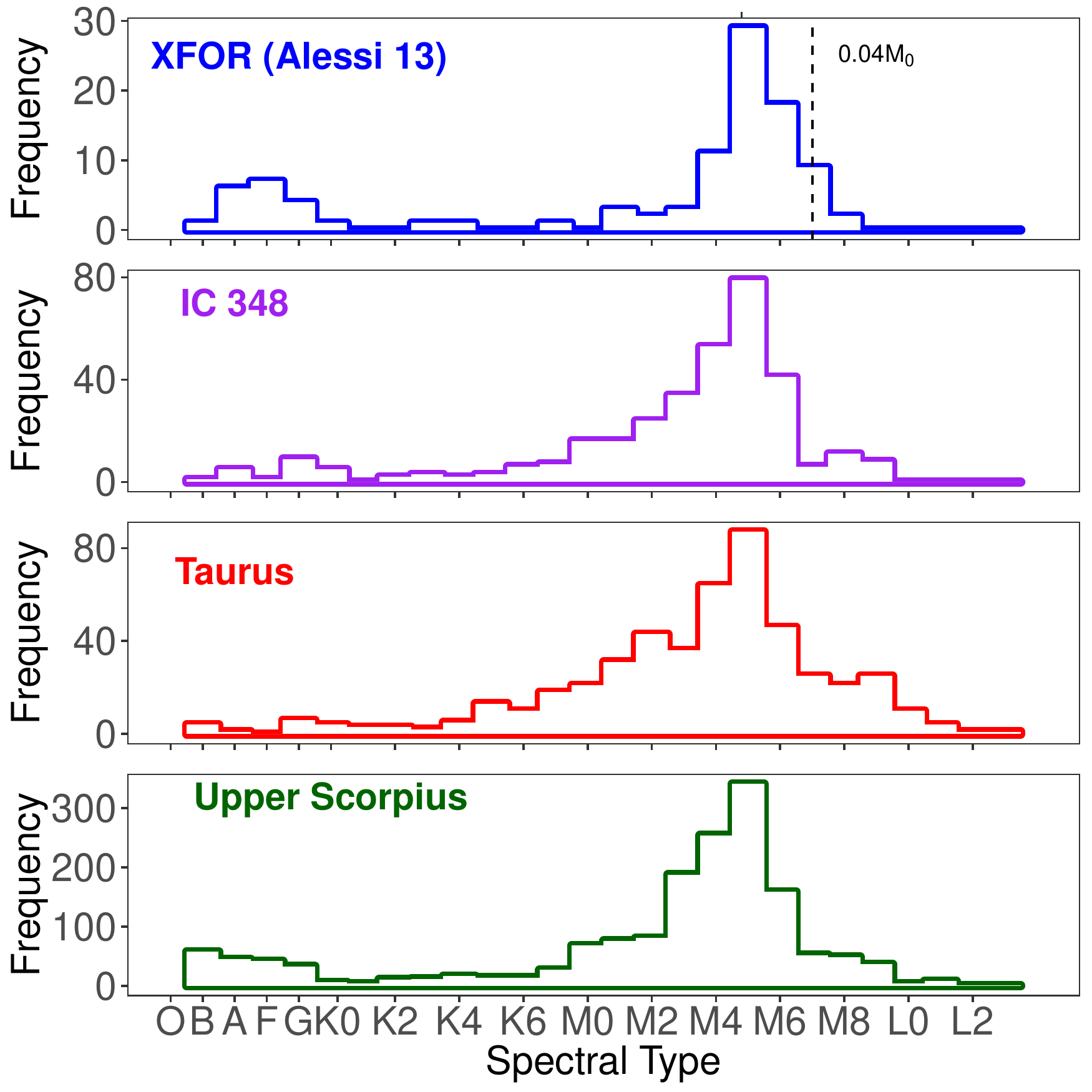}
\caption{
Distribution of spectral types of the samples of stars in the XFOR cluster (this paper), IC~348 \citep{Muench2007}, Taurus \citep{Esplin2019}, and Upper Scorpius \citep{Luhman2018}. The dashed line indicates the completeness limit of the XFOR sample. 
}\label{fig_IMF}
\end{center}
\end{figure}

In Figure~\ref{fig_IMF}, we illustrate the distribution of spectral types of the XFOR cluster and compare it to the samples of stars in IC~348 \citep{Muench2007}, Taurus \citep{Esplin2019}, and Upper Scorpius \citep{Luhman2018}. We show that both the XFOR cluster and the Upper Scorpius associations exhibit a surplus of early-type stars as compared to Taurus and IC~348. However, it is apparent that the distribution of the late-type stars in the XFOR cluster down to the completeness limit of our sample resembles the other young stellar groups. The similar distribution of spectral types that we observe among these young stellar groups suggests that the initial mass function most likely shows little variation. The Taurus and Upper Scorpius samples of stars include a more significant number of substellar objects, which are still missing in our sample of XFOR cluster members. These samples have been obtained from ancillary observations in the optical and infrared, in addition to the \textit{Gaia} data. We therefore argue that future studies in the XFOR cluster using additional material to complement the \textit{Gaia} catalogue will facilitate the detection of additional cluster members and potentially reveal a population of substellar objects in the cluster, extending the initial mass function down to the planetary-mass regime.  

\section{Conclusions}\label{section5}

In this study, we performed a new membership analysis of the XFOR cluster based on the astrometric and photometric data of the \textit{Gaia}-EDR3 catalogue. We applied a probabilistic method to infer membership probabilities of the stars over a field of more than 1\,000~deg$^{2}$ and selected 164 cluster members. We confirmed most of the cluster members that have been previously identified in the literature based on \textit{Gaia}-DR2 data and discovered new ones increasing the list of cluster members by about 60\%. We conducted low-resolution spectroscopy of the cluster members, derive the spectral type of the stars and use our new sample of members to revisit some properties of the cluster. 

Our results obtained in this study show that the XFOR cluster is located at $108.4\pm0.3$~pc and it is comoving with the THA, COL, and CAR young stellar associations. A dynamical traceback reveals that the cluster was significantly closer to these young stellar groups in the past, which suggests a common origin. We do not detect important expansion or rotation effects in the cluster and we argue that the observed velocity dispersion is most likely associated with random motions. The spatial distribution of the stars shows that the cluster is mass segregated and that the few disc-bearing stars in our sample are mostly located in the central portions of the cluster. The empirical isochrone of the cluster that we derive from our membership analysis is consistent with the age of 30~Myr. However, the observed scatter in the HR diagram, in particular for the least massive cluster members, along with the age differences among the various models, call for a revision using other age-dating techniques. We estimate our sample of cluster members to be complete down to about 0.04~M$_{\odot}$ and argue that the initial mass function of the cluster (inferred from the distribution of the spectral types of the stars) shows little variation compared to other young stellar groups. 

In this paper, we use the most recent data release delivered by the \textit{Gaia} space mission combined with archival spectra and our own observations to deliver a complete picture of the stellar population and properties of the cluster. We believe that future studies using ancillary material to complement the \textit{Gaia} catalogue have the potential to detect more substellar cluster members as done in other young stellar groups. The main limitation to the present study was the lack of high-resolution spectra. We therefore encourage astronomers to conduct high-resolution spectroscopy of the cluster that will allow us to investigate the kinematics of the stars in more detail, refine the age of the cluster, and prepare for exoplanet programmes. This will make the XFOR cluster a cornerstone for many studies related to star and planet formation at this young stage of stellar evolution.

\begin{acknowledgements}
This research has received funding from the European Research Council (ERC) under the European Union’s Horizon 2020 research 
and innovation programme (grant agreement No 682903, P.I. H. Bouy), and from the French State in the framework of the 
``Investments for the future” Program, IdEx Bordeaux, reference ANR-10-IDEX-03-02. This research has made use of the SIMBAD database, operated at CDS, Strasbourg, France. This work has made use of data from the European Space Agency (ESA) mission {\it Gaia} (\url{https://www.cosmos.esa.int/gaia}), processed by the {\it Gaia} Data Processing and Analysis Consortium (DPAC, \url{https://www.cosmos.esa.int/web/gaia/dpac/consortium}). Funding for the DPAC has been provided by national institutions, in particular the institutions participating in the {\it Gaia} Multilateral Agreement. This publication makes use of data products from the Wide-field Infrared Survey Explorer, which is a joint project of the University of California, Los Angeles, and the Jet Propulsion Laboratory/California Institute of Technology, funded by the National Aeronautics and Space Administration. This publication makes use of VOSA, developed under the Spanish Virtual Observatory project supported by the Spanish MINECO through grant AyA2017-84089. VOSA has been partially updated by using funding from the European Union’s Horizon 2020 Research and Innovation Programme, under Grant Agreement no 776403 (EXOPLANETS-A).
\end{acknowledgements}

\newpage
\bibliographystyle{aa} 
\bibliography{references} 

\begin{thebibliography}{79}
\expandafter\ifx\csname natexlab\endcsname\relax\def\natexlab#1{#1}\fi

\bibitem[{{Allard} {et~al.}(2012){Allard}, {Homeier}, \&
  {Freytag}}]{Allard2012}
{Allard}, F., {Homeier}, D., \& {Freytag}, B. 2012, Philosophical Transactions
  of the Royal Society of London Series A, 370, 2765

\bibitem[{{Allen} \& {Santillan}(1991)}]{Allen1991}
{Allen}, C. \& {Santillan}, A. 1991, \rmxaa, 22, 255

\bibitem[{{Asiain} {et~al.}(1999){Asiain}, {Figueras}, \& {Torra}}]{Asiain1999}
{Asiain}, R., {Figueras}, F., \& {Torra}, J. 1999, \aap, 350, 434

\bibitem[{{Baraffe} {et~al.}(2015){Baraffe}, {Homeier}, {Allard}, \&
  {Chabrier}}]{BHAC15}
{Baraffe}, I., {Homeier}, D., {Allard}, F., \& {Chabrier}, G. 2015, \aap, 577,
  A42

\bibitem[{{Baranne} {et~al.}(1996){Baranne}, {Queloz}, {Mayor}, {Adrianzyk},
  {Knispel}, {Kohler}, {Lacroix}, {Meunier}, {Rimbaud}, \& {Vin}}]{Baranne1996}
{Baranne}, A., {Queloz}, D., {Mayor}, M., {et~al.} 1996, \aaps, 119, 373

\bibitem[{{Bayo} {et~al.}(2008){Bayo}, {Rodrigo}, {Barrado Y Navascu{\'e}s},
  {Solano}, {Guti{\'e}rrez}, {Morales-Calder{\'o}n}, \& {Allard}}]{VOSA}
{Bayo}, A., {Rodrigo}, C., {Barrado Y Navascu{\'e}s}, D., {et~al.} 2008, \aap,
  492, 277

\bibitem[{{Bell} {et~al.}(2015){Bell}, {Mamajek}, \& {Naylor}}]{Bell2015}
{Bell}, C. P.~M., {Mamajek}, E.~E., \& {Naylor}, T. 2015, \mnras, 454, 593

\bibitem[{{Bergeron} {et~al.}(1995){Bergeron}, {Wesemael}, \&
  {Beauchamp}}]{Bergeron1995}
{Bergeron}, P., {Wesemael}, F., \& {Beauchamp}, A. 1995, \pasp, 107, 1047

\bibitem[{{Binks} \& {Jeffries}(2014)}]{Binks2014}
{Binks}, A.~S. \& {Jeffries}, R.~D. 2014, \mnras, 438, L11

\bibitem[{{Blanco-Cuaresma} {et~al.}(2014){Blanco-Cuaresma}, {Soubiran},
  {Heiter}, \& {Jofr{\'e}}}]{BlancoCuaresma2014}
{Blanco-Cuaresma}, S., {Soubiran}, C., {Heiter}, U., \& {Jofr{\'e}}, P. 2014,
  \aap, 569, A111

\bibitem[{{Bouy} {et~al.}(2013){Bouy}, {Bertin}, {Moraux}, {Cuillandre},
  {Bouvier}, {Barrado}, {Solano}, \& {Bayo}}]{Bouy2013}
{Bouy}, H., {Bertin}, E., {Moraux}, E., {et~al.} 2013, \aap, 554, A101

\bibitem[{{Bressan} {et~al.}(2012){Bressan}, {Marigo}, {Girardi}, {Salasnich},
  {Dal Cero}, {Rubele}, \& {Nanni}}]{PARSEC}
{Bressan}, A., {Marigo}, P., {Girardi}, L., {et~al.} 2012, \mnras, 427, 127

\bibitem[{{Buzzoni} {et~al.}(1984){Buzzoni}, {Delabre}, {Dekker}, {Dodorico},
  {Enard}, {Focardi}, {Gustafsson}, {Nees}, {Paureau}, \&
  {Reiss}}]{Buzzoni1984}
{Buzzoni}, B., {Delabre}, B., {Dekker}, H., {et~al.} 1984, The Messenger, 38, 9

\bibitem[{{Cantat-Gaudin} {et~al.}(2018){Cantat-Gaudin}, {Jordi}, {Vallenari},
  {Bragaglia}, {Balaguer-N{\'u}{\~n}ez}, {Soubiran}, {Bossini}, {Moitinho},
  {Castro-Ginard}, {Krone-Martins}, {Casamiquela}, {Sordo}, \&
  {Carrera}}]{CantatGaudin2018}
{Cantat-Gaudin}, T., {Jordi}, C., {Vallenari}, A., {et~al.} 2018, \aap, 618,
  A93

\bibitem[{{Castro-Ginard} {et~al.}(2020){Castro-Ginard}, {Jordi}, {Luri},
  {{\'A}lvarez Cid-Fuentes}, {Casamiquela}, {Anders}, {Cantat-Gaudin},
  {Mongui{\'o}}, {Balaguer-N{\'u}{\~n}ez}, {Sol{\`a}}, \&
  {Badia}}]{CastroGinard2020}
{Castro-Ginard}, A., {Jordi}, C., {Luri}, X., {et~al.} 2020, \aap, 635, A45

\bibitem[{{Chambers} {et~al.}(2016){Chambers}, {Magnier}, {Metcalfe},
  {Flewelling}, {Huber}, {Waters}, {Denneau}, {Draper}, {Farrow}, {Finkbeiner},
  {Holmberg}, {Koppenhoefer}, {Price}, {Rest}, {Saglia}, {Schlafly}, {Smartt},
  {Sweeney}, {Wainscoat}, {Burgett}, {Chastel}, {Grav}, {Heasley}, {Hodapp},
  {Jedicke}, {Kaiser}, {Kudritzki}, {Luppino}, {Lupton}, {Monet}, {Morgan},
  {Onaka}, {Shiao}, {Stubbs}, {Tonry}, {White}, {Ba{\~n}ados}, {Bell},
  {Bender}, {Bernard}, {Boegner}, {Boffi}, {Botticella}, {Calamida},
  {Casertano}, {Chen}, {Chen}, {Cole}, {Deacon}, {Frenk}, {Fitzsimmons},
  {Gezari}, {Gibbs}, {Goessl}, {Goggia}, {Gourgue}, {Goldman}, {Grant},
  {Grebel}, {Hambly}, {Hasinger}, {Heavens}, {Heckman}, {Henderson}, {Henning},
  {Holman}, {Hopp}, {Ip}, {Isani}, {Jackson}, {Keyes}, {Koekemoer}, {Kotak},
  {Le}, {Liska}, {Long}, {Lucey}, {Liu}, {Martin}, {Masci}, {McLean}, {Mindel},
  {Misra}, {Morganson}, {Murphy}, {Obaika}, {Narayan}, {Nieto-Santisteban},
  {Norberg}, {Peacock}, {Pier}, {Postman}, {Primak}, {Rae}, {Rai}, {Riess},
  {Riffeser}, {Rix}, {R{\"o}ser}, {Russel}, {Rutz}, {Schilbach}, {Schultz},
  {Scolnic}, {Strolger}, {Szalay}, {Seitz}, {Small}, {Smith}, {Soderblom},
  {Taylor}, {Thomson}, {Taylor}, {Thakar}, {Thiel}, {Thilker}, {Unger},
  {Urata}, {Valenti}, {Wagner}, {Walder}, {Walter}, {Watters}, {Werner},
  {Wood-Vasey}, \& {Wyse}}]{PanSTARRS}
{Chambers}, K.~C., {Magnier}, E.~A., {Metcalfe}, N., {et~al.} 2016, arXiv
  e-prints, arXiv:1612.05560

\bibitem[{{Choi} {et~al.}(2016){Choi}, {Dotter}, {Conroy}, {Cantiello},
  {Paxton}, \& {Johnson}}]{MIST}
{Choi}, J., {Dotter}, A., {Conroy}, C., {et~al.} 2016, \apj, 823, 102

\bibitem[{{Covey} {et~al.}(2007){Covey}, {Ivezi{\'c}}, {Schlegel},
  {Finkbeiner}, {Padmanabhan}, {Lupton}, {Ag{\"u}eros}, {Bochanski}, {Hawley},
  {West}, {Seth}, {Kimball}, {Gogarten}, {Claire}, {Haggard}, {Kaib},
  {Schneider}, \& {Sesar}}]{Covey2007}
{Covey}, K.~R., {Ivezi{\'c}}, {\v{Z}}., {Schlegel}, D., {et~al.} 2007, \aj,
  134, 2398

\bibitem[{{Cutri} {et~al.}(2003){Cutri}, {Skrutskie}, {van Dyk}, {Beichman},
  {Carpenter}, {Chester}, {Cambresy}, {Evans}, {Fowler}, {Gizis}, {Howard},
  {Huchra}, {Jarrett}, {Kopan}, {Kirkpatrick}, {Light}, {Marsh}, {McCallon},
  {Schneider}, {Stiening}, {Sykes}, {Weinberg}, {Wheaton}, {Wheelock}, \&
  {Zacarias}}]{2MASS}
{Cutri}, R.~M., {Skrutskie}, M.~F., {van Dyk}, S., {et~al.} 2003, VizieR Online
  Data Catalog, II/246

\bibitem[{{Dekker} {et~al.}(2000){Dekker}, {D'Odorico}, {Kaufer}, {Delabre}, \&
  {Kotzlowski}}]{UVES}
{Dekker}, H., {D'Odorico}, S., {Kaufer}, A., {Delabre}, B., \& {Kotzlowski}, H.
  2000, in Society of Photo-Optical Instrumentation Engineers (SPIE) Conference
  Series, Vol. 4008, Optical and IR Telescope Instrumentation and Detectors,
  ed. M.~{Iye} \& A.~F. {Moorwood}, 534--545

\bibitem[{{Dias} {et~al.}(2002){Dias}, {Alessi}, {Moitinho}, \&
  {L{\'e}pine}}]{Dias2002}
{Dias}, W.~S., {Alessi}, B.~S., {Moitinho}, A., \& {L{\'e}pine}, J.~R.~D. 2002,
  \aap, 389, 871

\bibitem[{{Elson} {et~al.}(1987){Elson}, {Fall}, \& {Freeman}}]{Elson1987}
{Elson}, R. A.~W., {Fall}, S.~M., \& {Freeman}, K.~C. 1987, \apj, 323, 54

\bibitem[{{Esplin} \& {Luhman}(2019)}]{Esplin2019}
{Esplin}, T.~L. \& {Luhman}, K.~L. 2019, \aj, 158, 54

\bibitem[{{Fabricius} {et~al.}(2021){Fabricius}, {Luri}, {Arenou}, {Babusiaux},
  {Helmi}, {Muraveva}, {Reyl{\'e}}, {Spoto}, {Vallenari}, {Antoja}, {Balbinot},
  {Barache}, {Bauchet}, {Bragaglia}, {Busonero}, {Cantat-Gaudin}, {Carrasco},
  {Diakit{\'e}}, {Fabrizio}, {Figueras}, {Garcia-Gutierrez}, {Garofalo},
  {Jordi}, {Kervella}, {Khanna}, {Leclerc}, {Licata}, {Lambert}, {Marrese},
  {Masip}, {Ramos}, {Robichon}, {Robin}, {Romero-G{\'o}mez}, {Rubele}, \&
  {Weiler}}]{Fabricius2021}
{Fabricius}, C., {Luri}, X., {Arenou}, F., {et~al.} 2021, \aap, 649, A5

\bibitem[{{Gagn{\'e}} \& {Faherty}(2018)}]{Gagne2018c}
{Gagn{\'e}}, J. \& {Faherty}, J.~K. 2018, \apj, 862, 138

\bibitem[{{Gagn{\'e}} {et~al.}(2021){Gagn{\'e}}, {Faherty}, {Moranta}, \&
  {Popinchalk}}]{Gagne2021}
{Gagn{\'e}}, J., {Faherty}, J.~K., {Moranta}, L., \& {Popinchalk}, M. 2021,
  \apjl, 915, L29

\bibitem[{{Gagn{\'e}} {et~al.}(2018{\natexlab{a}}){Gagn{\'e}}, {Mamajek},
  {Malo}, {Riedel}, {Rodriguez}, {Lafreni{\`e}re}, {Faherty}, {Roy-Loubier},
  {Pueyo}, {Robin}, \& {Doyon}}]{Gagne2018a}
{Gagn{\'e}}, J., {Mamajek}, E.~E., {Malo}, L., {et~al.} 2018{\natexlab{a}},
  \apj, 856, 23

\bibitem[{{Gagn{\'e}} {et~al.}(2018{\natexlab{b}}){Gagn{\'e}}, {Roy-Loubier},
  {Faherty}, {Doyon}, \& {Malo}}]{Gagne2018b}
{Gagn{\'e}}, J., {Roy-Loubier}, O., {Faherty}, J.~K., {Doyon}, R., \& {Malo},
  L. 2018{\natexlab{b}}, \apj, 860, 43

\bibitem[{{Gaia Collaboration} {et~al.}(2018){Gaia Collaboration}, {Brown},
  {Vallenari}, {Prusti}, {de Bruijne}, {Babusiaux}, {Bailer-Jones}, {Biermann},
  {Evans}, {Eyer}, {Jansen}, {Jordi}, {Klioner}, {Lammers}, {Lindegren},
  {Luri}, {Mignard}, {Panem}, {Pourbaix}, {Randich}, {Sartoretti}, {Siddiqui},
  {Soubiran}, {van Leeuwen}, {Walton}, {Arenou}, {Bastian}, {Cropper},
  {Drimmel}, {Katz}, {Lattanzi}, {Bakker}, {Cacciari}, {Casta{\~n}eda},
  {Chaoul}, {Cheek}, {De Angeli}, {Fabricius}, {Guerra}, {Holl}, {Masana},
  {Messineo}, {Mowlavi}, {Nienartowicz}, {Panuzzo}, {Portell}, {Riello},
  {Seabroke}, {Tanga}, {Th{\'e}venin}, {Gracia-Abril}, {Comoretto},
  {Garcia-Reinaldos}, {Teyssier}, {Altmann}, {Andrae}, {Audard},
  {Bellas-Velidis}, {Benson}, {Berthier}, {Blomme}, {Burgess}, {Busso},
  {Carry}, {Cellino}, {Clementini}, {Clotet}, {Creevey}, {Davidson}, {De
  Ridder}, {Delchambre}, {Dell'Oro}, {Ducourant},
  {Fern{\'a}ndez-Hern{\'a}ndez}, {Fouesneau}, {Fr{\'e}mat}, {Galluccio},
  {Garc{\'\i}a-Torres}, {Gonz{\'a}lez-N{\'u}{\~n}ez}, {Gonz{\'a}lez-Vidal},
  {Gosset}, {Guy}, {Halbwachs}, {Hambly}, {Harrison}, {Hern{\'a}ndez},
  {Hestroffer}, {Hodgkin}, {Hutton}, {Jasniewicz}, {Jean-Antoine-Piccolo},
  {Jordan}, {Korn}, {Krone-Martins}, {Lanzafame}, {Lebzelter}, {L{\"o}ffler},
  {Manteiga}, {Marrese}, {Mart{\'\i}n-Fleitas}, {Moitinho}, {Mora}, {Muinonen},
  {Osinde}, {Pancino}, {Pauwels}, {Petit}, {Recio-Blanco}, {Richards},
  {Rimoldini}, {Robin}, {Sarro}, {Siopis}, {Smith}, {Sozzetti}, {S{\"u}veges},
  {Torra}, {van Reeven}, {Abbas}, {Abreu Aramburu}, {Accart}, {Aerts},
  {Altavilla}, {{\'A}lvarez}, {Alvarez}, {Alves}, {Anderson}, {Andrei},
  {Anglada Varela}, {Antiche}, {Antoja}, {Arcay}, {Astraatmadja}, {Bach},
  {Baker}, {Balaguer-N{\'u}{\~n}ez}, {Balm}, {Barache}, {Barata}, {Barbato},
  {Barblan}, {Barklem}, {Barrado}, {Barros}, {Barstow}, {Bartholom{\'e}
  Mu{\~n}oz}, {Bassilana}, {Becciani}, {Bellazzini}, {Berihuete}, {Bertone},
  {Bianchi}, {Bienaym{\'e}}, {Blanco-Cuaresma}, {Boch}, {Boeche}, {Bombrun},
  {Borrachero}, {Bossini}, {Bouquillon}, {Bourda}, {Bragaglia}, {Bramante},
  {Breddels}, {Bressan}, {Brouillet}, {Br{\"u}semeister}, {Brugaletta},
  {Bucciarelli}, {Burlacu}, {Busonero}, {Butkevich}, {Buzzi}, {Caffau},
  {Cancelliere}, {Cannizzaro}, {Cantat-Gaudin}, {Carballo}, {Carlucci},
  {Carrasco}, {Casamiquela}, {Castellani}, {Castro-Ginard}, {Charlot},
  {Chemin}, {Chiavassa}, {Cocozza}, {Costigan}, {Cowell}, {Crifo}, {Crosta},
  {Crowley}, {Cuypers}, {Dafonte}, {Damerdji}, {Dapergolas}, {David}, {David},
  {de Laverny}, {De Luise}, {De March}, {de Martino}, {de Souza}, {de Torres},
  {Debosscher}, {del Pozo}, {Delbo}, {Delgado}, {Delgado}, {Di Matteo},
  {Diakite}, {Diener}, {Distefano}, {Dolding}, {Drazinos}, {Dur{\'a}n},
  {Edvardsson}, {Enke}, {Eriksson}, {Esquej}, {Eynard Bontemps}, {Fabre},
  {Fabrizio}, {Faigler}, {Falc{\~a}o}, {Farr{\`a}s Casas}, {Federici},
  {Fedorets}, {Fernique}, {Figueras}, {Filippi}, {Findeisen}, {Fonti},
  {Fraile}, {Fraser}, {Fr{\'e}zouls}, {Gai}, {Galleti}, {Garabato},
  {Garc{\'\i}a-Sedano}, {Garofalo}, {Garralda}, {Gavel}, {Gavras}, {Gerssen},
  {Geyer}, {Giacobbe}, {Gilmore}, {Girona}, {Giuffrida}, {Glass}, {Gomes},
  {Granvik}, {Gueguen}, {Guerrier}, {Guiraud}, {Guti{\'e}rrez-S{\'a}nchez},
  {Haigron}, {Hatzidimitriou}, {Hauser}, {Haywood}, {Heiter}, {Helmi}, {Heu},
  {Hilger}, {Hobbs}, {Hofmann}, {Holland}, {Huckle}, {Hypki}, {Icardi},
  {Jan{\ss}en}, {Jevardat de Fombelle}, {Jonker}, {Juh{\'a}sz}, {Julbe},
  {Karampelas}, {Kewley}, {Klar}, {Kochoska}, {Kohley}, {Kolenberg},
  {Kontizas}, {Kontizas}, {Koposov}, {Kordopatis}, {Kostrzewa-Rutkowska},
  {Koubsky}, {Lambert}, {Lanza}, {Lasne}, {Lavigne}, {Le Fustec}, {Le
  Poncin-Lafitte}, {Lebreton}, {Leccia}, {Leclerc}, {Lecoeur-Taibi},
  {Lenhardt}, {Leroux}, {Liao}, {Licata}, {Lindstr{\o}m}, {Lister}, {Livanou},
  {Lobel}, {L{\'o}pez}, {Managau}, {Mann}, {Mantelet}, {Marchal}, {Marchant},
  {Marconi}, {Marinoni}, {Marschalk{\'o}}, {Marshall}, {Martino}, {Marton},
  {Mary}, {Massari}, {Matijevi{\v{c}}}, {Mazeh}, {McMillan}, {Messina},
  {Michalik}, {Millar}, {Molina}, {Molinaro}, {Moln{\'a}r}, {Montegriffo},
  {Mor}, {Morbidelli}, {Morel}, {Morris}, {Mulone}, {Muraveva}, {Musella},
  {Nelemans}, {Nicastro}, {Noval}, {O'Mullane}, {Ord{\'e}novic},
  {Ord{\'o}{\~n}ez-Blanco}, {Osborne}, {Pagani}, {Pagano}, {Pailler},
  {Palacin}, {Palaversa}, {Panahi}, {Pawlak}, {Piersimoni}, {Pineau}, {Plachy},
  {Plum}, {Poggio}, {Poujoulet}, {Pr{\v{s}}a}, {Pulone}, {Racero}, {Ragaini},
  {Rambaux}, {Ramos-Lerate}, {Regibo}, {Reyl{\'e}}, {Riclet}, {Ripepi}, {Riva},
  {Rivard}, {Rixon}, {Roegiers}, {Roelens}, {Romero-G{\'o}mez}, {Rowell},
  {Royer}, {Ruiz-Dern}, {Sadowski}, {Sagrist{\`a} Sell{\'e}s}, {Sahlmann},
  {Salgado}, {Salguero}, {Sanna}, {Santana-Ros}, {Sarasso}, {Savietto},
  {Schultheis}, {Sciacca}, {Segol}, {Segovia}, {S{\'e}gransan}, {Shih},
  {Siltala}, {Silva}, {Smart}, {Smith}, {Solano}, {Solitro}, {Sordo}, {Soria
  Nieto}, {Souchay}, {Spagna}, {Spoto}, {Stampa}, {Steele},
  {Steidelm{\"u}ller}, {Stephenson}, {Stoev}, {Suess}, {Surdej}, {Szabados},
  {Szegedi-Elek}, {Tapiador}, {Taris}, {Tauran}, {Taylor}, {Teixeira},
  {Terrett}, {Teyssandier}, {Thuillot}, {Titarenko}, {Torra Clotet}, {Turon},
  {Ulla}, {Utrilla}, {Uzzi}, {Vaillant}, {Valentini}, {Valette}, {van Elteren},
  {Van Hemelryck}, {van Leeuwen}, {Vaschetto}, {Vecchiato}, {Veljanoski},
  {Viala}, {Vicente}, {Vogt}, {von Essen}, {Voss}, {Votruba}, {Voutsinas},
  {Walmsley}, {Weiler}, {Wertz}, {Wevers}, {Wyrzykowski}, {Yoldas},
  {{\v{Z}}erjal}, {Ziaeepour}, {Zorec}, {Zschocke}, {Zucker}, {Zurbach}, \&
  {Zwitter}}]{GaiaDR2}
{Gaia Collaboration}, {Brown}, A.~G.~A., {Vallenari}, A., {et~al.} 2018, \aap,
  616, A1

\bibitem[{{Gaia Collaboration} {et~al.}(2021){Gaia Collaboration}, {Brown},
  {Vallenari}, {Prusti}, {de Bruijne}, {Babusiaux}, {Biermann}, {Creevey},
  {Evans}, {Eyer}, {Hutton}, {Jansen}, {Jordi}, {Klioner}, {Lammers},
  {Lindegren}, {Luri}, {Mignard}, {Panem}, {Pourbaix}, {Randich}, {Sartoretti},
  {Soubiran}, {Walton}, {Arenou}, {Bailer-Jones}, {Bastian}, {Cropper},
  {Drimmel}, {Katz}, {Lattanzi}, {van Leeuwen}, {Bakker}, {Cacciari},
  {Casta{\~n}eda}, {De Angeli}, {Ducourant}, {Fabricius}, {Fouesneau},
  {Fr{\'e}mat}, {Guerra}, {Guerrier}, {Guiraud}, {Jean-Antoine Piccolo},
  {Masana}, {Messineo}, {Mowlavi}, {Nicolas}, {Nienartowicz}, {Pailler},
  {Panuzzo}, {Riclet}, {Roux}, {Seabroke}, {Sordo}, {Tanga}, {Th{\'e}venin},
  {Gracia-Abril}, {Portell}, {Teyssier}, {Altmann}, {Andrae}, {Bellas-Velidis},
  {Benson}, {Berthier}, {Blomme}, {Brugaletta}, {Burgess}, {Busso}, {Carry},
  {Cellino}, {Cheek}, {Clementini}, {Damerdji}, {Davidson}, {Delchambre},
  {Dell'Oro}, {Fern{\'a}ndez-Hern{\'a}ndez}, {Galluccio}, {Garc{\'\i}a-Lario},
  {Garcia-Reinaldos}, {Gonz{\'a}lez-N{\'u}{\~n}ez}, {Gosset}, {Haigron},
  {Halbwachs}, {Hambly}, {Harrison}, {Hatzidimitriou}, {Heiter},
  {Hern{\'a}ndez}, {Hestroffer}, {Hodgkin}, {Holl}, {Jan{\ss}en}, {Jevardat de
  Fombelle}, {Jordan}, {Krone-Martins}, {Lanzafame}, {L{\"o}ffler}, {Lorca},
  {Manteiga}, {Marchal}, {Marrese}, {Moitinho}, {Mora}, {Muinonen}, {Osborne},
  {Pancino}, {Pauwels}, {Petit}, {Recio-Blanco}, {Richards}, {Riello},
  {Rimoldini}, {Robin}, {Roegiers}, {Rybizki}, {Sarro}, {Siopis}, {Smith},
  {Sozzetti}, {Ulla}, {Utrilla}, {van Leeuwen}, {van Reeven}, {Abbas}, {Abreu
  Aramburu}, {Accart}, {Aerts}, {Aguado}, {Ajaj}, {Altavilla}, {{\'A}lvarez},
  {{\'A}lvarez Cid-Fuentes}, {Alves}, {Anderson}, {Anglada Varela}, {Antoja},
  {Audard}, {Baines}, {Baker}, {Balaguer-N{\'u}{\~n}ez}, {Balbinot}, {Balog},
  {Barache}, {Barbato}, {Barros}, {Barstow}, {Bartolom{\'e}}, {Bassilana},
  {Bauchet}, {Baudesson-Stella}, {Becciani}, {Bellazzini}, {Bernet}, {Bertone},
  {Bianchi}, {Blanco-Cuaresma}, {Boch}, {Bombrun}, {Bossini}, {Bouquillon},
  {Bragaglia}, {Bramante}, {Breedt}, {Bressan}, {Brouillet}, {Bucciarelli},
  {Burlacu}, {Busonero}, {Butkevich}, {Buzzi}, {Caffau}, {Cancelliere},
  {C{\'a}novas}, {Cantat-Gaudin}, {Carballo}, {Carlucci}, {Carnerero},
  {Carrasco}, {Casamiquela}, {Castellani}, {Castro-Ginard}, {Castro Sampol},
  {Chaoul}, {Charlot}, {Chemin}, {Chiavassa}, {Cioni}, {Comoretto}, {Cooper},
  {Cornez}, {Cowell}, {Crifo}, {Crosta}, {Crowley}, {Dafonte}, {Dapergolas},
  {David}, {David}, {de Laverny}, {De Luise}, {De March}, {De Ridder}, {de
  Souza}, {de Teodoro}, {de Torres}, {del Peloso}, {del Pozo}, {Delbo},
  {Delgado}, {Delgado}, {Delisle}, {Di Matteo}, {Diakite}, {Diener},
  {Distefano}, {Dolding}, {Eappachen}, {Edvardsson}, {Enke}, {Esquej}, {Fabre},
  {Fabrizio}, {Faigler}, {Fedorets}, {Fernique}, {Fienga}, {Figueras},
  {Fouron}, {Fragkoudi}, {Fraile}, {Franke}, {Gai}, {Garabato},
  {Garcia-Gutierrez}, {Garc{\'\i}a-Torres}, {Garofalo}, {Gavras}, {Gerlach},
  {Geyer}, {Giacobbe}, {Gilmore}, {Girona}, {Giuffrida}, {Gomel}, {Gomez},
  {Gonzalez-Santamaria}, {Gonz{\'a}lez-Vidal}, {Granvik},
  {Guti{\'e}rrez-S{\'a}nchez}, {Guy}, {Hauser}, {Haywood}, {Helmi}, {Hidalgo},
  {Hilger}, {H{\l}adczuk}, {Hobbs}, {Holland}, {Huckle}, {Jasniewicz},
  {Jonker}, {Juaristi Campillo}, {Julbe}, {Karbevska}, {Kervella}, {Khanna},
  {Kochoska}, {Kontizas}, {Kordopatis}, {Korn}, {Kostrzewa-Rutkowska},
  {Kruszy{\'n}ska}, {Lambert}, {Lanza}, {Lasne}, {Le Campion}, {Le Fustec},
  {Lebreton}, {Lebzelter}, {Leccia}, {Leclerc}, {Lecoeur-Taibi}, {Liao},
  {Licata}, {Lindstr{\o}m}, {Lister}, {Livanou}, {Lobel}, {Madrero Pardo},
  {Managau}, {Mann}, {Marchant}, {Marconi}, {Marcos Santos}, {Marinoni},
  {Marocco}, {Marshall}, {Martin Polo}, {Mart{\'\i}n-Fleitas}, {Masip},
  {Massari}, {Mastrobuono-Battisti}, {Mazeh}, {McMillan}, {Messina},
  {Michalik}, {Millar}, {Mints}, {Molina}, {Molinaro}, {Moln{\'a}r},
  {Montegriffo}, {Mor}, {Morbidelli}, {Morel}, {Morris}, {Mulone}, {Munoz},
  {Muraveva}, {Murphy}, {Musella}, {Noval}, {Ord{\'e}novic}, {Orr{\`u}},
  {Osinde}, {Pagani}, {Pagano}, {Palaversa}, {Palicio}, {Panahi}, {Pawlak},
  {Pe{\~n}alosa Esteller}, {Penttil{\"a}}, {Piersimoni}, {Pineau}, {Plachy},
  {Plum}, {Poggio}, {Poretti}, {Poujoulet}, {Pr{\v{s}}a}, {Pulone}, {Racero},
  {Ragaini}, {Rainer}, {Raiteri}, {Rambaux}, {Ramos}, {Ramos-Lerate}, {Re
  Fiorentin}, {Regibo}, {Reyl{\'e}}, {Ripepi}, {Riva}, {Rixon}, {Robichon},
  {Robin}, {Roelens}, {Rohrbasser}, {Romero-G{\'o}mez}, {Rowell}, {Royer},
  {Rybicki}, {Sadowski}, {Sagrist{\`a} Sell{\'e}s}, {Sahlmann}, {Salgado},
  {Salguero}, {Samaras}, {Sanchez Gimenez}, {Sanna}, {Santove{\~n}a},
  {Sarasso}, {Schultheis}, {Sciacca}, {Segol}, {Segovia}, {S{\'e}gransan},
  {Semeux}, {Shahaf}, {Siddiqui}, {Siebert}, {Siltala}, {Slezak}, {Smart},
  {Solano}, {Solitro}, {Souami}, {Souchay}, {Spagna}, {Spoto}, {Steele},
  {Steidelm{\"u}ller}, {Stephenson}, {S{\"u}veges}, {Szabados}, {Szegedi-Elek},
  {Taris}, {Tauran}, {Taylor}, {Teixeira}, {Thuillot}, {Tonello}, {Torra},
  {Torra}, {Turon}, {Unger}, {Vaillant}, {van Dillen}, {Vanel}, {Vecchiato},
  {Viala}, {Vicente}, {Voutsinas}, {Weiler}, {Wevers}, {Wyrzykowski}, {Yoldas},
  {Yvard}, {Zhao}, {Zorec}, {Zucker}, {Zurbach}, \& {Zwitter}}]{GaiaEDR3}
{Gaia Collaboration}, {Brown}, A.~G.~A., {Vallenari}, A., {et~al.} 2021, \aap,
  649, A1

\bibitem[{{Gaia Collaboration} {et~al.}(2016){Gaia Collaboration}, {Brown},
  {Vallenari}, {Prusti}, {de Bruijne}, {Mignard}, {Drimmel}, {Babusiaux},
  {Bailer-Jones}, {Bastian}, {Biermann}, {Evans}, {Eyer}, {Jansen}, {Jordi},
  {Katz}, {Klioner}, {Lammers}, {Lindegren}, {Luri}, {O'Mullane}, {Panem},
  {Pourbaix}, {Randich}, {Sartoretti}, {Siddiqui}, {Soubiran}, {Valette}, {van
  Leeuwen}, {Walton}, {Aerts}, {Arenou}, {Cropper}, {H{\o}g}, {Lattanzi},
  {Grebel}, {Holland}, {Huc}, {Passot}, {Perryman}, {Bramante}, {Cacciari},
  {Casta{\~n}eda}, {Chaoul}, {Cheek}, {De Angeli}, {Fabricius}, {Guerra},
  {Hern{\'a}ndez}, {Jean-Antoine-Piccolo}, {Masana}, {Messineo}, {Mowlavi},
  {Nienartowicz}, {Ord{\'o}{\~n}ez-Blanco}, {Panuzzo}, {Portell}, {Richards},
  {Riello}, {Seabroke}, {Tanga}, {Th{\'e}venin}, {Torra}, {Els},
  {Gracia-Abril}, {Comoretto}, {Garcia-Reinaldos}, {Lock}, {Mercier},
  {Altmann}, {Andrae}, {Astraatmadja}, {Bellas-Velidis}, {Benson}, {Berthier},
  {Blomme}, {Busso}, {Carry}, {Cellino}, {Clementini}, {Cowell}, {Creevey},
  {Cuypers}, {Davidson}, {De Ridder}, {de Torres}, {Delchambre}, {Dell'Oro},
  {Ducourant}, {Fr{\'e}mat}, {Garc{\'\i}a-Torres}, {Gosset}, {Halbwachs},
  {Hambly}, {Harrison}, {Hauser}, {Hestroffer}, {Hodgkin}, {Huckle}, {Hutton},
  {Jasniewicz}, {Jordan}, {Kontizas}, {Korn}, {Lanzafame}, {Manteiga},
  {Moitinho}, {Muinonen}, {Osinde}, {Pancino}, {Pauwels}, {Petit},
  {Recio-Blanco}, {Robin}, {Sarro}, {Siopis}, {Smith}, {Smith}, {Sozzetti},
  {Thuillot}, {van Reeven}, {Viala}, {Abbas}, {Abreu Aramburu}, {Accart},
  {Aguado}, {Allan}, {Allasia}, {Altavilla}, {{\'A}lvarez}, {Alves},
  {Anderson}, {Andrei}, {Anglada Varela}, {Antiche}, {Antoja}, {Ant{\'o}n},
  {Arcay}, {Bach}, {Baker}, {Balaguer-N{\'u}{\~n}ez}, {Barache}, {Barata},
  {Barbier}, {Barblan}, {Barrado y Navascu{\'e}s}, {Barros}, {Barstow},
  {Becciani}, {Bellazzini}, {Bello Garc{\'\i}a}, {Belokurov}, {Bendjoya},
  {Berihuete}, {Bianchi}, {Bienaym{\'e}}, {Billebaud}, {Blagorodnova},
  {Blanco-Cuaresma}, {Boch}, {Bombrun}, {Borrachero}, {Bouquillon}, {Bourda},
  {Bouy}, {Bragaglia}, {Breddels}, {Brouillet}, {Br{\"u}semeister},
  {Bucciarelli}, {Burgess}, {Burgon}, {Burlacu}, {Busonero}, {Buzzi}, {Caffau},
  {Cambras}, {Campbell}, {Cancelliere}, {Cantat-Gaudin}, {Carlucci},
  {Carrasco}, {Castellani}, {Charlot}, {Charnas}, {Chiavassa}, {Clotet},
  {Cocozza}, {Collins}, {Costigan}, {Crifo}, {Cross}, {Crosta}, {Crowley},
  {Dafonte}, {Damerdji}, {Dapergolas}, {David}, {David}, {De Cat}, {de Felice},
  {de Laverny}, {De Luise}, {De March}, {de Martino}, {de Souza}, {Debosscher},
  {del Pozo}, {Delbo}, {Delgado}, {Delgado}, {Di Matteo}, {Diakite},
  {Distefano}, {Dolding}, {Dos Anjos}, {Drazinos}, {Duran}, {Dzigan},
  {Edvardsson}, {Enke}, {Evans}, {Eynard Bontemps}, {Fabre}, {Fabrizio},
  {Faigler}, {Falc{\~a}o}, {Farr{\`a}s Casas}, {Federici}, {Fedorets},
  {Fern{\'a}ndez-Hern{\'a}ndez}, {Fernique}, {Fienga}, {Figueras}, {Filippi},
  {Findeisen}, {Fonti}, {Fouesneau}, {Fraile}, {Fraser}, {Fuchs}, {Gai},
  {Galleti}, {Galluccio}, {Garabato}, {Garc{\'\i}a-Sedano}, {Garofalo},
  {Garralda}, {Gavras}, {Gerssen}, {Geyer}, {Gilmore}, {Girona}, {Giuffrida},
  {Gomes}, {Gonz{\'a}lez-Marcos}, {Gonz{\'a}lez-N{\'u}{\~n}ez},
  {Gonz{\'a}lez-Vidal}, {Granvik}, {Guerrier}, {Guillout}, {Guiraud},
  {G{\'u}rpide}, {Guti{\'e}rrez-S{\'a}nchez}, {Guy}, {Haigron},
  {Hatzidimitriou}, {Haywood}, {Heiter}, {Helmi}, {Hobbs}, {Hofmann}, {Holl},
  {Holland}, {Hunt}, {Hypki}, {Icardi}, {Irwin}, {Jevardat de Fombelle},
  {Jofr{\'e}}, {Jonker}, {Jorissen}, {Julbe}, {Karampelas}, {Kochoska},
  {Kohley}, {Kolenberg}, {Kontizas}, {Koposov}, {Kordopatis}, {Koubsky},
  {Krone-Martins}, {Kudryashova}, {Kull}, {Bachchan}, {Lacoste-Seris}, {Lanza},
  {Lavigne}, {Le Poncin-Lafitte}, {Lebreton}, {Lebzelter}, {Leccia}, {Leclerc},
  {Lecoeur-Taibi}, {Lemaitre}, {Lenhardt}, {Leroux}, {Liao}, {Licata},
  {Lindstr{\o}m}, {Lister}, {Livanou}, {Lobel}, {L{\"o}ffler}, {L{\'o}pez},
  {Lorenz}, {MacDonald}, {Magalh{\~a}es Fernandes}, {Managau}, {Mann},
  {Mantelet}, {Marchal}, {Marchant}, {Marconi}, {Marinoni}, {Marrese},
  {Marschalk{\'o}}, {Marshall}, {Mart{\'\i}n-Fleitas}, {Martino}, {Mary},
  {Matijevi{\v{c}}}, {Mazeh}, {McMillan}, {Messina}, {Michalik}, {Millar},
  {Miranda}, {Molina}, {Molinaro}, {Molinaro}, {Moln{\'a}r}, {Moniez},
  {Montegriffo}, {Mor}, {Mora}, {Morbidelli}, {Morel}, {Morgenthaler},
  {Morris}, {Mulone}, {Muraveva}, {Musella}, {Narbonne}, {Nelemans},
  {Nicastro}, {Noval}, {Ord{\'e}novic}, {Ordieres-Mer{\'e}}, {Osborne},
  {Pagani}, {Pagano}, {Pailler}, {Palacin}, {Palaversa}, {Parsons}, {Pecoraro},
  {Pedrosa}, {Pentik{\"a}inen}, {Pichon}, {Piersimoni}, {Pineau}, {Plachy},
  {Plum}, {Poujoulet}, {Pr{\v{s}}a}, {Pulone}, {Ragaini}, {Rago}, {Rambaux},
  {Ramos-Lerate}, {Ranalli}, {Rauw}, {Read}, {Regibo}, {Reyl{\'e}}, {Ribeiro},
  {Rimoldini}, {Ripepi}, {Riva}, {Rixon}, {Roelens}, {Romero-G{\'o}mez},
  {Rowell}, {Royer}, {Ruiz-Dern}, {Sadowski}, {Sagrist{\`a} Sell{\'e}s},
  {Sahlmann}, {Salgado}, {Salguero}, {Sarasso}, {Savietto}, {Schultheis},
  {Sciacca}, {Segol}, {Segovia}, {Segransan}, {Shih}, {Smareglia}, {Smart},
  {Solano}, {Solitro}, {Sordo}, {Soria Nieto}, {Souchay}, {Spagna}, {Spoto},
  {Stampa}, {Steele}, {Steidelm{\"u}ller}, {Stephenson}, {Stoev}, {Suess},
  {S{\"u}veges}, {Surdej}, {Szabados}, {Szegedi-Elek}, {Tapiador}, {Taris},
  {Tauran}, {Taylor}, {Teixeira}, {Terrett}, {Tingley}, {Trager}, {Turon},
  {Ulla}, {Utrilla}, {Valentini}, {van Elteren}, {Van Hemelryck}, {van
  Leeuwen}, {Varadi}, {Vecchiato}, {Veljanoski}, {Via}, {Vicente}, {Vogt},
  {Voss}, {Votruba}, {Voutsinas}, {Walmsley}, {Weiler}, {Weingrill}, {Wevers},
  {Wyrzykowski}, {Yoldas}, {{\v{Z}}erjal}, {Zucker}, {Zurbach}, {Zwitter},
  {Alecu}, {Allen}, {Allende Prieto}, {Amorim}, {Anglada-Escud{\'e}},
  {Arsenijevic}, {Azaz}, {Balm}, {Beck}, {Bernstein}, {Bigot}, {Bijaoui},
  {Blasco}, {Bonfigli}, {Bono}, {Boudreault}, {Bressan}, {Brown}, {Brunet},
  {Bunclark}, {Buonanno}, {Butkevich}, {Carret}, {Carrion}, {Chemin},
  {Ch{\'e}reau}, {Corcione}, {Darmigny}, {de Boer}, {de Teodoro}, {de Zeeuw},
  {Delle Luche}, {Domingues}, {Dubath}, {Fodor}, {Fr{\'e}zouls}, {Fries},
  {Fustes}, {Fyfe}, {Gallardo}, {Gallegos}, {Gardiol}, {Gebran}, {Gomboc},
  {G{\'o}mez}, {Grux}, {Gueguen}, {Heyrovsky}, {Hoar}, {Iannicola}, {Isasi
  Parache}, {Janotto}, {Joliet}, {Jonckheere}, {Keil}, {Kim}, {Klagyivik},
  {Klar}, {Knude}, {Kochukhov}, {Kolka}, {Kos}, {Kutka}, {Lainey}, {LeBouquin},
  {Liu}, {Loreggia}, {Makarov}, {Marseille}, {Martayan}, {Martinez-Rubi},
  {Massart}, {Meynadier}, {Mignot}, {Munari}, {Nguyen}, {Nordlander}, {Ocvirk},
  {O'Flaherty}, {Olias Sanz}, {Ortiz}, {Osorio}, {Oszkiewicz}, {Ouzounis},
  {Palmer}, {Park}, {Pasquato}, {Peltzer}, {Peralta}, {P{\'e}turaud},
  {Pieniluoma}, {Pigozzi}, {Poels}, {Prat}, {Prod'homme}, {Raison}, {Rebordao},
  {Risquez}, {Rocca-Volmerange}, {Rosen}, {Ruiz-Fuertes}, {Russo}, {Sembay},
  {Serraller Vizcaino}, {Short}, {Siebert}, {Silva}, {Sinachopoulos}, {Slezak},
  {Soffel}, {Sosnowska}, {Strai{\v{z}}ys}, {ter Linden}, {Terrell}, {Theil},
  {Tiede}, {Troisi}, {Tsalmantza}, {Tur}, {Vaccari}, {Vachier}, {Valles}, {Van
  Hamme}, {Veltz}, {Virtanen}, {Wallut}, {Wichmann}, {Wilkinson}, {Ziaeepour},
  \& {Zschocke}}]{GaiaDR1}
{Gaia Collaboration}, {Brown}, A.~G.~A., {Vallenari}, A., {et~al.} 2016, \aap,
  595, A2

\bibitem[{{Galli} {et~al.}(2021){Galli}, {Bouy}, {Olivares}, {Miret-Roig},
  {Sarro}, {Barrado}, {Berihuete}, {Bertin}, \& {Cuillandre}}]{Galli2021}
{Galli}, P.~A.~B., {Bouy}, H., {Olivares}, J., {et~al.} 2021, \aap, 646, A46

\bibitem[{{Galli} {et~al.}(2020{\natexlab{a}}){Galli}, {Bouy}, {Olivares},
  {Miret-Roig}, {Sarro}, {Barrado}, {Berihuete}, \& {Brandner}}]{Galli2020a}
{Galli}, P.~A.~B., {Bouy}, H., {Olivares}, J., {et~al.} 2020{\natexlab{a}},
  \aap, 634, A98

\bibitem[{{Galli} {et~al.}(2020{\natexlab{b}}){Galli}, {Bouy}, {Olivares},
  {Miret-Roig}, {Vieira}, {Sarro}, {Barrado}, {Berihuete}, {Bertout}, {Bertin},
  \& {Cuillandre}}]{Galli2020b}
{Galli}, P.~A.~B., {Bouy}, H., {Olivares}, J., {et~al.} 2020{\natexlab{b}},
  \aap, 643, A148

\bibitem[{{Galli} {et~al.}(2019){Galli}, {Loinard}, {Bouy}, {Sarro},
  {Ortiz-Le{\'o}n}, {Dzib}, {Olivares}, {Heyer}, {Hernandez},
  {Rom{\'a}n-Z{\'u}{\~n}iga}, {Kounkel}, \& {Covey}}]{Galli2019}
{Galli}, P.~A.~B., {Loinard}, L., {Bouy}, H., {et~al.} 2019, \aap, 630, A137

\bibitem[{{Gentile Fusillo} {et~al.}(2019){Gentile Fusillo}, {Tremblay},
  {G{\"a}nsicke}, {Manser}, {Cunningham}, {Cukanovaite}, {Hollands}, {Marsh},
  {Raddi}, {Jordan}, {Toonen}, {Geier}, {Barstow}, \& {Cummings}}]{Fusillo2019}
{Gentile Fusillo}, N.~P., {Tremblay}, P.-E., {G{\"a}nsicke}, B.~T., {et~al.}
  2019, \mnras, 482, 4570

\bibitem[{{Gontcharov}(2006)}]{Gontcharov2006}
{Gontcharov}, G.~A. 2006, Astronomy Letters, 32, 759

\bibitem[{{Henden} {et~al.}(2015){Henden}, {Levine}, {Terrell}, \&
  {Welch}}]{APASS}
{Henden}, A.~A., {Levine}, S., {Terrell}, D., \& {Welch}, D.~L. 2015, in
  American Astronomical Society Meeting Abstracts, Vol. 225, American
  Astronomical Society Meeting Abstracts \#225, 336.16

\bibitem[{{H{\o}g} {et~al.}(2000){H{\o}g}, {Fabricius}, {Makarov}, {Urban},
  {Corbin}, {Wycoff}, {Bastian}, {Schwekendiek}, \& {Wicenec}}]{TYCHO2}
{H{\o}g}, E., {Fabricius}, C., {Makarov}, V.~V., {et~al.} 2000, \aap, 355, L27

\bibitem[{{Houk}(1982)}]{Houk1982}
{Houk}, N. 1982, {Michigan Catalogue of Two-dimensional Spectral Types for the
  HD stars. Volume\_3. Declinations -40 to -26.}

\bibitem[{{Irrgang} {et~al.}(2013){Irrgang}, {Wilcox}, {Tucker}, \&
  {Schiefelbein}}]{Irrgang2013}
{Irrgang}, A., {Wilcox}, B., {Tucker}, E., \& {Schiefelbein}, L. 2013, \aap,
  549, A137

\bibitem[{{Jackson} \& {Stoy}(1955)}]{Jackson1955}
{Jackson}, J. \& {Stoy}, R.~H. 1955, Annals of the Cape Observatory, 18, 0

\bibitem[{{Johnson} \& {Soderblom}(1987)}]{Johnson1987}
{Johnson}, D. R.~H. \& {Soderblom}, D.~R. 1987, \aj, 93, 864

\bibitem[{{Kaufer} {et~al.}(1999){Kaufer}, {Stahl}, {Tubbesing},
  {N{\o}rregaard}, {Avila}, {Francois}, {Pasquini}, \& {Pizzella}}]{FEROS}
{Kaufer}, A., {Stahl}, O., {Tubbesing}, S., {et~al.} 1999, The Messenger, 95, 8

\bibitem[{{Kesseli} {et~al.}(2017){Kesseli}, {West}, {Veyette}, {Harrison},
  {Feldman}, \& {Bochanski}}]{Kesseli2017}
{Kesseli}, A.~Y., {West}, A.~A., {Veyette}, M., {et~al.} 2017, \apjs, 230, 16

\bibitem[{{Kharchenko} {et~al.}(2013){Kharchenko}, {Piskunov}, {Schilbach},
  {R{\"o}ser}, \& {Scholz}}]{Kharchenko2013}
{Kharchenko}, N.~V., {Piskunov}, A.~E., {Schilbach}, E., {R{\"o}ser}, S., \&
  {Scholz}, R.~D. 2013, \aap, 558, A53

\bibitem[{{King}(1962)}]{King1962}
{King}, I. 1962, \aj, 67, 471

\bibitem[{{Koenig} \& {Leisawitz}(2014)}]{Koenig2014}
{Koenig}, X.~P. \& {Leisawitz}, D.~T. 2014, \apj, 791, 131

\bibitem[{{Kordopatis} {et~al.}(2013){Kordopatis}, {Gilmore}, {Steinmetz},
  {Boeche}, {Seabroke}, {Siebert}, {Zwitter}, {Binney}, {de Laverny},
  {Recio-Blanco}, {Williams}, {Piffl}, {Enke}, {Roeser}, {Bijaoui}, {Wyse},
  {Freeman}, {Munari}, {Carrillo}, {Anguiano}, {Burton}, {Campbell}, {Cass},
  {Fiegert}, {Hartley}, {Parker}, {Reid}, {Ritter}, {Russell}, {Stupar},
  {Watson}, {Bienaym{\'e}}, {Bland-Hawthorn}, {Gerhard}, {Gibson}, {Grebel},
  {Helmi}, {Navarro}, {Conrad}, {Famaey}, {Faure}, {Just}, {Kos},
  {Matijevi{\v{c}}}, {McMillan}, {Minchev}, {Scholz}, {Sharma}, {Siviero}, {de
  Boer}, \& {{\v{Z}}erjal}}]{Kordopatis2013}
{Kordopatis}, G., {Gilmore}, G., {Steinmetz}, M., {et~al.} 2013, \aj, 146, 134

\bibitem[{{Kraus} {et~al.}(2014){Kraus}, {Shkolnik}, {Allers}, \&
  {Liu}}]{Kraus2014}
{Kraus}, A.~L., {Shkolnik}, E.~L., {Allers}, K.~N., \& {Liu}, M.~C. 2014, \aj,
  147, 146

\bibitem[{{Lallement} {et~al.}(2019){Lallement}, {Babusiaux}, {Vergely},
  {Katz}, {Arenou}, {Valette}, {Hottier}, \& {Capitanio}}]{Lallement2019}
{Lallement}, R., {Babusiaux}, C., {Vergely}, J.~L., {et~al.} 2019, \aap, 625,
  A135

\bibitem[{{Lindegren} {et~al.}(2021{\natexlab{a}}){Lindegren}, {Bastian},
  {Biermann}, {Bombrun}, {de Torres}, {Gerlach}, {Geyer}, {Hern{\'a}ndez},
  {Hilger}, {Hobbs}, {Klioner}, {Lammers}, {McMillan}, {Ramos-Lerate},
  {Steidelm{\"u}ller}, {Stephenson}, \& {van Leeuwen}}]{Lindegren2021a}
{Lindegren}, L., {Bastian}, U., {Biermann}, M., {et~al.} 2021{\natexlab{a}},
  \aap, 649, A4

\bibitem[{{Lindegren} {et~al.}(2021{\natexlab{b}}){Lindegren}, {Klioner},
  {Hern{\'a}ndez}, {Bombrun}, {Ramos-Lerate}, {Steidelm{\"u}ller}, {Bastian},
  {Biermann}, {de Torres}, {Gerlach}, {Geyer}, {Hilger}, {Hobbs}, {Lammers},
  {McMillan}, {Stephenson}, {Casta{\~n}eda}, {Davidson}, {Fabricius},
  {Gracia-Abril}, {Portell}, {Rowell}, {Teyssier}, {Torra}, {Bartolom{\'e}},
  {Clotet}, {Garralda}, {Gonz{\'a}lez-Vidal}, {Torra}, {Abbas}, {Altmann},
  {Anglada Varela}, {Balaguer-N{\'u}{\~n}ez}, {Balog}, {Barache}, {Becciani},
  {Bernet}, {Bertone}, {Bianchi}, {Bouquillon}, {Brown}, {Bucciarelli},
  {Busonero}, {Butkevich}, {Buzzi}, {Cancelliere}, {Carlucci}, {Charlot},
  {Cioni}, {Crosta}, {Crowley}, {del Peloso}, {del Pozo}, {Drimmel}, {Esquej},
  {Fienga}, {Fraile}, {Gai}, {Garcia-Reinaldos}, {Guerra}, {Hambly}, {Hauser},
  {Jan{\ss}en}, {Jordan}, {Kostrzewa-Rutkowska}, {Lattanzi}, {Liao}, {Licata},
  {Lister}, {L{\"o}ffler}, {Marchant}, {Masip}, {Mignard}, {Mints}, {Molina},
  {Mora}, {Morbidelli}, {Murphy}, {Pagani}, {Panuzzo}, {Pe{\~n}alosa Esteller},
  {Poggio}, {Re Fiorentin}, {Riva}, {Sagrist{\`a} Sell{\'e}s}, {Sanchez
  Gimenez}, {Sarasso}, {Sciacca}, {Siddiqui}, {Smart}, {Souami}, {Spagna},
  {Steele}, {Taris}, {Utrilla}, {van Reeven}, \& {Vecchiato}}]{Lindegren2021b}
{Lindegren}, L., {Klioner}, S.~A., {Hern{\'a}ndez}, J., {et~al.}
  2021{\natexlab{b}}, \aap, 649, A2

\bibitem[{{Luhman} {et~al.}(2018){Luhman}, {Herrmann}, {Mamajek}, {Esplin}, \&
  {Pecaut}}]{Luhman2018}
{Luhman}, K.~L., {Herrmann}, K.~A., {Mamajek}, E.~E., {Esplin}, T.~L., \&
  {Pecaut}, M.~J. 2018, \aj, 156, 76

\bibitem[{{Luri} {et~al.}(2018){Luri}, {Brown}, {Sarro}, {Arenou},
  {Bailer-Jones}, {Castro-Ginard}, {de Bruijne}, {Prusti}, {Babusiaux}, \&
  {Delgado}}]{Luri2018}
{Luri}, X., {Brown}, A.~G.~A., {Sarro}, L.~M., {et~al.} 2018, \aap, 616, A9

\bibitem[{{Mamajek}(2016)}]{Mamajek2016}
{Mamajek}, E.~E. 2016, in Young Stars \& Planets Near the Sun, ed. J.~H.
  {Kastner}, B.~{Stelzer}, \& S.~A. {Metchev}, Vol. 314, 21--26

\bibitem[{{Meingast} {et~al.}(2019){Meingast}, {Alves}, \&
  {F{\"u}rnkranz}}]{Meingast2019}
{Meingast}, S., {Alves}, J., \& {F{\"u}rnkranz}, V. 2019, \aap, 622, L13

\bibitem[{{Miret-Roig} {et~al.}(2018){Miret-Roig}, {Antoja},
  {Romero-G{\'o}mez}, \& {Figueras}}]{MiretRoig2018}
{Miret-Roig}, N., {Antoja}, T., {Romero-G{\'o}mez}, M., \& {Figueras}, F. 2018,
  \aap, 615, A51

\bibitem[{{Miret-Roig} {et~al.}(2019){Miret-Roig}, {Bouy}, {Olivares}, {Sarro},
  {Tamura}, {Allen}, {Bertin}, {Serre}, {Berihuete}, {Beletsky}, {Barrado},
  {Hu{\'e}lamo}, {Cuillandre}, {Moraux}, \& {Bouvier}}]{Miret-Roig2019}
{Miret-Roig}, N., {Bouy}, H., {Olivares}, J., {et~al.} 2019, \aap, 631, A57

\bibitem[{{Miret-Roig} {et~al.}(2020){Miret-Roig}, {Galli}, {Brandner}, {Bouy},
  {Barrado}, {Olivares}, {Antoja}, {Romero-G{\'o}mez}, {Figueras}, \&
  {Lillo-Box}}]{MiretRoig2020}
{Miret-Roig}, N., {Galli}, P.~A.~B., {Brandner}, W., {et~al.} 2020, \aap, 642,
  A179

\bibitem[{{Muench} {et~al.}(2007){Muench}, {Lada}, {Luhman}, {Muzerolle}, \&
  {Young}}]{Muench2007}
{Muench}, A.~A., {Lada}, C.~J., {Luhman}, K.~L., {Muzerolle}, J., \& {Young},
  E. 2007, \aj, 134, 411

\bibitem[{{Olivares} {et~al.}(2019){Olivares}, {Bouy}, {Sarro}, {Miret-Roig},
  {Berihuete}, {Bertin}, {Barrado}, {Hu{\'e}lamo}, {Tamura}, {Allen},
  {Beletsky}, {Serre}, \& {Cuillandre}}]{Olivares2019}
{Olivares}, J., {Bouy}, H., {Sarro}, L.~M., {et~al.} 2019, \aap, 625, A115

\bibitem[{{Olivares} {et~al.}(2020){Olivares}, {Sarro}, {Bouy}, {Miret-Roig},
  {Casamiquela}, {Galli}, {Berihuete}, \& {Tarricq}}]{Olivares2020}
{Olivares}, J., {Sarro}, L.~M., {Bouy}, H., {et~al.} 2020, \aap, 644, A7

\bibitem[{{Pecaut} \& {Mamajek}(2013)}]{Pecaut2013}
{Pecaut}, M.~J. \& {Mamajek}, E.~E. 2013, \apjs, 208, 9

\bibitem[{{Perryman} {et~al.}(1998){Perryman}, {Brown}, {Lebreton}, {Gomez},
  {Turon}, {Cayrel de Strobel}, {Mermilliod}, {Robichon}, {Kovalevsky}, \&
  {Crifo}}]{Perryman1998}
{Perryman}, M.~A.~C., {Brown}, A.~G.~A., {Lebreton}, Y., {et~al.} 1998, \aap,
  331, 81

\bibitem[{{Riello} {et~al.}(2021){Riello}, {De Angeli}, {Evans}, {Montegriffo},
  {Carrasco}, {Busso}, {Palaversa}, {Burgess}, {Diener}, {Davidson}, {Rowell},
  {Fabricius}, {Jordi}, {Bellazzini}, {Pancino}, {Harrison}, {Cacciari}, {van
  Leeuwen}, {Hambly}, {Hodgkin}, {Osborne}, {Altavilla}, {Barstow}, {Brown},
  {Castellani}, {Cowell}, {De Luise}, {Gilmore}, {Giuffrida}, {Hidalgo},
  {Holland}, {Marinoni}, {Pagani}, {Piersimoni}, {Pulone}, {Ragaini}, {Rainer},
  {Richards}, {Sanna}, {Walton}, {Weiler}, \& {Yoldas}}]{Riello2021}
{Riello}, M., {De Angeli}, F., {Evans}, D.~W., {et~al.} 2021, \aap, 649, A3

\bibitem[{{Rivera} {et~al.}(2015){Rivera}, {Loinard}, {Dzib}, {Ortiz-Le{\'o}n},
  {Rodr{\'\i}guez}, \& {Torres}}]{Rivera2015}
{Rivera}, J.~L., {Loinard}, L., {Dzib}, S.~A., {et~al.} 2015, \apj, 807, 119

\bibitem[{{Roulston} {et~al.}(2020){Roulston}, {Green}, \&
  {Kesseli}}]{Roulston2020}
{Roulston}, B.~R., {Green}, P.~J., \& {Kesseli}, A.~Y. 2020, \apjs, 249, 34

\bibitem[{{Sarro} {et~al.}(2014){Sarro}, {Bouy}, {Berihuete}, {Bertin},
  {Moraux}, {Bouvier}, {Cuillandre}, {Barrado}, \& {Solano}}]{Sarro2014}
{Sarro}, L.~M., {Bouy}, H., {Berihuete}, A., {et~al.} 2014, \aap, 563, A45

\bibitem[{Schwarz(1978)}]{Schwarz1978}
Schwarz, G. 1978, The Annals of Statistics, 6, 461

\bibitem[{{Shkolnik} {et~al.}(2017){Shkolnik}, {Allers}, {Kraus}, {Liu}, \&
  {Flagg}}]{Shkolnik2017}
{Shkolnik}, E.~L., {Allers}, K.~N., {Kraus}, A.~L., {Liu}, M.~C., \& {Flagg},
  L. 2017, \aj, 154, 69

\bibitem[{{Siess} {et~al.}(2000){Siess}, {Dufour}, \& {Forestini}}]{SDF00}
{Siess}, L., {Dufour}, E., \& {Forestini}, M. 2000, \aap, 358, 593

\bibitem[{{Sim} {et~al.}(2019){Sim}, {Lee}, {Ann}, \& {Kim}}]{Sim2019}
{Sim}, G., {Lee}, S.~H., {Ann}, H.~B., \& {Kim}, S. 2019, Journal of Korean
  Astronomical Society, 52, 145

\bibitem[{{Torres} {et~al.}(2006){Torres}, {Quast}, {da Silva}, {de La Reza},
  {Melo}, \& {Sterzik}}]{Torres2006}
{Torres}, C.~A.~O., {Quast}, G.~R., {da Silva}, L., {et~al.} 2006, \aap, 460,
  695

\bibitem[{{Wang} \& {Chen}(2019)}]{Wang2019}
{Wang}, S. \& {Chen}, X. 2019, \apj, 877, 116

\bibitem[{{Wright} {et~al.}(2010){Wright}, {Eisenhardt}, {Mainzer}, {Ressler},
  {Cutri}, {Jarrett}, {Kirkpatrick}, {Padgett}, {McMillan}, {Skrutskie},
  {Stanford}, {Cohen}, {Walker}, {Mather}, {Leisawitz}, {Gautier}, {McLean},
  {Benford}, {Lonsdale}, {Blain}, {Mendez}, {Irace}, {Duval}, {Liu}, {Royer},
  {Heinrichsen}, {Howard}, {Shannon}, {Kendall}, {Walsh}, {Larsen}, {Cardon},
  {Schick}, {Schwalm}, {Abid}, {Fabinsky}, {Naes}, \& {Tsai}}]{AllWISE}
{Wright}, E.~L., {Eisenhardt}, P. R.~M., {Mainzer}, A.~K., {et~al.} 2010, \aj,
  140, 1868

\bibitem[{{Yen} {et~al.}(2018){Yen}, {Reffert}, {Schilbach}, {R{\"o}ser},
  {Kharchenko}, \& {Piskunov}}]{Yen2018}
{Yen}, S.~X., {Reffert}, S., {Schilbach}, E., {et~al.} 2018, \aap, 615, A12

\bibitem[{{Zacharias} {et~al.}(2013){Zacharias}, {Finch}, {Girard}, {Henden},
  {Bartlett}, {Monet}, \& {Zacharias}}]{Zacharias2013}
{Zacharias}, N., {Finch}, C.~T., {Girard}, T.~M., {et~al.} 2013, \aj, 145, 44

\bibitem[{{Zuckerman} {et~al.}(2019){Zuckerman}, {Klein}, \&
  {Kastner}}]{Zuckerman2019}
{Zuckerman}, B., {Klein}, B., \& {Kastner}, J. 2019, \apj, 887, 87 (ZKK2019)

\end{thebibliography}

\begin{appendix}
\begin{landscape}

\section{Tables (online material)}\label{appendix_tables}

\begin{table}[!h]
\caption{Properties of the 164 cluster members selected from our membership analysis. (This table will be available in its entirety in machine-readable form.) }\label{tab_members}
\scriptsize{
\begin{tabular}{lccccccccccccccccc}
\hline\hline
Source identifier&Other name&$\alpha$&$\delta$&$\mu_{\alpha}\cos\delta$&$\mu_{\delta}$&$\varpi$&Prob.&$V_{r}$&Ref.&ST&Ref.&$d$&$U$&$V$&$W$&SED\\
&&(h:m:s) &($^{\circ}$ $^\prime$ $^\prime$$^\prime$)&(mas/yr)&(mas/yr)&(mas)&&(km/s)&&&&(pc)&(km/s)&(km/s)&(km/s)&\\
\hline\hline

\textit{Gaia} EDR3 5062101682598708864 & 023851-345729 & 02 38 51.45 & -34 57 28.9 & $ 42.021 \pm 0.007 $& $ -1.253 \pm 0.011 $& $ 10.494 \pm 0.011 $& $ 0.86 $& $ 18.91 \pm 1.27 $& 1& \nodata & \nodata & $ 95.3 ^{+ 0.1 }_{ -0.1 } $& $ -15.7 ^{+ 0.3 }_{ -0.3 } $& $ -19.6 ^{+ 0.5 }_{ -0.5 } $& $ -9.4 ^{+ 1.2 }_{ -1.2 } $& III \\
\textit{Gaia} EDR3 5050023753524273792 & 023933-354315 & 02 39 33.13 & -35 43 14.7 & $ 44.876 \pm 0.015 $& $ -11.135 \pm 0.020 $& $ 10.915 \pm 0.021 $& $ 0.89 $& \nodata & \nodata & \nodata & \nodata & $ 91.6 ^{+ 0.2 }_{ -0.2 } $& \nodata & \nodata & \nodata & III \\
\textit{Gaia} EDR3 4953577617192474880 & 024133-372137 & 02 41 33.12 & -37 21 37.4 & $ 36.154 \pm 0.077 $& $ -5.297 \pm 0.096 $& $ 9.812 \pm 0.096 $& $ 0.99 $& \nodata & \nodata & \nodata & \nodata & $ 102.0 ^{+ 1.0 }_{ -1.0 } $& \nodata & \nodata & \nodata & III \\
\textit{Gaia} EDR3 4948563569292191104 & 024322-415108 & 02 43 21.86 & -41 51 08.3 & $ 34.097 \pm 0.012 $& $ -2.023 \pm 0.012 $& $ 9.171 \pm 0.012 $& $ 0.93 $& $ 14.57 \pm 0.68 $& 1& G3 & 1 & $ 109.1 ^{+ 0.2 }_{ -0.2 } $& $ -12.1 ^{+ 0.1 }_{ -0.1 } $& $ -18.7 ^{+ 0.3 }_{ -0.3 } $& $ -5.3 ^{+ 0.6 }_{ -0.6 } $& III \\
\textit{Gaia} EDR3 5049075081147607936 & 024941-360443 & 02 49 40.89 & -36  04 42.7 & $ 42.409 \pm 0.061 $& $ -8.961 \pm 0.085 $& $ 10.391 \pm 0.079 $& $ 0.97 $& \nodata & \nodata & \nodata & \nodata & $ 96.3 ^{+ 0.7 }_{ -0.7 } $& \nodata & \nodata & \nodata & III \\
\textit{Gaia} EDR3 4949158198924394496 & 025048-395556 & 02 50 47.86 & -39 55 56.2 & $ 41.888 \pm 0.023 $& $ -9.956 \pm 0.026 $& $ 10.383 \pm 0.027 $& $ 0.99 $& \nodata & \nodata & A3 & 1 & $ 96.3 ^{+ 0.3 }_{ -0.3 } $& \nodata & \nodata & \nodata & III \\
\textit{Gaia} EDR3 4949158198924393600 & 025051-395611 & 02 50 50.51 & -39 56 10.9 & $ 42.846 \pm 0.027 $& $ -6.021 \pm 0.033 $& $ 10.436 \pm 0.031 $& $ 0.99 $& \nodata & \nodata & M6 & 1 & $ 95.9 ^{+ 0.3 }_{ -0.3 } $& \nodata & \nodata & \nodata & \nodata \\
\textit{Gaia} EDR3 4755198025592857216 & 025251-435845 & 02 52 51.09 & -43 58 44.9 & $ 39.978 \pm 0.030 $& $ -4.600 \pm 0.038 $& $ 10.066 \pm 0.033 $& $ 1.00 $& \nodata & \nodata & M7 & 1 & $ 99.4 ^{+ 0.4 }_{ -0.4 } $& \nodata & \nodata & \nodata & \nodata \\
\textit{Gaia} EDR3 4755198025592857088 & 025251-435842 & 02 52 51.19 & -43 58 41.9 & $ 40.918 \pm 0.019 $& $ -4.950 \pm 0.025 $& $ 10.087 \pm 0.021 $& $ 1.00 $& \nodata & \nodata & M5 & 1 & $ 99.2 ^{+ 0.2 }_{ -0.2 } $& \nodata & \nodata & \nodata & \nodata \\
\textit{Gaia} EDR3 5048737565438297600 & 025255-370822 & 02 52 54.94 & -37 08 22.2 & $ 43.005 \pm 0.097 $& $ -7.333 \pm 0.121 $& $ 10.443 \pm 0.119 $& $ 0.97 $& \nodata & \nodata & M8 & 1 & $ 95.8 ^{+ 1.1 }_{ -1.1 } $& \nodata & \nodata & \nodata & III \\
\textit{Gaia} EDR3 4754698091399430400 & 025259-460940 & 02 52 59.31 & -46  09 40.2 & $ 31.094 \pm 0.012 $& $ 0.674 \pm 0.015 $& $ 9.280 \pm 0.013 $& $ 0.77 $& \nodata & \nodata & \nodata & \nodata & $ 107.8 ^{+ 0.2 }_{ -0.2 } $& \nodata & \nodata & \nodata & III \\
\textit{Gaia} EDR3 4949081572411575552 & 025352-401139 & 02 53 51.88 & -40 11 38.7 & $ 42.205 \pm 0.015 $& $ -6.464 \pm 0.016 $& $ 10.387 \pm 0.017 $& $ 0.97 $& $ 18.12 \pm 0.91 $& 1& F7 & 1 & $ 96.3 ^{+ 0.2 }_{ -0.2 } $& $ -12.3 ^{+ 0.2 }_{ -0.2 } $& $ -22.6 ^{+ 0.4 }_{ -0.4 } $& $ -6.6 ^{+ 0.8 }_{ -0.8 } $& III \\
\textit{Gaia} EDR3 4949081507988380800 & 025353-401149 & 02 53 53.01 & -40 11 48.6 & $ 40.668 \pm 0.020 $& $ -7.844 \pm 0.024 $& $ 10.356 \pm 0.023 $& $ 1.00 $& \nodata & \nodata & M5 & 1 & $ 96.6 ^{+ 0.2 }_{ -0.2 } $& \nodata & \nodata & \nodata & III \\
\textit{Gaia} EDR3 5051952743596712832 & 025518-334445 & 02 55 18.06 & -33 44 44.6 & $ 42.754 \pm 0.015 $& $ -9.715 \pm 0.019 $& $ 10.512 \pm 0.021 $& $ 0.99 $& \nodata & \nodata & M5 & 1 & $ 95.1 ^{+ 0.2 }_{ -0.2 } $& \nodata & \nodata & \nodata & III \\
\textit{Gaia} EDR3 4755745925980823680 & 025902-423220 & 02 59 01.54 & -42 32 20.5 & $ 42.227 \pm 0.026 $& $ -5.219 \pm 0.030 $& $ 10.606 \pm 0.025 $& $ 0.99 $& \nodata & \nodata & M6 & 1 & $ 94.3 ^{+ 0.2 }_{ -0.3 } $& \nodata & \nodata & \nodata & II \\
\textit{Gaia} EDR3 4755745925980823296 & 025903-423245 & 02 59 03.31 & -42 32 45.0 & $ 42.134 \pm 0.016 $& $ -4.867 \pm 0.018 $& $ 10.625 \pm 0.015 $& $ 0.99 $& \nodata & \nodata & M5 & 1 & $ 94.1 ^{+ 0.2 }_{ -0.2 } $& \nodata & \nodata & \nodata & III \\
\textit{Gaia} EDR3 5046046953109557504 & 025927-371340 & 02 59 27.28 & -37 13 40.5 & $ 39.541 \pm 0.065 $& $ -6.660 \pm 0.074 $& $ 9.796 \pm 0.084 $& $ 0.99 $& \nodata & \nodata & \nodata & \nodata & $ 102.1 ^{+ 0.9 }_{ -0.8 } $& \nodata & \nodata & \nodata & III \\
\textit{Gaia} EDR3 5044692977555851904 & 030416-391913 & 03 04 15.71 & -39 19 12.9 & $ 40.744 \pm 0.032 $& $ -5.553 \pm 0.038 $& $ 10.277 \pm 0.033 $& $ 1.00 $& \nodata & \nodata & M6 & 1 & $ 97.3 ^{+ 0.3 }_{ -0.3 } $& \nodata & \nodata & \nodata & III \\
\textit{Gaia} EDR3 5058672202751604480 & 030555-311128 & 03 05 54.63 & -31 11 27.7 & $ 38.004 \pm 0.032 $& $ -8.605 \pm 0.045 $& $ 9.543 \pm 0.049 $& $ 1.00 $& \nodata & \nodata & M7 & 1 & $ 104.8 ^{+ 0.6 }_{ -0.5 } $& \nodata & \nodata & \nodata & III \\
\textit{Gaia} EDR3 5058788888423003648 & 030602-301821 & 03 06 02.00 & -30 18 20.8 & $ 37.416 \pm 0.038 $& $ -8.050 \pm 0.052 $& $ 9.239 \pm 0.058 $& $ 0.99 $& \nodata & \nodata & M7 & 1 & $ 108.2 ^{+ 0.7 }_{ -0.6 } $& \nodata & \nodata & \nodata & III \\

\hline
\end{tabular}
\tablefoot{For each star, we provide the \textit{Gaia}-EDR3 identifier, an internal identifier based on the stellar position, position, proper motion and parallax (not corrected for zero-point offset) from the \textit{Gaia}-EDR3 catalogue, membership probability, radial velocity with reference, spectral type with reference, distance derived from Bayesian inference, UVW spatial velocity, and the SED subclass. References for radial velocities are: (1)~\textit{Gaia}-EDR3 \citep{GaiaEDR3}, (2)~\citet{Gontcharov2006}, (3)~This paper, (4)~RAVE \citep{Kordopatis2013}, and (5)~\citet{Shkolnik2017}. References for spectral type are: (1)~This paper, (2)~\citet{Houk1982}, (3)~\citet{Jackson1955}, (4)~\citet{Torres2006}, and (5)~\citet{Shkolnik2017}.}
}
\end{table}
\end{landscape}
\clearpage

\begin{table*}
\centering
\caption{Membership probability of all sources in the field investigated in our membership analysis using different $p_{in}$ values. (This table will be available in its entirety in machine-readable form.)
\label{tab_prob}}
\begin{tabular}{cccccc}
\hline\hline
Source Identifier&probability&probability&probability&probability&probability\\
&($p_{in}=0.5$)&($p_{in}=0.6$)&($p_{in}=0.7$)&($p_{in}=0.8$)&($p_{in}=0.9$)\\
\hline\hline

\textit{Gaia} EDR3 4956996823477431680&2.0E-285&1.5E-286&1.2E-295&9.7E-296&9.7E-296\\
\textit{Gaia} EDR3 4956997712534324480&2.0E-52&3.1E-52&6.5E-52&3.4E-59&6.4E-75\\
\textit{Gaia} EDR3 4956996853540851456&1.3E-94&7.6E-95&4.7E-96&1.0E-100&6.2E-116\\
\textit{Gaia} EDR3 4956998163507251584&6.2E-162&7.1E-163&3.6E-166&1.8E-173&6.1E-223\\
\textit{Gaia} EDR3 4956998197866989824&1.5E-110&1.1E-111&2.3E-114&1.6E-120&1.4E-156\\
\textit{Gaia} EDR3 4956996647382420864&3.4E-139&3.9E-140&1.6E-141&1.6E-149&4.2E-168\\
\textit{Gaia} EDR3 4956996892196908544&1.3E-58&4.9E-60&1.7E-60&1.1E-66&2.1E-70\\
\textit{Gaia} EDR3 4956996892196932352&8.0E-116&4.1E-116&1.8E-116&3.3E-125&9.8E-133\\
\textit{Gaia} EDR3 4956998129146322816&2.2E-13&2.4E-13&2.1E-13&8.2E-16&7.1E-16\\
\textit{Gaia} EDR3 4956998129145886208&5.0E-40&6.8E-40&6.7E-40&3.5E-44&2.9E-47\\
\textit{Gaia} EDR3 4956997265857719168&6.6E-76&1.3E-76&2.0E-77&6.0E-82&5.2E-89\\
\textit{Gaia} EDR3 4956997304513792640&1.4E-154&6.5E-155&5.2E-159&2.7E-168&2.8E-246\\
\textit{Gaia} EDR3 4957021833070690048&5.9E-32&4.9E-32&3.0E-32&7.1E-35&1.1E-35\\
\textit{Gaia} EDR3 4956997201434577408&2.8E-50&6.1E-50&1.1E-49&1.2E-56&2.3E-68\\
\textit{Gaia} EDR3 4957022146604630784&2.1E-101&3.7E-101&1.4E-101&8.2E-113&3.5E-127\\
\textit{Gaia} EDR3 4956997201433854720&4.5E-36&1.7E-36&2.1E-36&1.6E-41&6.6E-49\\
\textit{Gaia} EDR3 4956997201434577792&4.4E-134&1.6E-134&2.6E-136&5.6E-144&9.2E-171\\
\textit{Gaia} EDR3 4957021936149913216&7.8E-172&1.0E-172&1.8E-178&1.0E-183&7.9E-280\\
\textit{Gaia} EDR3 4956997231497990272&1.0E-88&1.1E-88&1.2E-88&2.9E-97&2.0E-104\\
\textit{Gaia} EDR3 4957021871726725120&3.3E-73&3.1E-73&1.1E-74&1.3E-83&4.7E-153\\

\hline\hline
\end{tabular}
\end{table*}

\begin{table*}
\centering
\caption{Empirical isochrone inferred from our membership analysis. (This table will be available in its entirety in machine-readable form.)
\label{tab_isochrone}}
\begin{tabular}{cccc}
\hline\hline
$G_{RP}$&$G-G_{RP}$&$G_{abs}$&ST\\
(mag)&(mag)&(mag)\\
\hline\hline

5.062&-0.144&-0.257&O7.6\\
5.096&-0.139&-0.218&O7.6\\
5.129&-0.134&-0.180&O7.6\\
5.162&-0.129&-0.142&O7.6\\
5.195&-0.124&-0.103&O7.6\\
5.229&-0.119&-0.065&O7.6\\
5.262&-0.114&-0.027&O7.6\\
5.295&-0.108&0.012      &O7.6\\
5.329&-0.103&0.050      &O7.6\\
5.362&-0.098&0.088      &O7.6\\
5.395&-0.093&0.127      &O7.7\\
5.428&-0.088&0.165      &O7.7\\
5.462&-0.083&0.203      &O7.7\\
5.495&-0.078&0.242      &O7.8\\
5.528&-0.073&0.280      &O7.8\\
5.561&-0.068&0.319      &O7.9\\
5.595&-0.063&0.357      &O8.0\\
5.628&-0.058&0.395      &O8.0\\
5.661&-0.052&0.434      &O8.1\\
5.695&-0.047&0.472      &O8.2\\

\hline\hline
\end{tabular}
\tablefoot{We provide the apparent magnitude in the $G_{RP}$-band and colour used in the membership analysis (see Sect.~\ref{section2}), the absolute magnitude in the $G$-band computed at the distance of the cluster (see Sect.~\ref{section4.2}), and the spectral type (ST). We converted the colours into spectral types (based on our observations presented in Sect.~\ref{section3}) from a fourth-order polynomial fit with coefficients $c_{0}= 29.4\pm3.8$, $c_{1}=-98.6\pm11.4$, $c_{2}=86.7\pm10.7$, $c_{3}=25.3\pm3.5$, and $c_{4}=19.2\pm0.4$. We explored different degrees of the polynomial to fit the data and chose the one with lowest BIC value.}
\end{table*}

\begin{table*}
\centering
\caption{Physical properties of the cluster member derived from the SED fit with VOSA. (This table will be available in its entirety in machine-readable form.)
\label{tab_phyiscal_properties}}
\begin{tabular}{ccccc}
\hline\hline
Source identifier&$\alpha$&$\delta$&Teff&$\log(L/L_{\odot})$\\
&(h:m:s)&($^{\circ}$ $^\prime$ $^\prime$$^\prime$)&(K)&\\
\hline\hline

\textit{Gaia} EDR3 5062101682598708864 & 02 38 51.45 & -34 57 28.9 & $ 3800 \pm 50 $& $ -1.203 ^{+ 0.002 }_{ -0.002 }$ \\
\textit{Gaia} EDR3 5050023753524273792 & 02 39 33.13 & -35 43 14.7 & $ 3000 \pm 50 $& $ -1.630 ^{+ 0.004 }_{ -0.004 }$ \\
\textit{Gaia} EDR3 4953577617192474880 & 02 41 33.12 & -37 21 37.4 & $ 2800 \pm 50 $& $ -2.617 ^{+ 0.011 }_{ -0.011 }$ \\
\textit{Gaia} EDR3 4948563569292191104 & 02 43 21.86 & -41 51 08.3 & $ 6500 \pm 50 $& $ 0.138 ^{+ 0.002 }_{ -0.002 }$ \\
\textit{Gaia} EDR3 5049075081147607936 & 02 49 40.89 & -36  04 42.7 & $ 2800 \pm 50 $& $ -2.400 ^{+ 0.008 }_{ -0.009 }$ \\
\textit{Gaia} EDR3 4949158198924394496 & 02 50 47.86 & -39 55 56.2 & $ 9400 \pm 100 $& $ 1.432 ^{+ 0.003 }_{ -0.003 }$ \\
\textit{Gaia} EDR3 4949158198924393600 & 02 50 50.51 & -39 56 10.9 & $ 2900 \pm 50 $& $ -1.897 ^{+ 0.004 }_{ -0.004 }$ \\
\textit{Gaia} EDR3 4755198025592857088 & 02 52 51.19 & -43 58 41.9 & $ 3100 \pm 50 $& $ -1.673 ^{+ 0.005 }_{ -0.005 }$ \\
\textit{Gaia} EDR3 5048737565438297600 & 02 52 54.94 & -37 08 22.2 & $ 2600 \pm 50 $& $ -2.218 ^{+ 0.012 }_{ -0.012 }$ \\
\textit{Gaia} EDR3 4754698091399430400 & 02 52 59.31 & -46  09 40.2 & $ 3700 \pm 50 $& $ -1.302 ^{+ 0.003 }_{ -0.003 }$ \\
\textit{Gaia} EDR3 4949081572411575552 & 02 53 51.88 & -40 11 38.7 & $ 6700 \pm 50 $& $ 0.560 ^{+ 0.003 }_{ -0.003 }$ \\
\textit{Gaia} EDR3 4949081507988380800 & 02 53 53.01 & -40 11 48.6 & $ 3000 \pm 50 $& $ -1.677 ^{+ 0.004 }_{ -0.004 }$ \\
\textit{Gaia} EDR3 5051952743596712832 & 02 55 18.06 & -33 44 44.6 & $ 2800 \pm 50 $& $ -1.541 ^{+ 0.004 }_{ -0.004 }$ \\
\textit{Gaia} EDR3 4755745925980823680 & 02 59 01.54 & -42 32 20.5 & $ 3000 \pm 50 $& $ -1.817 ^{+ 0.004 }_{ -0.005 }$ \\
\textit{Gaia} EDR3 4755745925980823296 & 02 59 03.31 & -42 32 45.0 & $ 3200 \pm 50 $& $ -1.390 ^{+ 0.004 }_{ -0.004 }$ \\
\textit{Gaia} EDR3 5046046953109557504 & 02 59 27.28 & -37 13 40.5 & $ 2600 \pm 50 $& $ -2.375 ^{+ 0.010 }_{ -0.010 }$ \\
\textit{Gaia} EDR3 5044692977555851904 & 03 04 15.71 & -39 19 12.9 & $ 3000 \pm 50 $& $ -2.041 ^{+ 0.005 }_{ -0.005 }$ \\
\textit{Gaia} EDR3 5058672202751604480 & 03 05 54.63 & -31 11 27.7 & $ 2900 \pm 50 $& $ -1.958 ^{+ 0.006 }_{ -0.006 }$ \\
\textit{Gaia} EDR3 5058788888423003648 & 03 06 02.00 & -30 18 20.8 & $ 2800 \pm 50 $& $ -2.100 ^{+ 0.007 }_{ -0.008 }$ \\
\textit{Gaia} EDR3 4754269934699753216 & 03 09 27.16 & -45 17 18.9 & $ 3800 \pm 50 $& $ -1.221 ^{+ 0.003 }_{ -0.003 }$ \\

\hline\hline
\end{tabular}
\tablefoot{We provide for each star the \textit{Gaia}-EDR3 source identifier, position, effective temperature, and bolometric luminosity. 
}
\end{table*}

\end{appendix}
\end{document}